\documentclass[fleqn,usenatbib]{mnras}

\usepackage{newtxtext,newtxmath}

\usepackage{etoolbox}

\usepackage[T1]{fontenc}
\usepackage{ae,aecompl}

\usepackage{graphicx}	
\usepackage{amsmath}	
\usepackage{import}

\usepackage{euclid}
\usepackage{comment}
\usepackage{tablefootnote}
\usepackage[acronyms,nogroupskip,nopostdot]{glossaries}
\usepackage{glossary-mcols}
\usepackage{appendix}
\usepackage{xcolor}
\usepackage{graphicx}
\usepackage{ulem}

\usepackage{todonotes}

\newglossary[cod]{codes}{cod}{ntn}{Codes}
\makeglossaries
\newglossaryentry{code:CAMB}
{
  name=\texttt{CAMB},
  description={Code for anisotropies in the microwave background},
}
\newglossaryentry{code:CLASS}
{
  name=\texttt{CLASS},
  description={Cosmological linear anisotropy solving system},
}

\newglossaryentry{code:classy}
{
  name=\texttt{classy},
  description={Python wrapper for CLASS},
}

\newglossaryentry{code:CONCEPT}
{
  name=\texttt{CO\textit{N}CEPT},
  description={Cosmological \textit{N}-body code in Python},
}

\newglossaryentry{code:CosmicEmu}
{
  name=\texttt{CosmicEmu},
  description={Cosmic emulator based on the Mira-Titan cosmological simulation suite (successor of FrankenEmu based on the Coyote simulation suite).},
}

\newglossaryentry{code:EuclidEmulator1}
{
  name=\texttt{EuclidEmulator1},
  description={Version 1 of EuclidEmulator. This is a code to predict the nonlinear corrections to DM power spectra for $w_0$CDM cosmologies.},
}

\newglossaryentry{code:EuclidEmulator2}
{
  name=\texttt{EuclidEmulator2},
  description={Version 2 of EuclidEmulator. This is a code to predict the nonlinear corrections to DM power spectra for $w_0wa$CDM$+\sum m_\nu$.},
}

\newglossaryentry{code:e2py}
{
  name=\texttt{e2py},
  description={Python wrapper for EuclidEmulator},
}

\newglossaryentry{code:HACC}
{
  name=\texttt{HACC},
  description={Hardware/Hybrid Accelerated Cosmology Code},
}

\newglossaryentry{code:HALOFIT}
{
  name=\texttt{HALOFIT},
  description={Analytical code to produce nonlinear power spectra},
}

\newglossaryentry{code:HMCode}
{
  name=\texttt{HMCode},
  description={Fast predictor for the nonlinear power spectrum based on the halomodel approach.},
}

\newglossaryentry{code:NGenHalofit}
{
  name=\texttt{NGenHalofit},
  description={Code to produce non-linear power spectra using a semi-analytical approach for large and a smoothing-spline-fit model for small scales},
}

\newglossaryentry{code:PKDGRAV3}
{
  name=\texttt{PKDGRAV3},
  description={parallel k-D tree gravity code (version 3); Cosmological N-body tree code},
}
\newglossaryentry{code:UQLab}
{
  name=\texttt{UQLab},
  description={Matlab-based uncertainty quantification framework},
}
\newacronym{AP}{AP}{Alcock-Paczy\'nsky}
\newacronym{BAO}{BAO}{baryon accoustic oscillations}
\newacronym{CDM}{CDM}{cold dark matter}
\newacronym{CMB}{CMB}{cosmic microwave background}
\newacronym{DE}{DE}{dark energy}
\newacronym{DM}{DM}{dark matter}
\newacronym{ED}{ED}{experimental design}
\newacronym{EE}{EE}{EuclidEmulator}
\newacronym{EFS2}{EFS2}{Euclid Flagship v2.0 Simulation}
\newacronym{EFHR}{EFHR}{Euclid Flagship High Resolution Simulation}
\newacronym{EOE}{EOE}{emulation-only error}
\newacronym{EoS}{EoS}{equation of state}
\newacronym{FMM}{FMM}{fast multi-pole method}
\newacronym{GR}{GR}{general theory of relativity}
\newacronym{GRF}{GRF}{Gaussian random field}
\newacronym{GPU}{GPU}{graphics processing unit}
\newacronym{HOD}{HOD}{halo occupation distribution}
\newacronym{IC}{IC}{initial condition}
\newacronym{oneLPT}{1LPT}{first order Lagrangian perturbation theory}
\newacronym{twoLPT}{2LPT}{second order Lagrangian perturbation theory}
\newacronym{LAR}{LAR}{Least angle regression}
\newacronym{LH}{LH}{Latin hypercube}
\newacronym{LHS}{LHS}{Latin hypercube sampling}
\newacronym{LOO}{LOO}{leave-one-out}
\newacronym{LV}{LV}{Large volume}
\newacronym{MCMC}{MCMC}{Markov chain Monte Carlo}
\newacronym{ML}{ML}{machine learning}
\newacronym{MSE}{MSE}{mean squared error}
\newacronym{NLC}{NLC}{nonlinear correction}
\newacronym{PC}{PC}{principal component}
\newacronym{PCA}{PCA}{principal component analysis}
\newacronym{PCE}{PCE}{polynomial chaos expansion}
\newacronym{PF}{P+F}{paired-and-fixed}
\newacronym{PCS}{PCS}{piece-wise cubic spline}
\newacronym{PM}{PM}{particle-mesh}
\newacronym{PPF}{PPF}{parametrized post-Friedmann}
\newacronym{RCF}{RCF}{resolution correction factor}
\newacronym{RMSE}{RMSE}{root mean squared error}
\newacronym{rRMSE}{rRMSE}{relative root mean squared error}
\newacronym{RSDs}{RSDs}{redshift space distortions}
\newacronym{SPCE}{SPCE}{sparse polynomial chaos expansion}
\newacronym{THF}{THF}{Takahashi HALOFIT}
\newacronym{UQ}{UQ}{uncertainty quantification}
\newacronym{ZA}{ZA}{Zel'dovich approximation}

\title[The EuclidEmulator2]{\Euclid preparation: IX. EuclidEmulator2 -- Power spectrum emulation with massive neutrinos and self-consistent dark energy perturbations}

\author[Euclid Collaboration]{
\parbox{\linewidth}{
Euclid Collaboration: M.~Knabenhans$^{1}$\href{https://orcid.org/0000-0002-2886-9838}
{\includegraphics[scale=0.75]{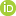}}\thanks{mischak@physik.uzh.ch}, J.~Stadel$^{1}$\href{https://orcid.org/0000-0001-7565-8622}{\includegraphics[scale=0.75]{Img/orcid_16x16}}, D.~Potter$^{1}$\,\href{https://orcid.org/0000-0002-0757-5195}{\includegraphics[scale=0.75]{Img/orcid_16x16}}, J.~Dakin$^{2}$\href{https://orcid.org/0000-0002-2915-0315}{\includegraphics[scale=0.75]{Img/orcid_16x16}},
S.~Hannestad$^{2}$\href{https://orcid.org/0000-0002-4922-9645}{\includegraphics[scale=0.75]{Img/orcid_16x16}}, T.~Tram$^{2}$\href{https://orcid.org/0000-0002-2411-063X}{\includegraphics[scale=0.75]{Img/orcid_16x16}}, S.~Marelli$^{3}$\href{https://orcid.org/0000-0002-9268-9014}{\includegraphics[scale=0.75]{Img/orcid_16x16}}, A.~Schneider$^{1,4}$\href{https://orcid.org/0000-0001-7055-8104}{\includegraphics[scale=0.75]{Img/orcid_16x16}}, R.~Teyssier$^{1}$\href{https://orcid.org/0000-0001-7689-0933}{\includegraphics[scale=0.75]{Img/orcid_16x16}},  S.~Andreon$^{5}$, N.~Auricchio$^{6}$, C.~Baccigalupi$^{7,8,9}$, A.~Balaguera-Antol\'inez$^{10,11}$, M.~Baldi$^{12,13,14}$, S.~Bardelli$^{6}$, P.~Battaglia$^{15}$, R.~Bender$^{16,17}$, A.~Biviano$^{9,18}$, C.~Bodendorf$^{17}$
% ...
}}

\date{Accepted XXX. Received YYY; in original form ZZZ}

\pubyear{2019}

\newcommand{\BAO}{\gls{BAO}}
\newcommand{\BAOs}{\glspl{BAO}}
\newcommand{\CDM}{\gls{CDM}}
\newcommand{\CMB}{\gls{CMB}}
\newcommand{\DM}{\gls{DM}}
\newcommand{\DE}{\gls{DE}}
\newcommand{\ED}{\gls{ED}}
\newcommand{\EoS}{\gls{EoS}}
\newcommand{\FMM}{\gls{FMM}}

\newcommand{\GRF}{\gls{GRF}}

\newcommand{\IC}{\gls{IC}}
\newcommand{\oneLPT}{\gls{oneLPT}}
\newcommand{\twoLPT}{\gls{twoLPT}}
\newcommand{\LAR}{\gls{LAR}}
\newcommand{\LH}{\gls{LH}}
\newcommand{\LHS}{\gls{LHS}}
\newcommand{\LOO}{\gls{LOO}}
\newcommand{\MCMC}{\gls{MCMC}}
\newcommand{\ML}{\gls{ML}}
\newcommand{\MSE}{\gls{MSE}}
\newcommand{\NLC}{\gls{NLC}}
\newcommand{\PF}{\gls{PF}}
\newcommand{\RCF}{\gls{RCF}}

\newcommand{\rRMSE}{\gls{rRMSE}}
\newcommand{\LCDM}{$\Lambda$CDM}
\newcommand{\PCE}{\gls{PCE}}
\newcommand{\PCA}{\gls{PCA}}
\newcommand{\UQ}{\gls{UQ}}
\newcommand{\wCDM}{$w_0$CDM}
\newcommand{\EFStwo}{\gls{EFS2}}

\newcommand{\Omm}{$\Omega_{\mathrm m}$}

\newcommand{\Omb}{$\Omega_{\mathrm b}$}

\newcommand{\Omrad}{$\Omega_{\rm rad}$}

\newcommand{\Omcdm}{$\Omega_{\rm CDM}$}

\newcommand{\Omdm}{$\Omega_{\rm DM}$}

\newcommand{\OmDE}{$\Omega_{\rm DE}$}

\newcommand{\Omnu}{$\Omega_\nu$}

\newcommand{\Sumnu}{$\Sigma m_{\nu}$}
\newcommand{\ns}{$n_{\mathrm s}$}
\newcommand{\h}{$h$}

\newcommand{\As}{$A_{\mathrm s}$}
\newcommand{\zini}{$z_{\mathrm{ini}}$}
\newcommand{\zintm}{$z_{\mathrm{intrm}}$}
\newcommand{\zfin}{$z_{\mathrm{fin}}$}

\newcommand{\mOmm}{\Omega_{\mathrm m}}

\newcommand{\mOmb}{\Omega_{\mathrm b}}

\newcommand{\mOmrad}{\Omega_{\rm rad}}

\newcommand{\mOmcdm}{\Omega_{\rm CDM}}

\newcommand{\mOmdm}{\Omega_{\rm DM}}

\newcommand{\mOmDE}{\Omega_{\rm DE}}

\newcommand{\mOmnu}{\Omega_\nu}

\newcommand{\mns}{n_{\mathrm s}}

\newcommand{\mAs}{A_{\mathrm s}}

\newcommand{\CAMB}{\gls{code:CAMB}}
\newcommand{\CLASS}{\gls{code:CLASS}}
\newcommand{\CONCEPT}{\gls{code:CONCEPT}}
\newcommand{\CosmicEmu}{\gls{code:CosmicEmu}}

\newcommand{\EEone}{\gls{code:EuclidEmulator1}}
\newcommand{\EEtwo}{\gls{code:EuclidEmulator2}}
\newcommand{\etwopy}{\gls{code:e2py}}

\newcommand{\Halofit}{\gls{code:HALOFIT}}
\newcommand{\HMCode}{\gls{code:HMCode}}

\newcommand{\NGenHF}{\gls{code:NGenHalofit}}
\newcommand{\PKDGRAV}{\gls{code:PKDGRAV3}}

\newcommand{\UQLab}{\gls{code:UQLab}}

\newcommand{\PC}{\mathrm{PC}}
\newcommand{\hompc}{\,h\,{\rm Mpc}^{-1}}
\newcommand{\mpcoh}{\,h^{-1}\,{\rm Mpc}}
\newcommand{\gpcoh}{\,h^{-1}\,{\rm Gpc}}
\newcommand{\gpcohcubed}{\,h^{-3}\,{\rm Gpc}^3}
\newcommand{\linv}{\ell^{-1}}

\newcommand{\corrOne}[1]{#1}

\newcommand{\strikethrough}[1]{}
\newcommand{\corrMischa}[1]{#1}
\newcommand{\corrViel}[1]{#1}
\newcommand{\corrPeacock}[1]{#1}
\newcommand{\corrCarvalho}[1]{#1}
\newcommand{\corrNtelis}[1]{#1}

\newcommand{\sub}[1]{_{\mathrm{#1}}}

\makeatletter
\let\oldtheequation\theequation
\renewcommand\tagform@[1]{\maketag@@@{\ignorespaces#1\unskip\@@italiccorr}}
\renewcommand\theequation{(\oldtheequation)}
\makeatother
\definecolor{mygreen}{RGB}{26,120,34}
\definecolor{myred}{RGB}{200,0,33}
\definecolor{myorange}{RGB}{233,133,46}
\definecolor{mypurple}{RGB}{120,0,100}
\definecolor{mycyan}{RGB}{56,200,252}
\definecolor{mymagenta}{RGB}{208,102,142}

\begin{document}
\label{firstpage}
\maketitle

\clearpage
\begin{abstract}
We present a new, updated version of the \texttt{EuclidEmulator} (called {\EEtwo}), a fast and accurate predictor for the nonlinear correction of the matter power spectrum. Percent-level accurate emulation is now supported in the eight-dimensional parameter space of $w_0w_a$CDM$+\sum m_\nu$ models between redshift $z=0$ and \corrCarvalho{$z=3$} for spatial scales within the range $0.01 \hompc\leq k \leq 10\hompc$. In order to achieve this level of accuracy, we have had to improve the quality of the underlying N-body simulations used as training data: (i) we use self-consistent linear evolution of non-dark matter species such as massive neutrinos, photons, dark energy and the metric field, (ii) we perform the simulations in the so-called N-body gauge, \corrPeacock{which} allows \corrOne{one} to interpret the results in the framework of general relativity, (iii) we run over 250 high-resolution simulations with $3000^3$ particles in boxes of \corrPeacock{$1(h^{-1}\;{\rm Gpc})^3$} volumes based on paired-and-fixed initial conditions and (iv) we \corrMischa{provide} a resolution correction \corrMischa{that can be applied to emulated results as a post-processing step} in order to drastically reduce systematic biases on small scales due to residual resolution effects in the simulations. We find that the inclusion of the dynamical dark energy parameter $w_a$ significantly increases the complexity and expense of creating the emulator. The high fidelity of {\EEtwo} is tested in various comparisons against N-body simulations as well as alternative fast predictors like {\Halofit}, {\HMCode} and {\CosmicEmu}. A blind test is successfully performed against the Euclid Flagship v2.0 simulation. Nonlinear correction factors emulated with {\EEtwo} are accurate at the level of $1\%$ or better for $0.01 \hompc\leq k \leq 10\hompc$ and $z\leq3$ compared to high-resolution dark matter only simulations. {\EEtwo} is publicly available at \url{https://github.com/miknab/EuclidEmulator2}.

\end{abstract}



\section{Introduction}
Ongoing and \corrPeacock{forthcoming} cosmological surveys such as DESI\footnote{www.desi.lbl.gov/category/announcements/} \citep{DESICollaboration2016}, LSST\footnote{www.lsst.org/lsst} \citep{LSSTScienceCollaboration2009}, \corrOne{{\Euclid}}\footnote{sci.esa.int/euclid} \citep{Laureijs2011EuclidReport}, and WFIRST\footnote{wfirst.gsfc.nasa.gov} \citep{Akeson2019} have the potential to shed light on {\DE}, \corrOne{{\DM}}, and the neutrino masses. 
It has been confirmed by \corrOne{Solar} and atmospheric neutrino experiments that neutrinos have finite mass (e.g. \citealt{Valle2005NeutrinoOscillations, Schwetz2008}), yet, attempts to pin down the total mass and the mass hierarchy of the neutrino flavour states have so far not been successful. The ongoing neutrino experiment KATRIN \citep{Weinheimer2002KATRINScale, Fraenkle2008, Wolf2010TheExperiment} has been launched in order to tighten the neutrino mass bounds with particle physics.
The equation of state parameter $w$ describing \corrOne{{\DE}} is also poorly constrained. While the current $\Lambda$CDM concordance cosmological model with a value of $w=-1$ is favoured, the error bars coming from \corrOne{{\Planck}} data alone are of order $50\%$. They shrink to $\sim10\%$ if more \corrOne{{\DE}} sensitive probes such as cluster counts, weak lensing and supernovae type Ia are additionally considered \citep{PlanckCollaboration:P.A.R.Ade2015}. Past and current surveys have a hard time constraining the \corrOne{{\DE}} parameter more accurately: Surveys like \corrOne{{\Planck}} \citep{PlanckCollaboration2006ThePlanck} probe the {\CMB} only, which is not very sensitive to \corrOne{{\DE}}. On the other hand, several surveys analysed either only spectroscopic probes (e.g. BOSS, \citealt{Dawson2013TheSDSS-III}) or only photometric probes (e.g. KiDS-450, \corrOne{\citealt{Hildebrandt2017KiDS-450:Lensing}}). In contrast, \corrOne{large-scale} hybrid photometric and spectroscopic experiments like \corrOne{{\Euclid}} \citep{Laureijs2011EuclidReport} will be able to reduce the error bars on the \corrOne{{\DE}} parameters, as the combination of weak lensing with galaxy clustering provides a promising handle on \corrOne{{\DE}} phenomena.

A common approach for the inference of cosmological parameters is to use Bayesian techniques. One specific possibility is to \corrOne{sample the likelihood function in a {\MCMC} approach}, compare the predicted observable (e.g. the power spectrum) to the one actually measured in an observation\corrOne{,} and extract the maximum likelihood estimator values for the cosmological parameters. However, this requires a large number of accurate theoretical predictions. For studying the nonlinear regime of cosmic structure formation, N-body simulations, still the most accurate tool available today, are numerically too expensive to be used for Bayesian inference and hence there is a high demand for more efficient methods. While halo models for massive neutrinos are investigated by several research groups (see e.g. \citealt{Massara2014} or \citealt{Hannestad2020}), surrogate models for N-body simulations, so-called emulators, have been shown to be very promising candidates. Several emulators are available already. Examples are \texttt{FrankenEmu} \citep{Heitmann2009, Heitmann2010, Heitmann2013}, {\CosmicEmu} \citep{Heitmann2016, Lawrence2017}, the emulators of the Aemulus project \citep{DeRose2019TheCosmology, McClintock2019TheFunction, Zhai2019}, {\NGenHF} \citep{Smith2019PrecisionUniverse}, {\EEone} \citep{Knabenhans2019EuclidSpectrum}, the \texttt{Dark quest} emulator \citep{Nishimichi2019DarkClustering} and \texttt{BE-HaPPY}\citep{Valcin2019BE-HaPPY:Neutrinos}.

In this paper we will introduce an update of {\EEone}. While {\EEone} was able to efficiently estimate the {\NLC} to the matter power spectrum for {\wCDM} cosmologies (\corrOne{with the time-variable {\DE} equation of state parameter $w_a$ set to 0}), {\EEtwo} can handle cosmologies with dynamical \corrOne{{\DE}} and massive neutrinos. In addition to a bigger parameter space, {\EEtwo} also pushes the upper limit of the $k$-range to $k\sub{max}\sim10\hompc$. The motivation for this is the same as described in \citet{Knabenhans2019EuclidSpectrum}: While clearly the dominant source of uncertainties on such small spatial scales is due to baryons, it is important to have best possible control over the dark-matter physics in this regime in order to avoid additional (and avoidable) uncertainties due to the dark sector. As we describe in \citet{Knabenhans2019EuclidSpectrum}, under certain assumptions it is possible to add baryonic and other corrections as a subsequent step in the pipeline, necessitating that theoretical precision is maintained in the high-$k$ regime of \corrOne{{\DM}} clustering. In \citet{Schneider2019, Schneider2019b} the authors follow this strategy and emulate the effect due to baryons on weak lensing observables on top of the underlying nonlinear \corrOne{{\DM}} physics. 

This paper is structured as follows\corrCarvalho{.} In \autoref{sec:theory}, theoretical aspects regarding massive neutrinos, dynamical {\DE} and the N-body gauge are discussed. In section \autoref{sec:NbodySims} we introduce our approach to \corrCarvalho{include} massive neutrinos in N-body simulations using {\PKDGRAV}. In \autoref{sec:Convergence} we report on an extensive convergence series that we have performed in order \corrCarvalho{to} estimate the uncertainties in the input simulations used for the construction of {\EEtwo}, considering volume and resolution effects. \corrPeacock{We will} find systematics in this \corrPeacock{section} whose treatment will then be discussed subsequently in \autoref{sec:rescorr}. In \autoref{sec:hfmockemu}, we investigate a prototype emulator based on {\Halofit} data (as we have already done in \citealt{Knabenhans2019EuclidSpectrum}). In \autoref{sec:EE2}, insights taken from this prototype emulator are used to construct the fully simulation-based {\EEtwo}, whose performance is ultimately assessed in \autoref{sec:performance}. We then gain some insight into parameter degeneracies at the level of the matter power spectrum within the $w_0w_a${\CDM}$+\sum m_\nu$ models. Our conclusions are found in \autoref{sec:Conclusion}.

\section{Theoretical Background}
\label{sec:theory}
\subsection{Massive neutrinos}
In oscillation experiments studying solar and terrestrial neutrinos, compelling evidence has been found that the three flavor eigenstates of neutrinos (\corrOne{$e$}, $\mu$ and $\tau$) can be mixed \citep{Becker-Szendy1992Electron-andFlux, Fukuda1998MeasurementsDays,Fukuda1998EvidenceNeutrinos, Ahmed2004, 2015arXiv151206148D}. This implies that neutrinos must have finite mass eigenstates, a fact lying outside of the current standard model of particle physics. While those experiments were able to measure the differences between the squared neutrinos masses, they cannot measure the absolute mass scale. Currently, we only have bounds on the sum of the neutrino masses, see e.g. \cite{ParticleDataGroup2016IntroductionListings}. On cosmological scales, light neutrinos are very abundant and hence are expected to have a significant effect on \corrOne{large-scale} structure. As we are considering neutrinos of masses smaller than 1 eV, they have become non-relativistic only after the electron-nucleon recombination and hence they have imprinted only a small signal into the cosmic microwave background (CMB). However, neutrinos constitute a fraction of the \corrOne{{\DM}} in our Universe and hence the \corrOne{{\DM}} power spectrum is the key quantity to look at when trying to constrain light neutrino masses.

For the rest of this paper we are restricting ourselves to three light mass eigenstates, i.e. we neglect the possibility of heavy sterile neutrinos. Additionally, we consider only the case of three degenerate neutrino masses, i.e. we neglect that the measured squared differences for their masses is given by \citep{Tanabashi2018REVIEWPHYSICS}
\begin{align}
\delta m_{21}^2 &= m_2^2-m_1^2 = (7.37^{+0.197}_{-0.146})\times 10^{-5} \corrOne{{\rm eV}}^2\,,\\
\delta m_{31}^2 &= \vert m_3^2-m_1^2 \vert = (2.56^{+0.043}_{-0.037})\times 10^{-3} \corrOne{{\rm eV}}^2\,,
\end{align}
or alternatively
\begin{equation}
\delta m_{23}^2 = \vert m_2^2-m_3^2 \vert = (2.54\pm0.04)\times 10^{-3} \corrOne{{\rm eV}}^2\,,  
\end{equation}
depending on the considered neutrino hierarchy. This simplification is justified as the difference in the resulting power spectra between a degenerate and a realistic (normal or inverted) mass hierarchy is expected to be well sub-per cent. Neutrinos have a non-vanishing mass while at the same time they have a high \corrOne{velocity dispersion}, \corrCarvalho{hence} they do not belong to the category of {\CDM}. In this paper we focus on cosmologies where there are non-zero contributions from both cold and non-cold \corrOne{{\DM}} particles. Such a {\CDM} plus massive neutrino (CDM$+\sum m_\nu$)\footnote{In this paper we use the notation CDM$+\sum m_\nu$ in order to avoid confusion with mixed \corrOne{{\DM}} (MDM) models with significant contributions from more exotic warm or even hot \corrOne{{\DM}} species.} cosmology also serves as the new \corrOne{{\Euclid}} reference cosmology (see \autoref{tab:parbox}), in contrast to a more standard pure CDM cosmology.

While neutrinos are still relativistic, their free-streaming length is of the size of the Hubble scale. Only after the transition to the non-relativistic phase, the comoving free-streaming scale of neutrinos starts to shrink. As a result, neutrino perturbations get washed out on scales below the free-streaming scale. Due to gravitational interaction, this also suppresses the clustering of {\CDM} on small enough scales. The strength of the effect for a specific $k$-mode depends on both redshift and mass of the neutrinos. But even for small (but finite) neutrino masses we expect a several percent suppression signal in the \corrOne{{\DM}} power spectrum due to these effects. The main effect is on the background expansion of the Universe: If only massless neutrinos are considered, {\Omdm} is identical to {\Omcdm}. However, in order to maintain spatial flatness, even when a non-vanishing {\Omnu} is present, {\Omdm} remains unaltered and thus this requires {\Omcdm} to be decreased accordingly:
\begin{equation}
\label{eq:flatness}
\Omega = \mOmDE+\mOmdm+\mOmb+\mOmrad = \mOmDE+(\mOmcdm+\mOmnu)+\mOmb+\mOmrad\,\corrOne{.}
\end{equation}
The {\CDM} density \corrCarvalho{is} thus decreased by {\Omnu} (while the \corrOne{{\DM}} density parameter \corrCarvalho{does} not change). As {\CDM} and neutrinos evolve differently over the history of the Universe, this leads to a suppression of the {\CDM}+baryon power spectrum. \corrPeacock{This power suppression can help us constrain the sum of the neutrino masses (see e.g. \citealt{Ichiki2009, Coulton2019, Copeland2020})}.

Neutrinos at very high redshifts ($z\gtrsim 1000$) are still a relativistic species, and hence would primarily contribute to {\Omrad}. However, our simulations focus on the nonlinear growth of structure at ($z\lesssim 10$), at which time the contribution of massive neutrinos shifts to $\mOmdm$. It is for this reason that we consider {\Omnu} as a contribution to 
{\Omdm} in the above. Nonetheless, in our N-body simulations the transition between fully relativistic and non-relativistic neutrinos, including the full distribution function at any given redshift, is accounted for by the {\CLASS} Boltzmann solver. All neutrino effects are self-consistently included in our simulations at the linear level, such that this assignment of {\Omnu} to {\Omdm} is purely a convenient choice of parameterisation.

\subsection{Dynamical dark energy}
In the standard $\Lambda$CDM cosmology, \corrOne{{\DE}} is assumed to be a cosmological constant with a time-independent equation of state parameter given by $w\equiv-1$. This implies that $\rho_\Lambda(t)={\rm const}$ which can be seen from the following conservation equation:
\begin{align}
T^\nu_{\;0;\nu} = \dot{\rho}_\Lambda+3H(\rho_\Lambda+p) &= 0\,\corrOne{,}\\
\dot{\rho}_\Lambda+3H\rho_\Lambda(1+w)  &= 0\,\corrOne{,}\\
\dot{\rho}_\Lambda &= 0\;,
\end{align}
where $T_{\mu\nu}$ is the energy momentum tensor, $p$ the pressure, $\rho$ the density and the over-dot denotes a derivative w.r.t. cosmic time. Although this value is close to the best fitting value from supernova surveys \citep{Betoule2014ImprovedSamples}, \corrPeacock{\strikethrough{ the uncertainties on this value are still large and hence the nature of \corrOne{{\DE}} is very poorly constrained.}} \corrOne{{\DE}} with a slightly time-dependent equation of state parameter is not ruled out by the data currently available. The effects of \corrOne{{\DE}} perturbations become relevant only on the largest scales (usually at $k\lesssim 0.1 \hompc$) as has recently been studied carefully in \citet{Dakin2019DarkSimulations}. Nevertheless, the effects of \corrOne{{\DE}} on the matter power spectrum can become quite significant, primarily because changes in the \gls{DE} component have an impact on the scale factor $a(t)$ which in turn affects the power spectrum on all scales.

Here we shall just briefly recap the key aspects of the theory of time-dependent \corrOne{{\DE}} relevant in the context of \EEtwo. We shall follow closely the explanations given in \citet{Dakin2019DarkSimulations} where this topic has been reviewed in greater detail. 
There are two popular ways of describing \corrOne{{\DE}} in the setting of an effective theory: the fluid description and the \corrOne{so-called} \gls{PPF} description \citep{Hu2007ParametrizedGravity}. In the fluid description, {\DE} is considered a fluid with an equation-of-state parameter $w(a)$ ($a$ being the cosmic scale factor) and a constant rest-frame sound-speed $c_{\text{s}}$. We will adopt the widely used parameterisation $w(a)=w_0 + w_a(1-a)$. As can be seen in Equation (2.9) in \citet{Dakin2019DarkSimulations}, the Euler equation for a \corrCarvalho{{\DE} fluid} features a factor $(1+w)^{-1}$, leading to divergences for cosmologies with a {\DE} {\EoS} parameter that evolves across $1+w=0$ over time. This is a manifestation of the fact that such a {\DE} is gravitationally unstable for the lack of additional internal degrees of freedom \citep{Fang2008CrossingEnergy}. The case $w=-1.0$ is sometimes referred to as the ``phantom divide'' or ``phantom barrier'' and models crossing it are called ``phantom-crossing'' cosmologies. It follows, unfortunately, that the fluid description of {\DE} is not well-suited to describe phantom-crossing cosmologies and yet there is no reason why these models should not be taken into account in our analysis.

As described in \citet{Fang2008CrossingEnergy}, common approaches to deal with this problem are either to ignore the {\DE} perturbations in an ad hoc manner or to simply turn them off which violates the energy-momentum conservation for non-$\Lambda$ models and leads to inconsistencies \corrCarvalho{among} the Einstein equations. For dynamical {\DE} models it is hence crucial to thoroughly address the issue of phantom-crossing.

The parameterisation of minimally coupled {\DE} is not a complete system of equations but rather it requires two closure conditions. The relation between density and pressure fluctuations of {\DE}\corrCarvalho{,} giving rise to its {\EoS} parameter and sound speed\corrCarvalho{,} serves as one of these closure relations in the fluid description. The \gls{PPF} formalism on the other hand takes a direct relation between the momentum density of {\DE} and that of {\DM} on large scales to close the system of equations \citep{Fang2008CrossingEnergy}, thus \corrPeacock{circumventing} the divergence of sound speed at phantom-crossing. This describes the {\DE} momentum perturbations on large scales. To describe them on small scales, an effective sound speed $c_\Gamma$ is introduced which is related to the scale below which the {\DE} field becomes sufficiently homogeneous compared \corrCarvalho{to} the matter field. From an interpolation between these two limits one can obtain an evolution equation for the potential of {\DE} in its rest frame:
\begin{equation}
    \partial_\tau\Gamma = \frac{\partial_\tau a}{a}\left[S\left(1+\frac{c^2_\Gamma k^2}{\mathcal {H}^2}\right)^{-1}-\Gamma\left(1+\frac{c^2_\Gamma k^2}{\mathcal {H}^2}\right)\right]\,,
\end{equation}
with $\mathcal{H}$ the conformal Hubble parameter \corrOne{and $\partial_\tau$ denoting the derivative w.r.t. conformal time $\tau$}. $\Gamma$ is related to the \corrOne{rest-frame} {\DE} density field \corrOne{$\delta\rho_{\rm DE}^{\rm rest}$} via the Poisson equation
\begin{equation}
    k^2\Gamma = -4\pi Ga^2\delta\rho_{\rm DE}^{\rm rest}\,,
\end{equation}
$c_\Gamma$ \corrOne{being} the effective sound speed and $S$ being defined as
\begin{equation}
    S\equiv\frac{4\pi Ga^2}{\mathcal{H}}(\rho_{\rm DE}+p_{\rm DE})\frac{\theta^{\rm N}_{\rm t}}{k^2}\,.
\end{equation}
In the last equation, in turn, $\theta^{\rm N}_{\rm t}$ \corrPeacock{denotes} the velocity divergence field of all other species than {\DE} in the conformal Newtonian gauge. For a more complete discussion we refer the reader to \citet{Dakin2019DarkSimulations}, section 2.1.2.

\subsection{General relativity in N-body simulations: The N-body gauge}
\label{sec:nbodygauge}
Traditionally, the formation of \corrOne{large-scale} structure is simulated with Newtonian N-body codes. There are two ways to bring the \gls{GR} into these simulations, in order to study its effects on structure formation\corrCarvalho{.} The first option is to replace the Newtonian equations of motion of structure formation by their general relativistic \corrPeacock{counterpart}. However, this is not an easy task, in particular because scientists have optimised N-body codes for Newtonian physics over the last decades. \corrPeacock{The second option is hence to still use these optimised Newtonian N-body simulation codes which is allowed under certain circumstances.} As has been shown in \corrPeacock{\citet{Chisari2011}, Newtonian simulations predict the clustering properties of {\DM} and galaxies with negligible errors even on very large scales if non-relativistic components or relativistic but non-clustering components are modelled and if a proper set of coordinates (i.e. gauge) is chosen. According to \citet{Fidler2016RelativisticFormation,Fidler2017RelativisticSimulations} it suffices to} add a relativistic correction to the Newtonian potential $\phi$ in the Euler and the Poisson equation (for more practical detail see e.g. \citealt{Dakin2019DarkSimulations}):
\begin{align}
(\partial_\tau + \mathcal{H})\mathbf{v}_{\rm CDM+b}^{\rm Nb} &= -\nabla\phi+\nabla\gamma^{\rm Nb}\,,\\
\nabla^2\phi &= 4\pi G a^2\delta\rho_{\rm tot}^{\rm Nb}\,,
\end{align}
where $\tau$ is the conformal time and \corrOne{$\delta\rho_{\rm tot}^{\rm Nb}$} is the total density perturbation from all species,
\begin{equation}
    \delta\rho_{\rm tot} = \delta\rho_{\rm CDM} + \delta\rho_{\rm b} + \delta\rho_{\rm photon} + \delta\rho_{\nu} + \delta\rho_{\rm DE}\,, 
\end{equation}
where we note that $\delta\rho_{\rm DE} \neq 0$ for $w \neq -1$. The \gls{GR} correction potential $\gamma^{\rm Nb}$ is built from \corrCarvalho{any} other gravitating quantity not already accounted for by $\delta\rho_{\rm tot}^{\rm Nb}$, i.e.\ the momentum density, pressure and shear of photons, neutrinos and \corrOne{{\DE}}. We further parametrise this as
\begin{equation}
    \nabla^2\gamma^{\rm Nb} = -4\pi G a^2\delta\rho_{\rm metric}\,,
\end{equation}
where $\delta\rho_{\rm metric}$ is a fictitious density perturbation, the Newtonian gravity \corrCarvalho{of} which implements all general relativistic effects of $\nabla^2\gamma^{\rm Nb}$.

Finally, the full effective potential is split into $\phi - \gamma^{\rm Nb} \equiv \phi_{\rm sim} + \phi_{\rm lin}$, with $\nabla^2\phi_{\rm sim}$ formally equal to  $4\pi G a^2(\delta\rho_{\rm{CDM}}^{\rm Nb} + \delta\rho_{\rm{b}}^{\rm Nb})$ but in the simulation computed through usual N-body (particle) techniques, while
\begin{equation}
    \nabla^2\phi_{\rm lin} \equiv 4\pi G a^2\bigl(\delta\rho_{\rm photon}^{\rm Nb} + \delta\rho_{\nu}^{\rm Nb} + \delta\rho_{\rm DE}^{\rm Nb} + \delta\rho_{\rm metric} \bigr) \label{eq:phi_lin}
\end{equation}
is solved on a grid using Fourier techniques and then applied to the particles as a correction force to the main particle gravity. Notice that $\phi_{\rm lin}$ is the object called $\phi_{\rm GR}$ in other publications such as e.g. \citet{Tram2019FullySimulations} or \citet{Dakin2019DarkSimulations, Dakin2019FullySimulations}.

Note that the continuity equation is formally not affected by this additional \gls{GR} potential $\gamma^{\rm Nb}$. This then can be considered writing the full general relativistic evolution equations of the N-body simulations in a very special gauge, the \corrOne{so-called} \textit{N-body gauge} (indicated by the superscript ``Nb''). Doing so, it is possible to apply a gauge transformation of the output in a post-processing step in order to re-obtain general relativistic results in any \corrOne{observationally relevant} gauge up to first order.
\section{Cosmological Simulations}
\label{sec:NbodySims}
Cosmic emulators are based on training data (also known as the experimental design) generated by N-body simulations. Unsurprisingly, the quality of any emulator hence crucially depends on the quality of the training data. We invested great efforts into optimising the quality of the training set simulations. The construction of {\EEtwo} training data relies mainly on three codes: {\PKDGRAV}, which is the main simulation code, and {\CONCEPT} which in turn depends on the Einstein-Boltzmann solver {\CLASS}. All three codes had to be optimised in order to be able to fully self-consistently treat cosmologies with massive neutrinos and dynamical {\DE}. In this section we shall describe how these codes \corrCarvalho{were} optimised to generate a suite of simulations used as the experimental design for {\EEtwo} and thereby put emphasis on the differences with respect to the generation of the experimental design of {\EEone}.

\subsection{Pipeline overview}
We shall give a quick overview over the implemented pipeline in order to facilitate the understanding of the steps involved in generating the training data for the emulator. More detail is given for each step in dedicated sections below.

\paragraph*{Pre-computation of cosmological quantities:} We have designed a pipeline in which all cosmological background and linear quantities, such as the time dependence of the Hubble parameter $H(z)$, are computed before running any simulation. This is done through {\CONCEPT} with {\CLASS} in order to take all relevant physics into account (e.g.\ this approach allows for a fully relativistic treatment). In this step, transfer functions are computed to linear level at many different redshifts for all species contributing to $\phi_{\rm lin}$, that are later used to provide the small corrective force contributions in the N-body simulations. In the case of {\EEone}, however, we only computed the transfer function at $z=0$, which we later used for {\IC} generation via the usual back-scaling approach, and used an analytical form for $H(z)$ to compute the background evolution.

\paragraph*{Gauge transformation:} This step was entirely missing in {\EEone} where we used the transfer function computed in synchronous gauge to set up the {\IC}s. {\CLASS}\corrCarvalho{,} as well as {\CAMB}\corrCarvalho{,} compute quantities either in the synchronous or in the conformal Newtonian gauge. However, in order to make Newtonian N-body codes consistent with \gls{GR}, their input transfer functions have to be mapped to the N-body gauge described in \autoref{sec:nbodygauge}. This transformation is performed by the {\CONCEPT} code. The results are then stored into look-up tables inside HDF5 files that can be queried by the N-body code.

\paragraph*{N-body simulation:} Both the initial condition generation and the actual N-body simulations are performed with {\PKDGRAV}. \corrOne{\Gls{PF}} \citep{Angulo2016} \corrOne{{\oneLPT}} initial conditions are set up at redshift $z=99$ (see also \autoref{sec:Convergence} for a more in-depth discussion). The nonlinear evolution of {\DM} particles is computed with a binary \corrOne{tree-based} {\FMM}. To these tree forces we add a \gls{PM} field for the fluctuations due to massive neutrinos, photons, \gls{DE} and the metric field in order to study their effect on cosmological structure growth at linear level \corrViel{(for simulations treating neutrinos fully nonlinearly see e.g. \citealt{Banerjee2018, Bird2018})}. This aspect is one of the key differences between {\EEone} and {\EEtwo}.

\paragraph*{Post-processing:} From the simulations we obtain the power spectra for each realisation of the fixed-IC simulation pair. In order to get the final {\PF} power spectrum we compute the average of those individual realisations. As ultimately we are primarily interested in the {\NLC}, we compute it w.r.t. the linear theory power spectrum (in our case computed by {\CLASS}) at each redshift\corrOne{. Explicitly, the {\NLC} $B(k,z)$ is the dimension-less quantity defined through the relation
\begin{equation}
\label{eq:Bdef}
    P_{\rm nl}(k,z) = P_{\rm LinTh}(k,z) B(k,z)\,,
\end{equation}
where the sub-script $\rm nl$ stands for ``nonlinear'' and the sub-script $\rm LinTh$ for ``linear theory''.} Notice that, in contrast, in \citet{Knabenhans2019EuclidSpectrum} the {\NLC} was defined slightly differently: we computed it via division by the {\IC} power spectrum of the simulation.
We also compute an additional factor that corrects for power suppression at small scales caused by resolution effects in the simulations. \corrOne{This factor we shall refer to as the {\RCF} and it will be discussed in \autoref{sec:rescorr}}. 

\paragraph*{Emulator construction:} The obtained data matrix is then principal component analysed. The emulator then predicts the vector of principal component weights.

\subsection{Pre-processing with CO{\textit N}CEPT: Transfer functions}
For generation of particle initial conditions inside {\PKDGRAV}, matter density and velocity transfer functions $\delta_{\rm{CDM}+\rm{b}}(a_{\rm ini}, k)$ and $\theta_{\rm{CDM}+\rm{b}}(a_{\rm ini}, k)$ are required, where $a_{\rm ini}$ is the scale factor at the start of the simulation. As we seek to carry out the simulation in N-body gauge, this is also the gauge of these transfer functions. Note that\corrCarvalho{,} due to the contributions from species other than CDM, we cannot obtain accurate particle velocities from only $\delta_{\rm{CDM}+\rm{b}}$.

To keep the simulation in N-body gauge, we need to repeatedly apply the linear GR correction force $-\nabla\phi_{\rm lin}$. From \eqref{eq:phi_lin} we see that we are thus in need of $\delta\rho_{\rm photon}(a, k)$, $\delta\rho_{\nu}(a, k)$, $\delta\rho_{\rm DE}(a, k)$ and $\delta\rho_{\rm metric}(a, k)$. Obtaining these N-body gauge transfer functions, the {\CONCEPT} code has been run in advance. This is a full N-body code in its own right, with the added trait of very tight integration with the {\CLASS} code. Internally, {\CONCEPT} takes the synchronous gauge output from {\CLASS}, converts it to N-body gauge and uses it for both particle initial conditions and GR corrections. The relevant gauge transformations are
\begin{align}
    \delta\rho_{\alpha}^{\rm Nb} &= \delta\rho_{\alpha}^{\rm s} + 3\mathcal{H}(1 + w_{\alpha})\bar{\rho}_\alpha \frac{\theta_{\rm tot}^{\rm s}}{k^2}\,, \\
    \theta_\alpha^{\rm Nb} &= \theta_\alpha^{\rm s} + \partial_\tau\biggl( \frac{h_{\rm mp}}{2} - 3\mathcal{H} \frac{\theta_{\rm tot}^{\rm s}}{k^2} \biggr) \,, 
\end{align}
where superscripts `Nb' and `s' indicate the N-body and synchronous gauge respectively, $\alpha$ labels the species, \corrOne{$h_{\rm mp}$} is the trace of the metric perturbation in synchronous gauge and $\theta_{\rm tot}$ is the total velocity divergence of all species.

\begin{figure}
	\includegraphics[width=\columnwidth]{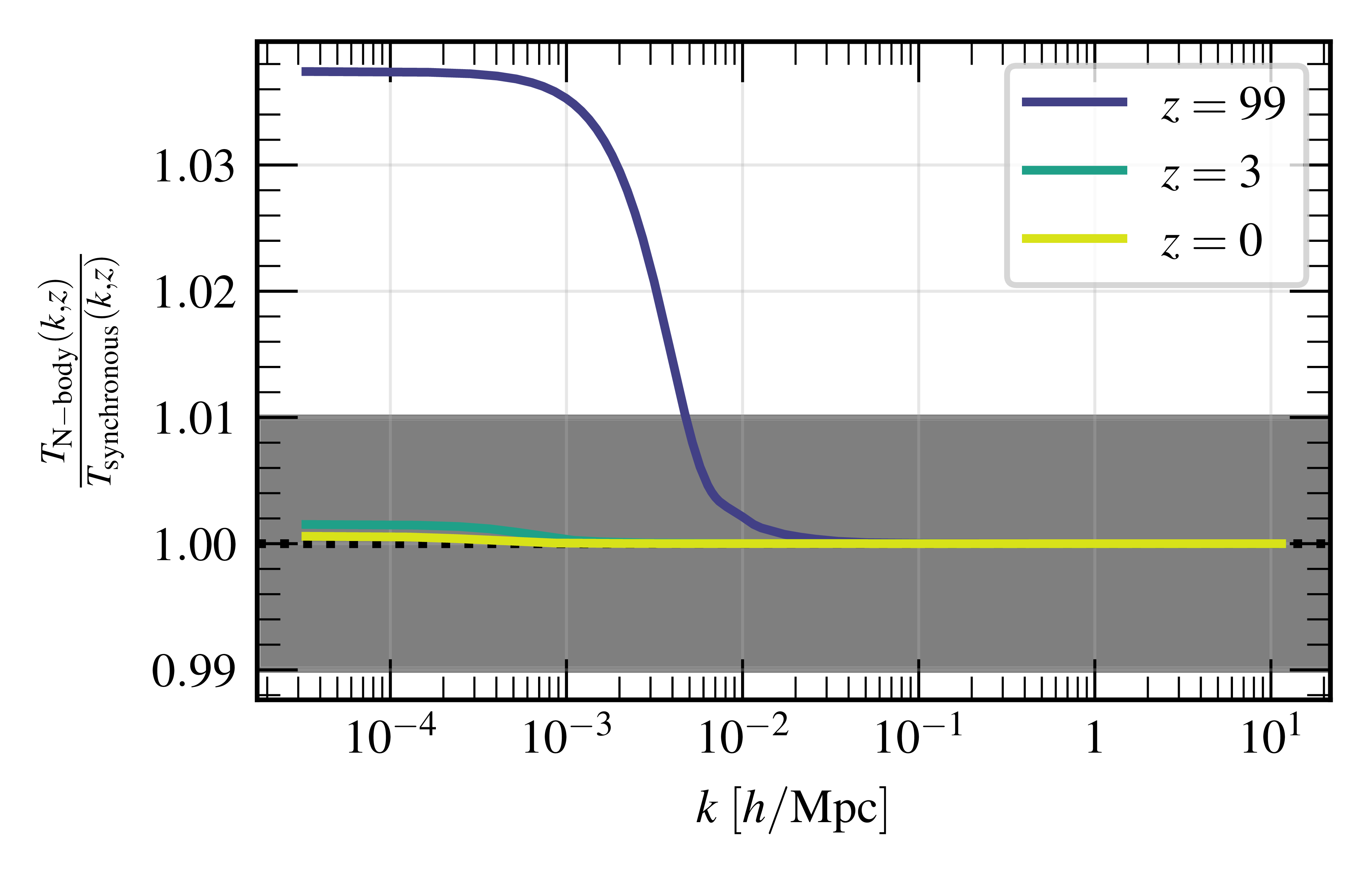}
	\caption{\corrPeacock{Ratio between the {\CDM}+baryon transfer functions in the N-body gauge (as used for the N-body simulations in this work) and the synchronous gauge at three different redshifts. At redshift $z=99$ where the initial conditions are realised, the two gauges lead to transfer functions differing by almost 4\% at very large scales. At the $k\sim 0.006 \hompc$ corresponding to the fundamental $k$-mode of the simulations that will be used to construct {\EEtwo} (see \autoref{sec:Convergence}), the effect is just about at the 1\% level.}}
    \label{fig:gauge}
\end{figure}

Though more complicated, the $\delta\rho_{\rm metric}(a, k)$ transfer function is similarly constructed from various {\CLASS} \corrPeacock{outputs}, some of which are only available in the {\CLASS} version that ships with {\CONCEPT} (see \citealt{Dakin2019FullySimulations} for more details). We also note that the computation of the DE pressure perturbation within {\CLASS}, needed for $\delta\rho_{\rm metric}$, is wrong in the standard version of {\CLASS}, but fixed in the version shipping with {\CONCEPT} (see \citealt{Dakin2019DarkSimulations} for details).

For use with {\PKDGRAV}, the so-called {\CLASS}-utility was added to {\CONCEPT}, which saves the N-body gauge transfer functions to an external HDF5 file, which is then read in by {\PKDGRAV}. In the HDF5 all requested transfer functions \corrViel{(i.e. on top of the {\CDM}+baryon transfer function also that of the photons, the massive neutrinos, the {\DE} and the metric perturbations)} are stored on a global $(a, k)$ grid, where special attention must be given to the interpolations performed within this grid in order to achieve the required precision. Various background quantities\corrOne{, for example \ $H(a)$, that are} computed by {\CLASS} are also stored in the HDF5 and used by {\PKDGRAV}, as these generally are non-trivial to compute in the presence of massive neutrinos.

\subsection{Initial condition generation}
In order to get the initial conditions for the particle positions, we generate a regular grid of particles which we displace from their grid points using first\corrCarvalho{-}order Lagrangian perturbation theory. We imprint an initial power spectrum $P_{\rm ini}$ with the pairing-and-fixing technique \citep{Angulo2016}, i.e.  
\begin{equation}
\label{eq:P.png}
P(\vert\delta_{i,\rm lin}\vert, \theta_i) = \frac{1}{2\pi}\delta_{\rm D}(\vert\delta_{i,\rm lin}\vert-\sqrt{P_{\rm ini}})\,,
\end{equation}
with $\delta_{\rm D}$ being the Dirac delta function and the index $i$ labelling the Fourier modes (for more info about how we use the pairing-and-fixing technique to generate initial conditions, please refer to \citealt{Knabenhans2019EuclidSpectrum}). The initial power spectrum is computed based on the {\CDM}+baryon overdensity field $\delta_{\rm m}$ pre-computed by {\CLASS}+{\CONCEPT} at $z_{\rm ini} = 99$. While {\PKDGRAV} has no further use for $\delta_{\rm m}(z < z_{\rm ini})$ we still tabulate these, as they are used when computing the {\NLC} factors. We then use
\begin{equation}
P_{\rm m,\;ini}(k,z_{\rm ini}) = \zeta^2(k)\delta_{\rm m}^2(k,z_{\rm ini})\,\corrOne{,}
\end{equation} where the $\zeta$-function is defined as
\begin{equation}
    \zeta(k) = \pi\sqrt{2A_{\rm s}}k^{-3/2}\left(\frac{k}{k_\ast}\right)^\frac{n_{\rm s} - 1}{2}\exp\left[\frac{\alpha_{\rm s}}{4}\ln\left(\frac{k}{k_\ast}\right)^2\right]\,\corrOne{.}
\end{equation}
Here, {\As} is the spectral amplitude, {\ns} is the spectral index\corrOne{,} and the running $\alpha_{\rm s} =0$ in all cases. The pivot scale is set to its standard \corrOne{{\Planck}} value, $k_\ast=0.05\;{\rm Mpc}^{-1}$.
The resulting initial power spectrum is used to displace the particles from their regular grid points.

\subsection{Nonlinear evolution}
\label{subsec:nonlinevol}
Once the initial condition is generated, the {\DM} particles are evolved by the tree code {\PKDGRAV} using {\FMM} and a multi-timestepping approach. For further technical details about the gravity evolution of {\DM} in {\PKDGRAV}, we refer to \citet{Potter2016}.

What is newly introduced in the version of {\PKDGRAV} that is used for this work (which is also new compared \corrCarvalho{to} the version used to construct {\EEone}) is the interaction with massive neutrinos as well as other linearly evolved species, namely photons, \corrOne{{\DE},} and the metric field. For this, {\PKDGRAV} fetches the linear evolution of all these linearly evolved species from the pre-computed {\CLASS} transfer functions at every base \corrCarvalho{time step}. Via the associated power spectrum, the transfer function can be converted into a density field that is realised on a grid which ultimately leads to a weak corrective mesh force. This particle-mesh interaction provides an additional gravity source term to the particle-particle interaction such that the {\DM} field is evolved taking the linear species into account. 

During the early \corrOne{Universe}, it is important to capture the effects of the \corrOne{high-frequency} oscillations in the linear fields, particularly in the metric field, otherwise we would see a slight offset of the power spectrum at linear scales. The linear evolution may only be done when the simulation is ``time \corrPeacock{synchronised}'' which happens at the start of a base time step. One approach would be to take sufficient base time steps to capture this effect at high redshift, but it would be computationally wasteful at lower redshifts. Instead we adopted the following scheme\corrCarvalho{.} Each N-body simulation was started at redshift \zini$=99$ and evolved in 60 time steps to \zintm$=10$. From \zintm we continued each simulation down to \zfin$=0$ in another 100 time steps resulting in a total number of base time steps $n_z=160$. By employing this approach we achieve \corrPeacock{agreement between the linear evolution of all particle species and {\CLASS}} over the entire simulation starting from \corrOne{{\zini} all} the way to $z=0$.

\section{Convergence Tests}
\label{sec:Convergence}
We have performed extensive convergence tests for the power spectrum and the {\NLC} in different dimensions. These convergence test results serve to update those presented in \citet{Knabenhans2019EuclidSpectrum}.
\subsection{Volume}
\label{subsec:volconv}
We start by determining the minimal volume required to achieve convergence at the 1\% level. We compare a series of box size of edge lengths $L\in\{512, 1024, 2048, 4096\} \mpcoh$ to a reference volume $V=L^3=(8192 \mpcoh)^3$. In this process we fix the resolution parameter to \corrOne{$\linv\equiv N/L=1/3 \hompc$.\footnote{Notice that in this paper we use $N$ for the number of particles per dimension.}}

As our new pipeline (described in \autoref{sec:NbodySims}) allows us to recover the linear scales very accurately at all redshifts, we could try to actually perform the convergence comparison directly against linear theory itself. However, as for simulation box sizes $L<2048 \mpcoh$ even the smallest $k$ modes are already slightly nonlinear, the finite volume effects overlap with the pre-virialisation dip such that particularly at low redshifts no clear conclusion can be drawn.

\begin{figure*}
	\includegraphics[width=\textwidth]{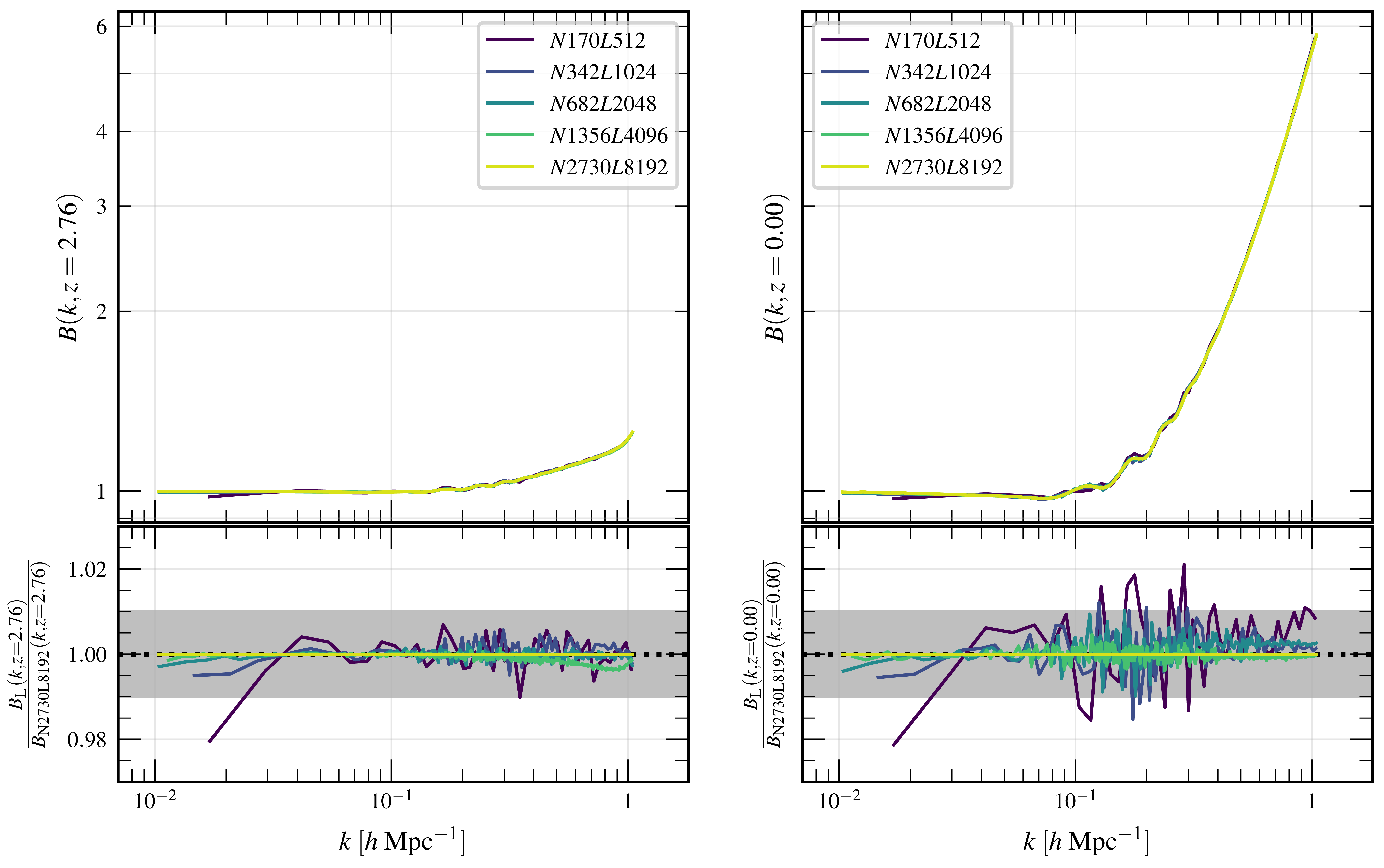}
	\caption{Convergence test results for volume of the {\NLC} factor \corrPeacock{B(k,z)} of simulations with $L\in\{512, 1024, 2048, 4096, 8192\} \mpcoh$. We show results for two different redshifts. Apart from individual $k$ modes on mildly nonlinear scales, simulation boxes with $L=1024\mpcoh$ are converged at the 1\% level.}
    \label{fig:VolConv_B}
\end{figure*}

For this reason, we performed additionally the volume convergence test based on the {\NLC} factor (i.e. we compare to the {\NLC} factor of the simulation in the \corrOne{$L=8192 \mpcoh$ box}). The result is shown in \autoref{fig:VolConv_B}. In this test the pre-virialisation dip is cancelled out such that we are only left with the finite volume effects. We find that a simulation box of $L=1024 \mpcoh$ is converged at the level of $\sim1\%$ (the cosmic variance of only a few individual mildly nonlinear $k$ modes \corrCarvalho{exceeds} the 1\% limit). Notice that this result is unchanged if one would perform the comparison to the largest box at the power spectrum level\corrCarvalho{.} As the {\NLC} is computed with respect to linear theory, dividing two {\NLC}s by each other leads to cancellation of the linear theory out of the expression such that one is left with a ratio of power spectra only:
\begin{equation}
    \frac{B_L}{B_{L=8192}}=\frac{P_L}{P_{\rm LinTh}}\frac{P_{\rm LinTh}}{P_{L=8192}}=\frac{P_L}{P_{L=8192}}\,\corrOne{.}
\end{equation}
Based on this result we choose the simulation box side length for our simulations to be $L=1\gpcoh$.

\subsection{Resolution}
\label{subsec:resconv}
For the investigation of the mass resolution required to get convergence of the power spectrum at a desired level, we are interested in the minimal value of the resolution parameter $\linv$ of N-body simulations. In \citet{Knabenhans2019EuclidSpectrum} we claimed that the power spectra are converged at the level of 1\% up to $k\sim 5 \hompc$ for $\linv\geq1.6 \hompc$. We state clearly that this statement was overly optimistic. We underestimated the minimal $\linv$ because we compared to a simulation with $\linv=4\hompc$ which at that time was the best we could do. For the present work we were able to double the resolution parameter of our reference simulation to $\linv=8\hompc$. In turn, this increases our current estimate of the minimal $\linv$ value required to achieve convergence at the 1\% level.

\begin{figure*}
	\includegraphics[width=\textwidth]{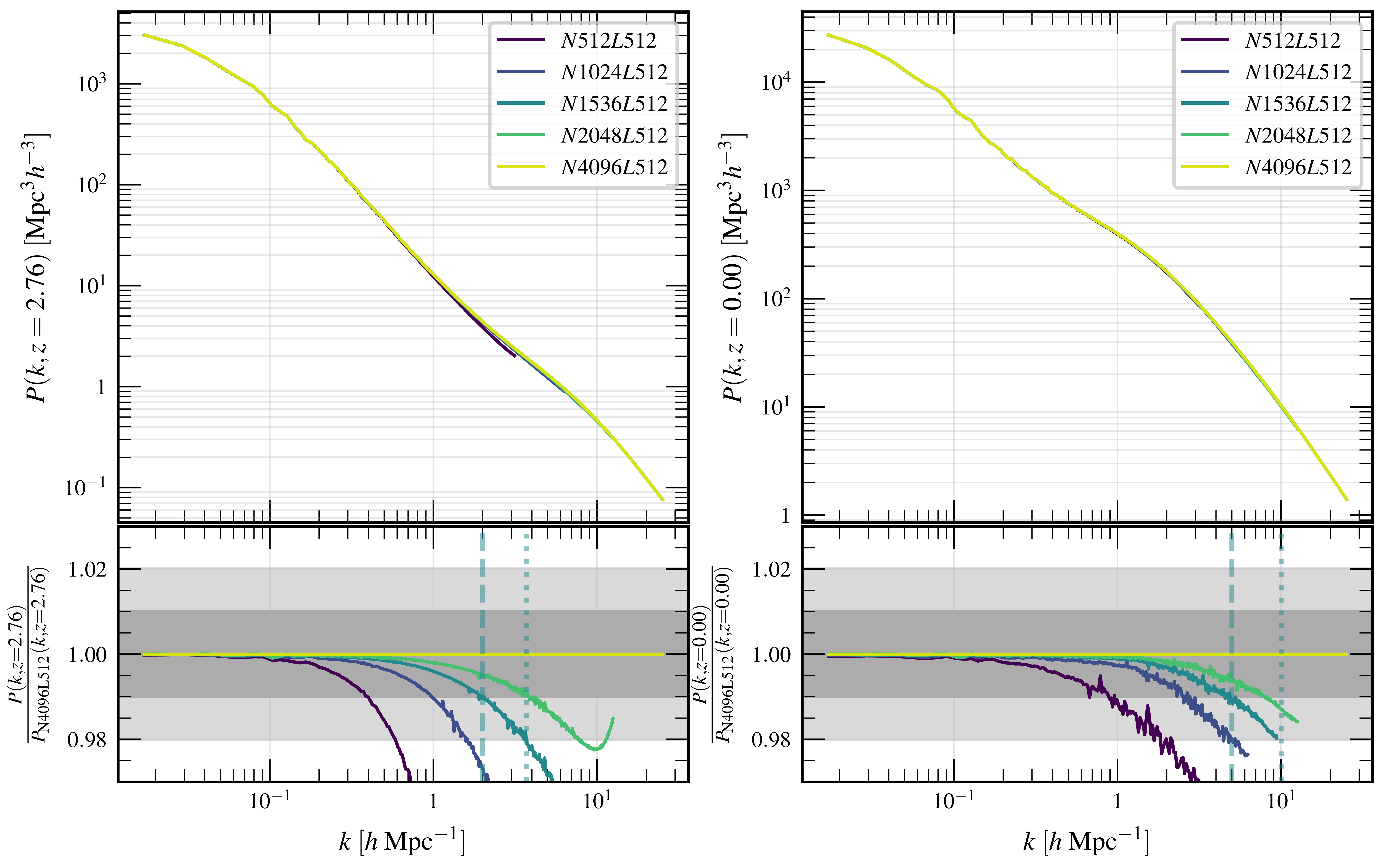}
	\caption{Convergence test results for mass resolution of full nonlinear power spectra showing results for simulations with \corrOne{$\linv\in\{1,2,3,4,8\} \hompc$}. We show results for two different redshifts. The faint, dark green vertical lines correspond to the $k$ values for which the dark green, solid curves (corresponding to the $N1536L512$-case) deviates more than 1\% (dashed lines) and 2\% (dotted lines), respectively, from the reference $N4096L512$ (\corrOne{$\linv=8\hompc$}) simulation. At $z=2.76$, we find that the \corrOne{$\linv=3 \hompc$} simulation is converged within 1\% up to $k\sim 2 \hompc$ and within 2\% up to $k\sim3.5\hompc$. At $z=0$, however, the same simulation is converged within 2\% all the way up to $k=k_{\rm max}\sim10\hompc$ and even stays within 1\% from the reference up to $k\sim 5\hompc$.}
    \label{fig:ResConv}
\end{figure*}

From \autoref{fig:ResConv} one clearly observes that simulations with $\linv\geq4\hompc$ are required to be converged at the 1\% level at $k=10 \hompc$ with respect to the $\linv=8\hompc$ simulation at $z=0$ and the resolution needs to be even higher if 1\%-convergence at higher redshifts is required. Given the minimal volume found in \autoref{subsec:volconv}, such high $\linv$ values imply prohibitively large particle numbers for our simulations. Taking our computational budget into account, we decided to run simulations with $\linv=3\hompc$ thereby doubling the resolution parameter compared \corrCarvalho{to} {\EEone}. This means that our simulations, according to the currently best estimate available, are converged at redshift $z=0$ up to $k_{2\%}=9.42\hompc$ at the level of 2\% and up to $k_{1\%}\sim 5\hompc$ at the level of 1\%. Further results showing how the values for $k_{1\%}$ and $k_{2\%}$ evolve with redshift are shown in \autoref{tab:convk}.

\begin{table}
\centering%
\caption{Dependency on the redshift of the $k$ modes at which the $\linv=3\hompc$ simulation are converged at 1\% and 2\%, respectively. For redshifts 0, 0.5 and 1 the $k_{2\%}$ values correspond to the highest $k$ mode obtained from the simulations and hence should be understood as lower bounds.}
\begin{tabular}{lcc}
& $k_{1\%}\;[\hompc]$ & $k_{2\%}\;[\hompc]$ \\
\hline
$z=0$&4.87&$\geq$9.42\\
$z=0.5$&4.36&$\geq$9.42\\
$z=1$&3.99&$\geq$9.42\\
$z=2.76$&1.97&3.57\\
$z=10$&1.34&1.99
\label{tab:convk}
\end{tabular}
\end{table}


Notice that the suppression of power due to low mass resolutions is very systematic. We investigate this further in \autoref{sec:rescorr}.

Further we summarise that a compromise between the requirements estimated from the convergence tests and our computational budget leads to the following specifications for our N-body simulations\corrCarvalho{.} We evolve $3000^3$ particles in boxes with a volume of $1\gpcohcubed$. 

\subsection{Paired-and-fixed vs. Gaussian random field initial conditions}
We also re--evaluate the comparison between {\PF} simulations and simulations based on {\GRF} initial conditions. For this comparison we ran one pair of fixed amplitude simulations and an ensemble of 50 different {\GRF} simulations. Over all redshifts of interest ($z\leq3$), {\PF} simulations approximate the {\GRF} ensemble average very well on all but the largest scales. However, on all scales the deviation of the {\PF} power spectrum w.r.t. the {\GRF} ensemble average lies well within the bound set by $\max(\sigma_{\rm GRF}, 1\%)$. This confirms the finding of \citet{Angulo2016}. It needs to be taken into account, though, that in that publication an ensemble of 300 {\GRF} simulations (i.e. six times larger than our ensemble) has been used for comparison. Based on this exploration and the results found already in \citet{Knabenhans2019EuclidSpectrum} we again use the pairing-and-fixing approach to efficiently reduce cosmic variance.

\subsection{Quantities not investigated in this convergence series}
Quantities we have not tested in this convergence series are the starting redshift, the main time-stepping parameter and the resolution of the PM-grid for the linear species. \corrPeacock{As we describe below, the softening is related to the mass resolution but its convergence was not tested independently either.} We set the softening parameter to default values of \gls{code:PKDGRAV3}: $\epsilon=1/(50N)$ and the time-stepping parameter $\eta=0.20$. The number of PM-grid cells for the linear species has been conservatively chosen to be a \corrCarvalho{quarter} of the CDM+b particle number $N$ per dimension, i.e. we use $N_{\rm lin}= 750$ PM-grid cells per dimension. The ratio $N/N_{\rm lin}=4$ has already been used in the generation of the {\EFStwo} simulation, where it has been \corrPeacock{proved} to be a more than adequate choice. We also follow the {\EFStwo} simulation for the choice of the initial redshift, $z_{\rm ini}=99$. Convergence of these quantities has been studied in \citet{Schneider2015}.
Further, we have not investigated {\twoLPT} {\IC}s for this work but use {\oneLPT} {\IC}s to set up our simulations. While {\twoLPT} is expected to improve the resolution convergence results presented above, {\PKDGRAV} does not yet support {\twoLPT} for multiple fluids.

\subsection{Results of this convergence analysis}
We have identified that one requires simulations with $3000^3$ particles and a resolution parameter of \corrOne{$\linv = 3\hompc$} \corrViel{(corresponding to a minimally resolved mass of $\sim 3.3\times 10^9 M_\odot/h$} \corrMischa{and a Nyquist frequency of $k_{\rm max}\sim9.4 \hompc$)} in order to achieve satisfactory accuracy on both large and small scales. A simulation of this size and resolution takes a bit more than 2000 node hours (on GPU accelerated nodes\footnote{The simulations were run on the Piz Daint supercomputer at the Swiss National Scientific Supercomputing Centre (CSCS)}).
\section{Resolution correction}
\label{sec:rescorr}
In \autoref{fig:ResConv} in \autoref{subsec:resconv} we show that too low a mass resolution leads to a suppression of power on small scales where the amplitude of this suppression grows both with a growing ratio $\linv_{\rm lowRes}/\linv_{\rm highRes}$ and as the redshift $z$ increases. However, as this effect is very systematic it is possible to correct for it in a post-processing step by compensating the suppression with a resolution correction factor (\RCF).

\subsection{Cosmology dependence of the resolution-induced power suppression}
\begin{figure*}
	\includegraphics[width=\textwidth]{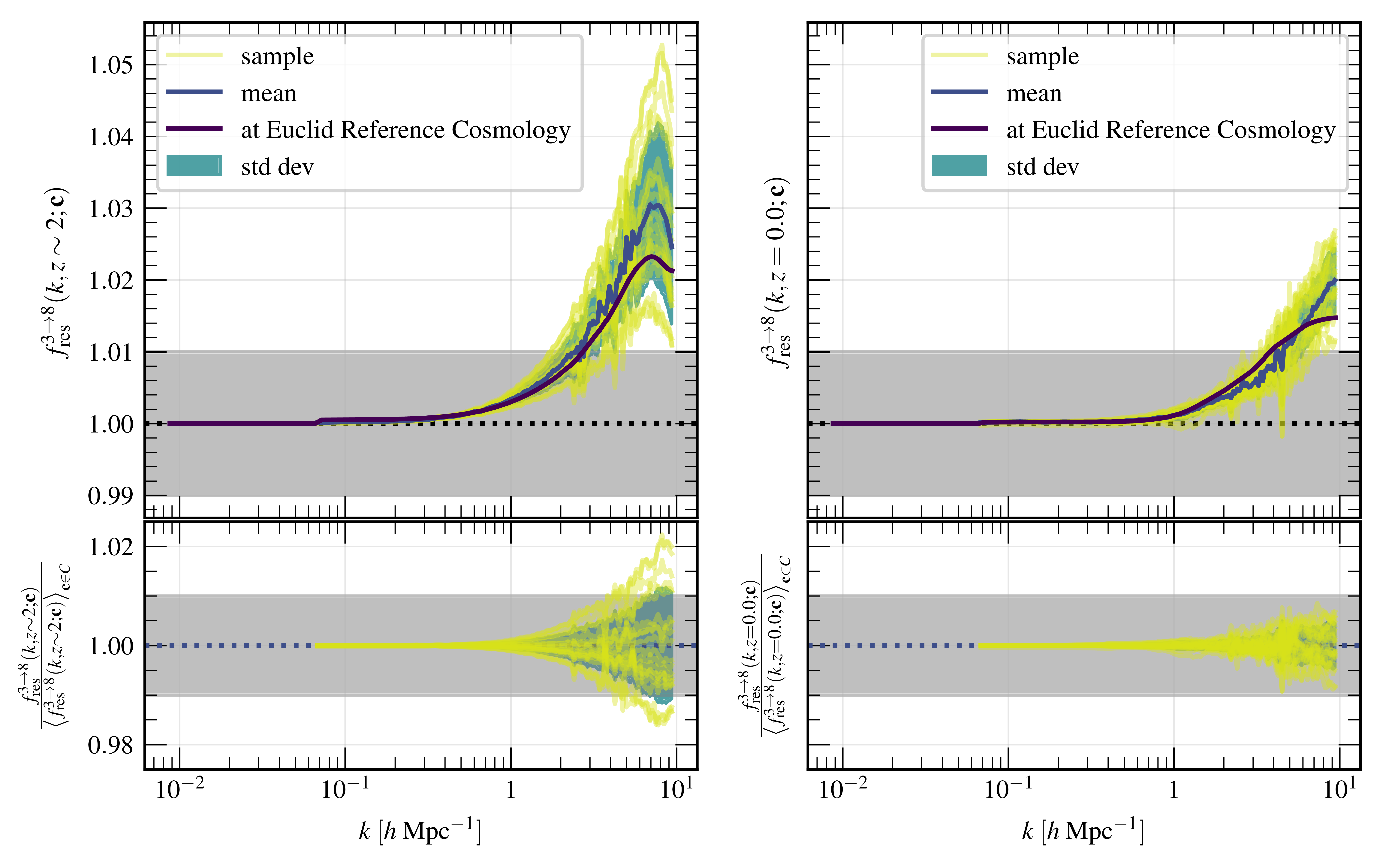}
	\caption{The resolution-induced power suppression is systematic and can be corrected for. In this plot we show the resolution corrections $f_{\rm res}^{3\to8}(k,z; \vec{c})$ computed from a sample of 20 cosmologies represented by the yellow curves. The newly introduced uncertainty due to variance in cosmology, which we neglect, is merely $0.5\%$ (standard deviation, green shaded region) or lower at $z=0$ and at the level of $1\%$ at $z\sim 2$. Given the fact that the bias itself is at the $5\%$ level at higher redshifts, correcting for the resolution has a positive net effect. The curves corresponding to the \corrOne{Euclid Reference Cosmology} are overplotted. We shall use this cosmology to compute the {\RCF} which we apply to all other cosmologies.}
    \label{fig:Rescorr}
\end{figure*}

In general it has to be assumed that the precise shape of the resolution-induced power suppression (and equivalently that of the {\RCF}) depends on cosmology. To test this statement we compute a series of simulations for twenty different cosmologies. We choose these cosmologies to be a subset of the {\ED} of {\EEtwo}. Of course, running multiple \corrOne{high-resolution} reference runs in large boxes is too expensive, so we perform our test in a smaller box\corrCarvalho{.} We run these simulations in boxes of $L=128\mpcoh$ side length with $N^3=384^3$ and $N^3=1024^3$ particles\corrCarvalho{, corresponding to resolution parameters of $\linv=3\hompc$ and $\linv=8\hompc$, respectively}. The {\RCF} is then simply defined as:
\begin{equation}
    f_{\rm res}^{3\to8}(k,z; \vec{c}) = \frac{P^{\linv=3\hompc}(k,z; \vec{c})}{P^{\linv=8\hompc}(k,z; \vec{c})}\corrOne{\,,}
\end{equation}
where $\vec{c}$ denotes a specific cosmology. We thus compare the same resolutions as we do in the case of the dark-green and yellow curves in \autoref{fig:ResConv}. We show the results of this analysis in \autoref{fig:Rescorr}. In this figure, the two upper panels correspond to the lower panels in \autoref{fig:ResConv}. The variability in the ratio of power spectra due to variability in the cosmology is then shown in the lower panels of \autoref{fig:Rescorr}. As is clearly visible, the cosmology dependence of this ratio is weak, particularly at very low redshifts. Also at high redshifts, the $1\sigma$-standard deviation is considerably smaller than the biases at the same $k$ modes for all tested cosmologies. Accordingly, we may simplify
\begin{equation}
    f_{\rm res}^{3\to8}(k,z; \vec{c}) \approx f_{\rm res}^{3\to8}(\vec{c}^\ast; k,z) :=f_{\rm res}^{3\to8}(k,z)\corrOne{\,,}
\end{equation}
where $\vec{c}^\ast$ denotes any reasonable cosmology not too different from the cosmologies for which low resolution simulations are run. In our case the \corrOne{Euclid Reference Cosmology}, the \corrOne{{\Planck}} 2015 best-fit cosmology or the central cosmology of the parameter box defined below in \autoref{tab:parbox} would all be viable choices for $\vec{c}^\ast$. 

From this we conclude that a cosmology-independent correction factor, although introducing a new source of uncertainty, improves the power spectrum measurement of the N-body simulation by a few percent at high $k$. Of course, in future work this should be improved even further by emulating the cosmology dependence of the resolution correction factor. This would greatly reduce the newly introduced uncertainty while still mostly removing the bias.

We do compute $f_{\rm res}^{3\to8}(k,z)$ at the \corrOne{Euclid Reference Cosmology} as defined in \autoref{tab:parbox}. The {\RCF} curve is shown in \autoref{fig:Rescorr} for different redshifts covering the entire redshift range of interest. Notice that values $f_{\rm res}^{3\to8}$ for $k<2\pi/(128 \mpcoh)$ are set to unity (see discussion below in \autoref{subsec:boxsizedep}).

\subsection{Dependence on simulation box size}
\label{subsec:boxsizedep}
For this approach to be a practical strategy, the {\RCF} must not depend strongly on simulation box sizes. Otherwise, the {\RCF} would itself rely on \corrOne{high-resolution} runs in large boxes which are exactly the simulations that are not affordable. Whether this is the case was tested by comparing the {\RCF} from a $L=128 \mpcoh$ box to the corresponding {\RCF} computed from a simulation run in a $L=512 \mpcoh$ box. As the latter curve (requiring a simulation with $N^3=4096^3$ particles) is already very expensive to produce, we did this test for only one single cosmology. Further, we tested if the {\RCF} $f_{\rm res}^{3\to8}(z,k)$ depends on the box size and we found that it does not in any other way than the fact that $k_{\rm min}=2\pi/L$ is of course changed. There is a limit to the minimally allowed box size, though\corrCarvalho{.} While computing an {\RCF} one must make sure that a box size is chosen such that $f_{\rm res}^{3\to8}(z,k_{\rm min})=1$ for all $z$ of interest such that the resulting {\RCF} can be safely extrapolated to larger scales by setting $f_{\rm res}^{3\to8}(z,k)\equiv1$ for all $k<k_{\rm min}$.

\subsection{Correction strategy}
Our suggested strategy to fight this resolution effect is hence as follows\corrCarvalho{.} Starting from a power spectrum with a resolution parameter of $\linv=3\hompc$, we can resolution-correct it by multiplying it with a \corrOne{$k$- and redshift-dependent} (but cosmology-independent) {\RCF} $f_{\rm res}^{3\to8}(k,z)$, i.e.
\begin{equation}
    P^{\linv=8\hompc}(k,z; \vec{c}) \approx f_{\rm res}^{3\to8}(k,z) P^{\linv=3\hompc}(k,z; \vec{c})\,.
\end{equation}
\corrViel{This corresponds to lowering the minimally resolved mass by roughly an order of magnitude from $\sim 3.3\times 10^9 M_\odot/h$ to ${\sim 1.7\times 10^8M_\odot/h}$} \corrMischa{(corresponding to a Nyquist frequency of ${k_{\rm max}\sim 25 \hompc}$)}.

As we define the {\NLC} for {\EEtwo} with respect to linear theory which is not affected by this resolution effect, the very same correction can be applied to those quantities:
\begin{equation}
    B^{\linv=8\hompc}(k,z; \vec{c}) \approx f_{\rm res}^{3\to8}(k,z) B^{\linv=3\hompc}(k,z; \vec{c})\,.
\end{equation}

\section{Projection Studies using CLASS-based Mock Emulators: training EuclidEmulator2}
\label{sec:hfmockemu}
N-body simulations of the matter field in $w_0w_a$CDM+$\sum m_{\nu}$ cosmologies are expensive, even when the mass resolution is low. It is hence not affordable to run thousands of simulations that would allow for an in-depth analysis of a given emulator that involves (potentially several) training, test and validation sets. In order to develop an understanding of the final emulation error and its dependence on the dimensionality of the parameter space as well as on the size of the experimental design, we construct mock emulators based on {\Halofit} \citep{Bird2012} data. We have followed this strategy already in \citet{Knabenhans2019EuclidSpectrum} where it has \corrPeacock{proved} to yield a reliable estimate for the performance of the real, simulation-based emulator.

In this section we first define the parameter space inside which the emulator is constructed. This parameter space is the same for the {\Halofit}-based emulator as for the actual simulation-based {\EEtwo}. In a next step we (approximately) optimise the hyperparameters of the model in order to use the resulting ``architecture'' for the computation of learning curves\corrCarvalho{,} etc. We shall then apply the findings to the training of the actual, simulation-based {\EEtwo}.

\subsection{Definition of the parameter space}
\label{DeltaRegion}
For {\EEtwo} we consider \corrOne{{\CDM}} models with dynamical \corrOne{{\DE}} and including massive neutrinos, often abbreviated as $w_0w_a$CDM+$\sum m_{\nu}$ models. More precisely, this means that we parametrise the considered cosmologies via the following eight parameters:
\begin{itemize}
    \item \Omb, the total baryon density \corrOne{parameter} in the Universe\,,
    \item \Omm, the total matter density \corrOne{parameter} in the Universe\,,
    \item \Sumnu, the sum of masses of all neutrino families\,,
    \item $h$, the dimensionless Hubble parameter\,,
    \item {\ns}, the spectral index\,,
    \item $w_0$, the time-independent part of the {\DE} {\EoS} parameter\,,
    \item $w_a$, the linearly \corrOne{scale factor}-dependent part of the {\DE} {\EoS} parameter\,,
    \item {\As}, the spectral amplitude\,.
\end{itemize}
The radiation density {\Omrad} is given by the \gls{CMB} temperature which we fix at 2.7255 K. The {\DE} density {\OmDE} is then inferred from the flatness condition \corrOne{given by \autoref{eq:flatness} with $\Omega=1$ on the left-hand side}.

We do not want to include any prior knowledge about a most likely cosmology into the construction of the emulator other than the assumption that a flat $w_0w_a$CDM+$\sum m_{\nu}$ model is sufficiently accurate in order to describe our Universe. In the context of {\EEtwo}, we explicitly ignore alternative gravity and other more exotic cosmological models. Emulators for such models have been published by other research groups as e.g. \citet{Winther2019EmulatorsCDM, Giblin2019OnCosmologies}. We thus apply flat (uniform) priors \corrCarvalho{to} each of the eight input parameters. In order to have a well defined, normalised prior probability distribution function we thus need to define compact intervals along each dimension over which the final emulator will be defined. Mathematically this means that we have to define intervals $[a_1,b_1], \dots, [a_8,b_8]$ such that the final parameter box $\Pi$ is the Cartesian product of all intervals:
\begin{equation}
    \Pi := \varprod_{i=1}^8[a_i,b_i]
\end{equation}
The choice of the interval boundaries is mostly arbitrary and depends mainly on the tasks that \corrPeacock{will} be tackled by the emulator. Without imposing any restrictions, we assume that {\EEtwo} will be mostly applied \corrCarvalho{to} {\MCMC} searches of the cosmological parameter space to solve the inverse problem of finding the parameter values best describing our Universe. It is reasonable to assume that these values are not too far away from the current best-fit values as published by modern cosmological experiments. We \corrPeacock{centre} the parameter box for {\EEtwo} around the cosmology ``EE2\corrPeacock{centre}'' which is defined in \autoref{tab:parbox}. Notice that this central cosmology is identical to the \corrOne{Euclid Reference Cosmology} (up to two decimal places) for all dimensions but the sum of the neutrino masses. We are thus left with the definition of the width of the intervals along each dimension in such a way that the resulting parameter box remains small enough that a sample of size $\sim200$ (corresponding to our computational budget) contains enough information to achieve a generalisation error of $\lesssim 1\%$. To determine these in a systematic way we run a number of full N-body simulations along each parameter axis, both below and above the central value. This allows us to determine by how much each individual parameter has to be varied in order to cause roughly a $\pm 10\%$ variation in the output {\NLC} factor (the emulation target). Following this prescription, the output variation is an order of magnitude larger than the uncertainty in the output ensuring significant discrimination power while keeping the parameter box, \autoref{tab:parbox}, reasonably small. Yet, for certain tasks (particularly in the field of weak gravitational lensing), the resulting parameter box may be too small. However, further away from the central cosmology\corrCarvalho{,} 1\% accuracy is \corrPeacock{no longer} necessary and {\EEtwo} could be extended via a multi-fidelity procedure. In addition, it is expected that cosmologies outside this parameter box can be ruled out by the linear theory power spectrum alone.

\begin{table}
\centering%
\caption{Parameter box for {\EEtwo} defined through its lower bounds (``min'') and its upper bounds (``max''). The central cosmology of the parameter box (\corrOne{``EE2\corrPeacock{centre}''}), which is {\it almost} identical to the \corrOne{Euclid Reference Cosmology}, is also given here. $\Omega_{\rm rad}$ is the same for all cosmologies, corresponding to $T_{\rm CMB}=2.7255$ K.}
\begin{tabular}{lllll}
& min & max & EE2\corrPeacock{centre} & Euclid Reference\\
\hline
$\Omega_{\rm b}$ & $0.04$ & $0.06$ & $0.05$ & $0.049$\\
$\Omega_{\rm m}$ & $0.24$ & $0.40$ & $0.32$ & $0.319$\\
$\sum m_\nu$ & $0.0$ eV & $0.15$ eV & $0.075$ eV &$0.058$ eV\\
$n_{\rm s}$ & $0.92$ & $1.00$ & $0.96$ &$0.96$\\
$h$ & $0.61$ & $0.73$ & $0.67$ & $0.67$\\
$w_0$ & $-1.3$ & $-0.7$ & $-1.0$ & $-1.0$\\
$w_a$ & $-0.7$ & $0.7$ & $0.0$ & $0.0$\\
$A_{\rm s}$ & $1.7\times 10^{-9}$ & $2.5\times 10^{-9}$ & $2.1\times 10^{-9}$& $2.1\times 10^{-9}$\\
\label{tab:parbox}
\end{tabular}
\end{table}

The range for neutrino masses allows for less than $10\%$ of output variability. We accept this compromise in order to improve emulation accuracy (due to a reduced volume of the parameter box). At the same time we do not expect this cut to have a large impact because the neutrino signal is expected to mostly affect the linear scales. Notice further that\corrCarvalho{,} for reasons discussed in \autoref{subsec:PCAEigval}, we trained {\EEtwo} only on cosmologies with $w_a<0.5$. This does, however, not change the parameter box over which the emulator is defined.

\subsection{\texttt{Halofit} mock data sets}
\label{subsec:mockemu}
For the {\Halofit}-based analyses we create multiple data sets for training and validation. The training data sets were sampled using {\LHS} \corrOne{first published in \citet{McKay1979ComparisonCode}}, just in the same way as reported on in \citet{Knabenhans2019EuclidSpectrum}.
We create a series of \corrOne{{\LH}} samples with different sizes, $n_{\rm ED}\in\{25, 50, 100, 200, 300, 400\}$. For each size we generate $5n_{\rm ED}$ sets and choose the realisation that maximises the \corrCarvalho{minimum} distance between all sampling points.

For validation we create much larger sets than for training. Notice that sampling large sets with {\LHS} is computationally demanding as it scales polynomially with the number of points to sample. Additionally, we are primarily interested in the performance of the emulator inside an axis-aligned ellipsoid inscribed in the parameter box (\autoref{tab:parbox}). {\LHS}, however, is designed to be space-filling and thus the high computational cost comes with a low efficiency as due to the high dimensionality of the parameter space most sample points lie outside that ellipsoid. We hence decided to sample the validation sets purely randomly and filter the sample with an ellipsoidal mask. Following this procedure we generate in total 30 validation sets with roughly 1500 sampling points each (resulting in $45\,000$ validation points).

In addition to these data sets, we have also created data sets for cosmologies organised in a grid of $50\times50$ points in each parameter plane, resulting in another $70\,000$ {\Halofit} evaluations. These sets are on the one hand used for analysis of the {\PCA} eigenvectors (see \autoref{subsec:PCAEigval}) and, on the other hand, also for validation purposes (see \corrOne{\autoref{subsec:learningcurves} and \ref{subsec:hfperformance}}).

For each sampled cosmology in all of the sets mentioned above, we run {\Halofit} and evaluate the matter power spectrum at redshift $z=0$. This result is then divided by the linear matter power spectrum in order to get the {\NLC} factors which ultimately form the data sets of interest.

\subsection{Emulation strategy: PCE}
\label{subsec:PCE}
We use \corrNtelis{a supervised regression technique called} {\PCE} to emulate the {\NLC} factor. We use the implementation of this method in the MATLAB package {\UQLab}\footnote{\url{www.uqlab.com}} \citep{Marelli2014, Marelli2017UQLabExpansion, Marelli2017UQLabAnalysis}. {\PCE} in its generality is well documented in several publications such as \citet{Xiu2002,Blatman2009AdaptiveAnalysis, Blatman2009, Blatman2010, Blatman2011, Marelli2017UQLabExpansion, Marelli2018, Torre2018} and its application to cosmological emulation is discussed in \citet{Knabenhans2019EuclidSpectrum}. As a reminder we repeat that we express {\NLC} factors using {\PCA} and {\PCE} with the following emulation equation
\begin{equation}
    \label{eq:emueq}
    B(k,z; \vec{c}) \approx \mu_{\rm PCA}(k,z) + \sum_{j=1}^{n_{\rm PCA}}\sum_{\alpha\in{\mathcal A}_j^{p,q,r}} \hat\eta_{j,\alpha}\Psi_j^\alpha[f(\vec{c})]{\rm PC}_j(k,z)\,,
\end{equation}
where $\vec{c}$ stands for a vector of the eight cosmological parameters discussed here which is transformed through $f$ into the standard unit hypercube $[-1,1]^8$. The {\PCA} quantities are the mean $\mu_{\rm PCA}(k,z)$ and the eigenvectors ${\rm PC}_j(k,z)$. The actual {\PCE} is given by the inner sum\corrCarvalho{,} with $\Psi_j^\alpha$ being the {\PCE} basis functions, \corrOne{$\hat \eta_{j,\alpha}$} the coefficients and $\alpha$ being an element from a multi-index set ${\mathcal A}_j^{p,q,r}$.

We shall also stress again that this procedure of combining {\PCE} and {\PCA} is the standard approach for emulating vector-valued quantities with {\PCE} \citep{Blatman2013SparseQuantities}. This implies, however, that {\EEtwo}, having a target space of $n_{\rm PCA}$ dimensions, is actually a conglomerate of $n_{\rm PCA}$ single, \corrOne{scalar-valued} emulators.

The actual learning algorithm we use is a regularised (i.e. LASSO-type) version of a \corrCarvalho{least-squares} minimisation called {\LAR}\corrOne{, discussed in detail in \citet{Efron2004LEASTREGRESSION, Blatman2011}}. This regression algorithm minimises the loss function
\begin{equation}
    \hat \eta_{j,\alpha} = \underset{\eta_{j}\in\mathbb{R}^{\vert\mathcal{A}_j\vert}}{\mathrm{argmin}}\;\mathbb{E}\left[\left(\eta_{j}^\top\cdot\Psi_j(\vec{c})-w_{j, {\rm true}}(\vec{c})\right)^2+\lambda\sum_{\alpha\in\mathcal{A}_j}\vert \eta_{j,\alpha}\vert\right]\,.
\end{equation}
Here, \corrOne{$w_{j, {\rm true}}$} denotes the true value of the $j$-th principal component weight and we use $\mathcal{A}_j$ as a shorthand notation for $\mathcal{A}_j^{p,q,r}$ for the sake of readability.

The regularisation term enforces low-rank (i.e. sparse) solutions. This is motivated by the so-called \textit{sparsity-of-effects principle} according to which most of the variance of the underlying model is encoded in interaction terms \corrCarvalho{among} only a small number of parameters. Further, enforcing sparse basis representations serves the purpose of reducing the memory requirements of the emulator code. 

\subsection{Principal component analysis}
\label{subsec:PCAEigval}
\begin{figure}
	\includegraphics[width=\columnwidth]{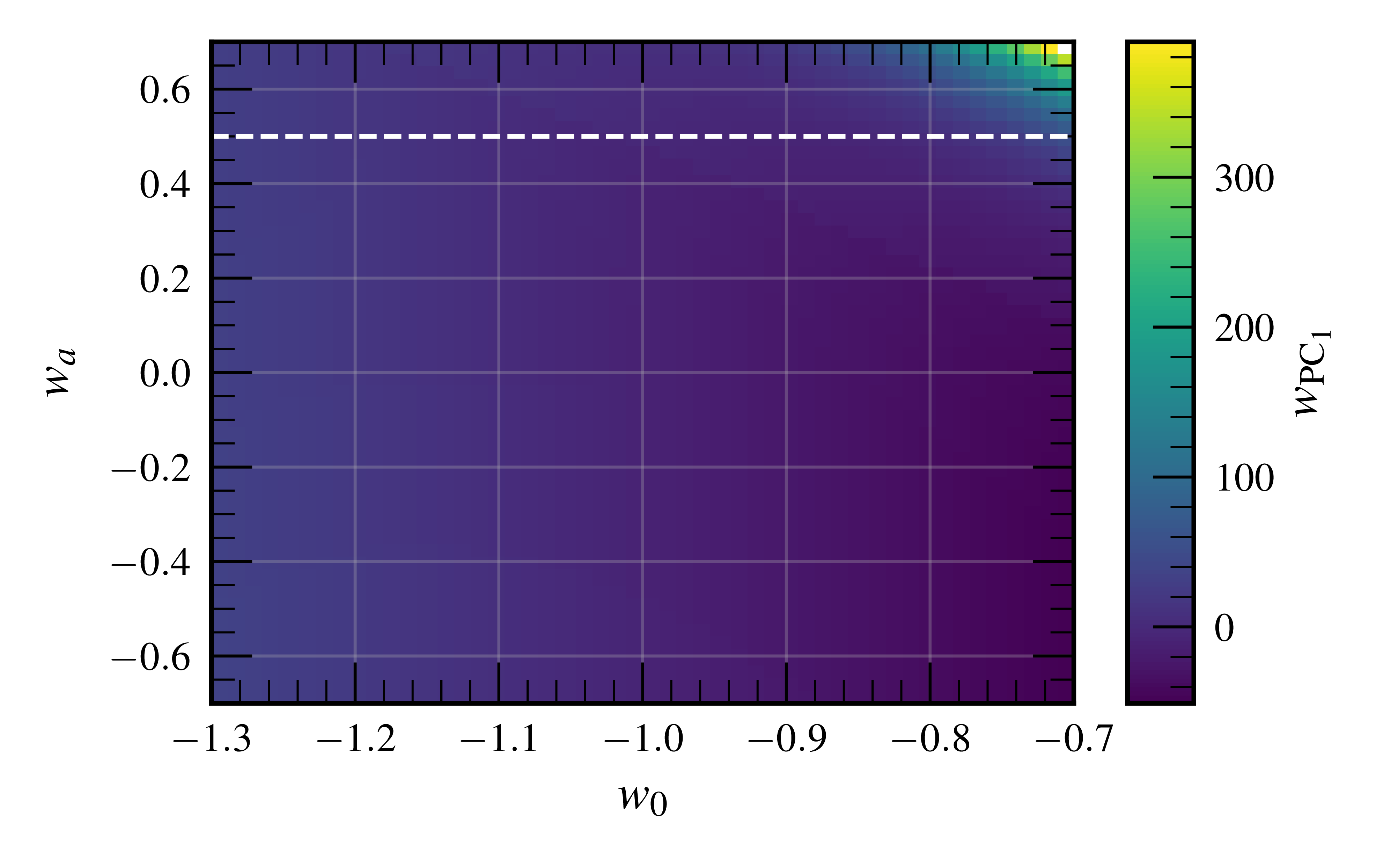}
	\caption{Evolution of the first order principal component weight $w_{{\rm PC}_1}$ as a function of $w_0$ and $w_a$. The exponentially increasing value is evident for $w_0+w_a\to 0$. The white, dashed line indicates where the cut is made in order to avoid the problematic region. We have investigated also less aggressive cuts all of which lead to worse performance of the emulator suggesting that the problematic region is even more extended in higher dimensions.}
    \label{fig:wPC1w0wa}
\end{figure}

Vanilla {\PCE} can only predict scalars. In order to create a {\PCE} emulator for a non-scalar quantity like {\NLC}\corrCarvalho{,} it is thus mandatory to decompose the full signal into a series such that only scalar coefficients have to be emulated. This is done in the standard way using {\PCA} (also used in other emulators based on different techniques than {\PCE}, such as e.g. \citealt{Heitmann2010, Nishimichi2019DarkClustering} and others). The coefficients of the principal components (also called weights or eigenvalues) are thus the quantities that are actually emulated. \corrPeacock{To our knowledge, the dependence of those coefficients on the
cosmological parameters has not been investigated in any of the
papers about cosmic emulators employing PCA published over the last
decade.}
As we use a polynomial regression method of finite order for emulation, we implicitly assume that the dependence of the PC weights on the cosmological parameters is sufficiently polynomial. To investigate if this assumption is justified we perform a {\PCE} on each set of the above mentioned 2\,500 data points sampled in the coordinate planes and plot the resulting first order coefficients $w_{\rm PC_1}$ as heatmaps. The $(w_0,w_a)$ plane stands out as for $w_0+w_a\to 0$ the corresponding first order coefficient grows exponentially (see \autoref{fig:wPC1w0wa}). From a physical point of view this does not come as a surprise as a cosmology with such a {\DE} {\EoS} is highly exotic as it implies a {\DE} with an almost matter-like nature shortly after the big bang. Such a cosmology is not of interest to us as it is highly unrealistic. We have thus identified a clear non-polynomial dependence in the functions we try to emulate. We mitigate this problem by masking out the critical region. In practice, we train our emulator based on all cosmologies within our training set but exclude the very 19 cosmologies that do not meet the condition:
\begin{equation}
    w_a < 0.5\,.
\end{equation}
As we will show in \autoref{subsec:learningcurves}, this modification of the training data set is crucial for the performance of the emulator (although its size is decreased from 127 to only 108 training cosmologies!). This does, however, not restrict the allowed input parameters of the resulting emulator in any way. Clearly, the generalisation performance of the emulator is considerably worse in the masked region than for cosmologies with $w_a<0.5$. We have tried other, less aggressive cuts, too (e.g. cutting along $w_0+w_a<0.5$) but have found that only the cut along $w_a<0.5$ leads to a satisfactory generalisation performance.

\subsection{About error measurements}
\label{subsec:errs}
As in all {\ML} and {\UQ} tasks, error quantifications play a central role in this work. Ultimately, we are primarily interested in creating an emulator that generalises well in an $L_1$ sense in the cosmological parameter space. To be more precise, we try to minimise the generalisation error of the emulator given by the maximum of the relative mean absolute error:
\begin{equation}
    \varepsilon_{\mathrm{maxrMAE}} = \underset{k,z}{\mathrm{max}} \left\langle\left\vert \frac{B_{\rm emu}(k,z; \vec{c})-B_{\rm true}(k,z; \vec{c})}{B_{\rm true}(k,z; \vec{c})}\right\vert\right\rangle_{\vec{c}\in C}\,.
\end{equation}
Notice that the mean indicated by the angle brackets is taken over the cosmologies $\vec{c}$ \corrOne{defined} in the parameter space $C$ given in \autoref{tab:parbox}. The maximisation, on the other hand, is performed over the non-regressed parameters $k$ and $z$. This generalisation error will be approximated by a validation error of the form 
\begin{equation}
\label{eq:defmaxrMAE}
    \hat{\varepsilon}_{\mathrm{maxrMAE}} = \underset{k,z}{\mathrm{max}} \frac{1}{N_{\rm val}}\sum_{\vec{c}\in C_{\rm val}}\left\vert \frac{B_{\rm emu}(k,z; \vec{c})-B_{\rm true}(k,z; \vec{c})}{B_{\rm true}(k,z; \vec{c})}\right\vert
\end{equation}
where $C_{\rm val}$ designates the set of validation cosmologies.

However, since an estimate of the above error requires a validation set which we do not have readily available in all situations, we also often use a cross-validation error metric as an alternative. This is given by the {\LOO} error defined as
\begin{equation}
    \varepsilon_{{\rm LOO}, j} = \frac{\sum_{i=1}^{n_{\rm ED}}\left[w_{{\rm true},j}\left(\vec{c}^{(i)}\right)-w_{{\rm emu},j}^{{\rm PCE}\backslash i}\left(\vec{c}^{(i)}\right)\right]^2}{\sum_{i=1}^{n_{\rm ED}}\left[w_{{\rm true},j}\left(\vec{c}^{(i)}\right)-\hat\mu_{ w_{j}}\right]^2}.
\end{equation}
Here, $w$ stands for the eigenvalues of the {\PCA} which are the quantities that are actually emulated in this work (for more details on this refer to \citealt{Knabenhans2019EuclidSpectrum}). Each $w_j$ corresponds to the inner sum in \corrOne{\autoref{eq:emueq}} running over the multi-index $\alpha$. To compute this error one trains a {\PCE} emulator on all training example but the $i$-th one \corrPeacock{(indicated by the superscript ${\rm PCE}\backslash i$)}. In this very example the emulator is then evaluated (second term in the numerator). The quantity is finally rescaled by the overall variance of the quantity $w_j$. More details on this quantity and how to compute it efficiently can be found in \citet{Marelli2017UQLabExpansion}. It shall be emphasised that the \corrCarvalho{subscript} $i$ runs over cosmologies in the \textit{training} set and no reference to any validation examples is made. Further, this metric measures the emulator performance not in the {\NLC} space but rather in the more abstract\corrCarvalho{,} associated principal component space (hence the subscript $j$ which refers to the order of the principal component).

\subsection{Hyperparameter optimisation using the \texttt{Halofit} mock data}
The hyperparameters of {\PCE} for emulation of non-scalar quantities are given by
\begin{itemize}
    \item the minimum percentage $a_{\rm PCA}$ of explained variance retained in the {\PCA} (strongly related to the number $n_{\rm PCA}$ of principal components taken into account),
    \item the polynomial order $p$ at which the {\PCE} is truncated,
    \item the maximum interaction $r$ (number of factors per monomial in {\PCE}),
    \item the $q$-norm.
\end{itemize}
Remember that actually there is a separate {\PCE} for each principal component (as described in \autoref{subsec:PCE}). As a result, the hyperparameters $p,q$ and $r$ can be chosen differently for each $i=1,\dots,n_{\rm PCA}$ and the accuracy parameter $a$ is the only hyperparameter that has to be chosen globally for obvious reasons. As training a {\PCE} is relatively cheap, there is no need for a sophisticated optimisation algorithm. Rather, we perform a (partially greedy) grid search over the grid given by
\begin{equation}
\label{eq:hyperparvals}
    \begin{split}
        a_{\rm PCA} &\in \{0.9, 0.99, \dots, 0.99999999\}\,,\\
        r &\in \{1,2,\dots,5\}\,,\\
        p &\in \{2,3,\dots,20\}\,,\\
        q &\in \{0.3, 0.35, 0.4,\dots, 0.9\}\,.
    \end{split}
\end{equation}
The small set of low numbers looked at for $r$ are motivated by the ``sparsity-of-effects'' principle \citep{Marelli2017UQLabExpansion}. Strictly speaking, there are two steps here\corrCarvalho{.} In \corrCarvalho{the} first step a vector of $a_{\rm PCA}$-values is created and\corrCarvalho{, in the second step, for each value of $a_{\rm PCA}$} a multitude of $(p,q,r)$-grids \corrCarvalho{is} searched (one for each principal component). This is important because the optimal point in the $(p,q,r)$-grid is chosen based on a different criterion than the optimal $a_{\rm PCA}$ value. The former is chosen based on a minimisation of the {\LOO} cross-validation error without ever seeing a validation point. This happens entirely on the level of principal component weights (i.e. eigenvalues of the covariance matrix) and thus this step is performed independently for each principal component. For identification of the (near-)optimal $a_{\rm PCA}$ value, in contrast, an emulator is trained for each value of $a_{\rm PCA}$ and evaluated on a validation set. We then aim to minimise $\hat\varepsilon_{\rm maxrMAE}$ as defined in \autoref{eq:defmaxrMAE}. Notice that\corrCarvalho{,} while the latter error is the correct quantity to look at when judging the overall performance of the emulator, this can only be investigated once a separate emulator is trained (and, as well, a separate set of hyperparameters is optimised) for each principal component. It is of paramount importance to understand that we go through this procedure really only to fix the values for $a_{\rm PCA}$. As $p$, $q$ and $r$ can be optimised without the need of a separate validation set, they can be optimised once the final emulator is being trained based on actual simulations without the risk of overfitting to a validation set. After evaluating the $\hat\varepsilon_{\rm maxrMAE}$ for all candidate values of $a_{\rm PCA}$, we conclude that
\begin{equation}
    \begin{split}
        a_{\rm PCA} &= 0.99999
    \end{split}
\end{equation}
is the optimal (i.e. loss minimising) value for this global hyperparameter. If we continue to increase $a_{\rm PCA}$, we find that we are attempting to capture the numerical noise in the simulations, leading to an increase again of $\hat\varepsilon_{\rm maxrMAE}$.

\subsection{Learning curves}
\label{subsec:learningcurves}
\begin{figure}
	\includegraphics[width=\columnwidth]{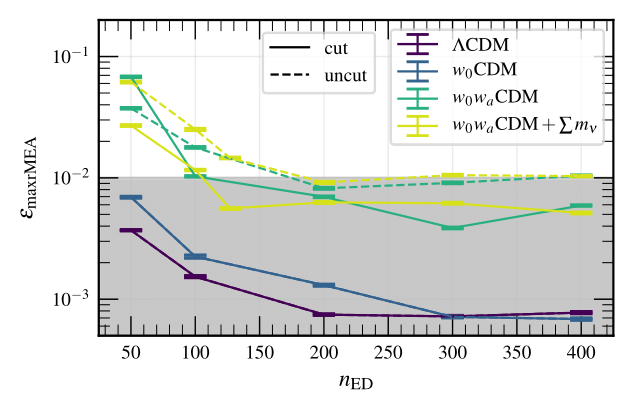}
	\caption{Learning curve for {\CLASS}-based emulators for four different cosmological models. On the $y$-axis the estimated mean of the validation error distribution over the 30 different validation sets is plotted. The error bars show the standard error of this estimated mean. Notice that we explicitly plot the point at $n_{\rm ED}=127$ for the $w_0w_a$CDM+$\sum m_\nu$ model as this point corresponds to our actual training set. The learning curves have been computed (i) taking all training examples into account (dashed lines, labelled ``uncut'') and (ii) ignoring those training examples with $w_a>0.5$ (solid lines, labelled ``cut''). For the {\LCDM} and {\wCDM} models this distinction makes no difference because for those models $w_a$ was fixed to 0.} Validation examples were sampled randomly inside axis-aligned hyperellipsoid inscribed in the parameter box (see \autoref{subsec:mockemu} for a description of the data sets).
    \label{fig:learningCurves}
\end{figure}

Now that we have defined the bounds for all parameters, have created the necessary data sets and optimised the global hyperparameter $a_{\rm PCA}$, it is natural to ask the following two questions:
\begin{itemize}
    \item How does the validation error of the emulator decrease as the number of training examples increases?
    \item Given a fixed number of training examples, how does the validation error increase as a consequence of adding the two additional dimensions compared \corrCarvalho{to} version 1 of \texttt{EuclidEmulator}?
\end{itemize}
To answer these questions we train emulators for different cosmological models with different dimensionalities, namely $\Lambda$CDM (5D), $w_0$CDM (6D, corresponding to {\EEone}), $w_0w_a$CDM (7D) and $w_0w_a$CDM+$\sum m_{\nu}$ (8D, corresponding to {\EEtwo}). Each model is trained on the series of training sets mentioned in \autoref{subsec:mockemu}. The emulators are then validated on the 30 ellipsoidal validation sets. In this context it becomes evident why we have produced so many validation sets: It allows \corrOne{one} to get statistics on the validation error (namely the standard error of the estimated mean error). Notice that such a representative test is by far beyond what is achievable with simulation data as several tens of thousands of simulations with at least moderate mass resolution would have to be run. The resulting learning curves are plotted in \autoref{fig:learningCurves}.

While it is easy to achieve validation errors $<1\%$ for the 5D and 6D models with only 50 training examples, the complexity of the 7D and 8D models is considerably higher, particularly if we train based on examples in the entire original parameter space without masking out the problematic region in the $(w_0,w_a)$ plane (see the discussion in \autoref{subsec:PCAEigval}). In this scenario, we would require $\gtrsim 400$ training examples to reach accuracies of $1\%$ or better. Masking out the region where $w_0+w_a\sim0$ in the training set, reduces the training set size to only 100 to 200 examples.

Notice that in \autoref{fig:learningCurves} we show the error estimate for $n_{\rm ED}=127$ for the case of $w_0w_a$CDM+$\sum m_{\nu}$ cosmologies. This measurement was added in hindsight because it turned out that the actual simulation-based emulator would achieve the target accuracy already with a training set of only 127 simulations selected from an {\LHS} of size 200 (see \autoref{subsec:EDsim} for a short discussion).

\subsection{Application to the training of \texttt{EuclidEmulator2}}

We shall now anticipate some training aspects of the actual, simulation-based {\EEtwo}. We use the same surrogate model in order to train the emulator as for the {\Halofit}-based mock emulator discussed above and for {\EEone}, i.e. sparse {\PCE} combined with {\PCA}. The number of principal components is defined through the threshold value for the minimally explained variance in the data set (which is independent of the size of the training set). As this \corrCarvalho{was} investigated in \autoref{subsec:mockemu} and found that $a_{\rm PCA}=0.99999$ is the optimal value, we can now use the same value for the training of the actual emulator. Having found in the learning curves analysis that a training set of 127 cosmologies is enough to achieve the targeted accuracy, this value for $a_{\rm PCA}$ corresponds to retaining $n_{\rm PCE}=14$ principal components (i.e. the target space of the full vector-valued emulator is 14-dimensional).

Also, \corrCarvalho{we found} for the {\Halofit}-based mock emulator that setting the maximal interaction $r_{\rm max}=4$, the maximal polynomial order $p_{\rm max}=20$, and varying the $q$-norm between $q_{\rm min}=0.3$ and $q_{\rm max}=0.9$ leads to convergence in the selection of terms in the expansion series. So we recycle this here, too. The parameters $p$ and $q$\corrCarvalho{,} as well as the actual interaction number $r$\corrCarvalho{,} are optimised individually for each principal component and listed in \autoref{tab:hyperparvals}. The full training process (i.e. optimising the values for $p$, $r$ and $q$ for each principal component and fitting the coefficients) takes only 9 seconds on a usual MacBook pro with a 2.8 GHz Intel Core i7 CPU. As a result we get a {\PCE} with a total of 574 terms (all other coefficients vanish). Notice that this corresponds to a very small number of terms, i.e. an extremely sparse {\PCE}, as in our case there were 3\,108\,105 terms without sparsification.
\begin{table}
    \centering
    \caption{Table with optimal hyperparameter values for all 14 scalar-valued {\PCE}s of {\EEtwo}. The resulting {\PCE} contains 574 non-trivial terms.}
    \begin{tabular}{cccc}
          PC order  & $p$ & $r$ & $q$\\ 
          \hline
          1&3&2&0.45\\
          2&3&2&0.45\\
          3&4&2&0.5\\
          4&4&2&0.45\\
          5&3&2&0.45\\
          6&4&2&0.4\\
          7&4&2&0.5\\
          8&4&2&0.45\\
          9&4&2&0.5\\
          10&16&3&0.4\\
          11&4&2&0.5\\
          12&12&4&0.5\\
          13&4&3&0.5\\
          14&15&2&0.4\\
    \end{tabular}
    \label{tab:hyperparvals}
\end{table}

\subsection{Performance estimation of the mock emulator}
\label{subsec:hfperformance}
From the learning curves presented above we can expect {\EEtwo} to be sub-1\% accurate at $z=0$ over the entire $k$ range of interest for the available training set of 108 simulations (being a subset of the originally planned set of 200 training examples). While this is the final goal, it is yet interesting to see how the error evolves as a function of cosmology. For this \corrCarvalho{we evaluated} the {\Halofit}-based mock emulator on all 70\,000 validation cosmologies sampled in the 28 parameter planes of the feature space. The results are 28 error maps that we show in Appendix \ref{app:errmaps}. It is clearly visible that the validation error is below $2\%$ (and hence at the same level as the mass resolution-related uncertainty in the simulations at small scales) for the vast majority of cosmologies lying inside the axis-aligned hyperellipsoid inscribed in the parameter box (indicated by a grey ellipse in the error maps). Only for cosmologies with a large value of $w_0+w_a$ the error grows to $\sim5\%$. We reiterate, however, that on average over the entire 8D-hyperellipsoid the error drops below $1\%$. Outside that ellipsoidal region, the errors sometimes exceed the $2\%$ limit. We also designate the $w_a=0.5$ boundary by a grey dashed line.

\section{The training set of EuclidEmulator2}
\label{sec:EE2}
\subsection{Experimental design: sampling}
\label{subsec:EDsampling}

We sample the points in the parameter space defined in \autoref{tab:parbox} using {\LHS}, as we have done already in \citet{Knabenhans2019EuclidSpectrum}. {\LHS} is a very \corrOne{straightforward} sampling technique that is widely used and accepted in the cosmological emulator community \citep{Heitmann2009, Heitmann2010, Heitmann2013, Nishimichi2019DarkClustering, DeRose2019TheCosmology, Gration2019DynamicalEmulation, Rogers2019} and \corrCarvalho{extensively presented} in the statistical sampling literature \citep{McKay1979ComparisonCode, Tang1993OrthogonalHypercubes, Liefvendahl2006AHypercubes, Crombecq2011StochasticsModeling, Damblin2013, Sheikholeslami2017ProgressiveModels, Yang2017ASampling, Garg2017, Swiler2006EvaluationApproximations}. Endowed with an additional optimisation step (we use a distance-based criterion), its main advantage is that it combines good space-filling properties with a high degree of randomness. For an in-depth explanation of the exact steps we go through to generate the sample we refer to \citet{Knabenhans2019EuclidSpectrum}.

We chose to generate a sample with 200 points based on the argument that in \autoref{fig:learningCurves} we show that a training set of this size should be large enough to achieve a validation error below 1\%, while a set of only 100 examples is expected to just miss this requirement in the 8D parameter box. In fact, we use exactly the same {\LHS} of size $n_{\rm ED}=200$ to run the simulations as we used in the investigation of the mock emulators in \autoref{sec:hfmockemu}.

\begin{figure*}
    \includegraphics[width=\textwidth]{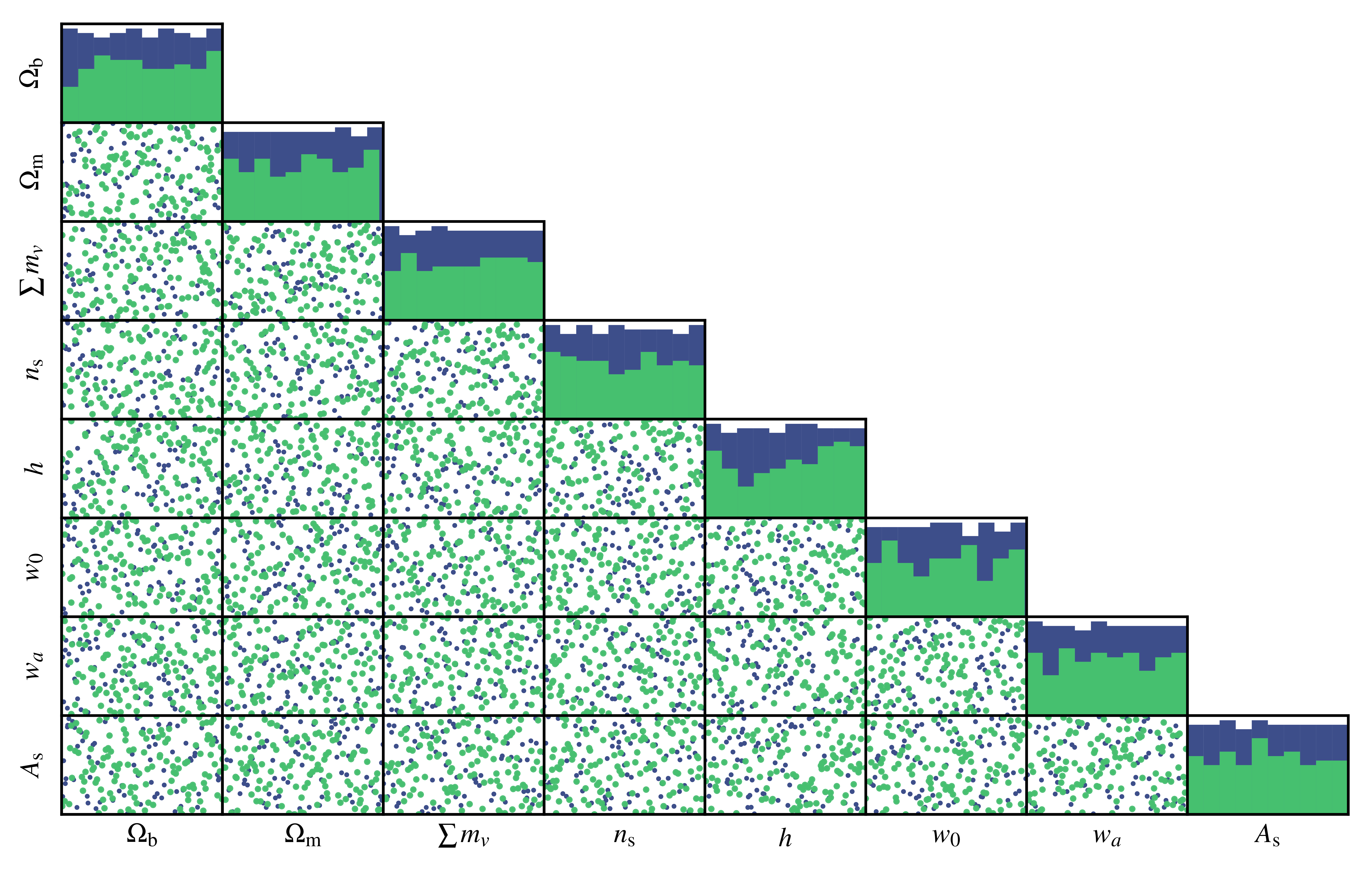}
	\caption{Distribution of {\LH}-sampled cosmologies represented in the coordinate planes of the parameter space. The blue circles and histograms show the full {\ED} of 200 data points while the green dots indicate the subset of 127 training examples actually used to create {\EEtwo}. The ranges of each subplot exactly matches that specified in \autoref{tab:parbox}.}
    \label{fig:ED}
\end{figure*}

The resulting sample of cosmologies is shown in blue in \autoref{fig:ED}.

\subsection{Experimental design: simulations}
\label{subsec:EDsim}
As the generation of each training example (i.e. each pair of simulations per cosmology) corresponds to an investment of about 4000 node hours of computation, \corrCarvalho{we tested} the performance of {\PCE}-based emulators along the way also before the planned training set of 200 cosmologies was completed. Doing so we noticed that the emulator achieved the targeted sub-percent accuracy when trained on only 127 examples. In this case the sample of course is \corrPeacock{no longer} an {\LHS} but rather resembles a random sampling. Consequently, there was no need to invest more time and effort in running the remaining 73 pairs of simulations. The completed set of 127 simulation pairs used for the training of {\EEtwo} is plotted in \autoref{fig:ED}. Notice that it was for this very reason why we also looked at the sample of size 127 when investigating the {\Halofit}-based mock emulator in \autoref{subsec:learningcurves}.

From this fact we can learn two important conclusions for future projects:
\begin{itemize}
    \item At least as long as the marginal distributions of sample points along all parameter dimensions do not have regions where there are no sampling points at all, {\LHS} is not a necessity for good performance of {\PCE} as random sampling works fine too. 
    
    \item When generating an {\ED} is related to large computational costs (and hence to a non-negligible risk of failing to generate a large sample in one go), it is advisable to choose an enrichable sampling technique such as e.g. the {\LHS}-based adaptive response surface method (LHS-ARSM, \citealt{Wang2003AdaptivePoints}) or active learning (as e.g. done in \citealt{Rogers2019}).
\end{itemize}

We thus use an experimental design of 127 (respectively 108 after applying the cut in $w_a$) {\PF} simulations randomly sampling the parameter box. Each simulation samples the power spectrum at 613 $k$-modes and 100 time-steps between \zintm$=10$ and \zfin$=0$. While all this data \corrCarvalho{are} used to compute the emulator, we only allow the user to emulate up to $z_{\rm max}=3.0$ because the overall accuracy decreases considerably for higher redshifts (primarily because the underlying simulations have not converged for higher redshifts as can be extrapolated from \autoref{fig:ResConv}).

\subsection{Post processing: Computation of the NLC}
We compute the {\NLC} for each power spectrum by dividing the nonlinear power spectrum resulting from the simulation by the linear theory power spectrum computed by {\CLASS}. Notice that this is different from what is done for {\EEone} where the {\NLC} was computed via a division by the re-scaled power spectrum measured from the simulation particle realisation at the initial condition. For training the emulator, the {\NLC} is converted into log space because we have shown in \corrOne{\citet{Knabenhans2019EuclidSpectrum}} that this improves the generalisation of the emulator. \corrMischa{We stress that the resolution correction factor introduced in \autoref{sec:rescorr} is \textit{not} applied to the training data set such that users of {\EEtwo} can decide individually whether they want to apply this correction or not to the emulated result.} We compile the {\NLC} data into a data matrix ${\mathcal{D}}^{\rm{CDM+b}} \in {\mathbb{R}}^{n_{\rm{ED}}\times n_z n_k}$. This data matrix is then decomposed into its principal component basis $\{{\rm{PC}}_i\vert i\leq n_{\rm{PCA}}\}$ where $n_{\rm{PCA}}$ denotes the number of principal components taken into account. As a result, the \corrPeacock{$m$}-th row in ${\mathcal{D}}$ can be represented as follows:
\begin{equation}
    {\mathcal{D}}_m = \sum_{i=1}^{n_{\rm PCA}} w_i(c_m) \PC_i(k,z)\,,
\end{equation}
where the argument \corrPeacock{$c_m$} of the PC weight stands for the vector of parameters defining the \corrPeacock{$m$}-th cosmology and the arguments of the principal components are the $k$ mode and \corrCarvalho{the} redshift. We hence build $n_{\rm{PCA}}$ individual training sets defined by
\begin{equation}
    {\mathcal T}_i = \left\{w_i(c_m)\;\vert\;m\leq n_{\rm{ED}}\right\},\;\forall \;i\in\{i,\dots,n_{\rm{PCA}}\}
\end{equation}
that are used to train the $n_{\rm{PCA}}$ individual, scalar-valued emulators.

\section{Emulator Performance, Errors and Sensitivity to Parameters}
\label{sec:performance}
\subsection{Sensitivity analysis}
\label{Sensitivity}
As for {\EEone}, we have again performed a Sobol' analysis to investigate the relative importance of each cosmological parameter on the final {\NLC}. Notice that since the parameterisation of the emulator changed significantly from {\EEone}, it cannot be expected that the Sobol' indices remain unaltered. Clearly, for {\EEtwo}, the matter density parameter {\Omm} dominates the behaviour of the resulting {\NLC}. At the same time, to first principal component order, $\sum m_\nu$ and $n_{\rm s}$ are almost entirely negligible. While in the case of neutrino masses this does not come as a surprise (the effect of massive neutrinos is mostly captured by the linear signal already), one might not have guessed that for the spectral index. This is resolved when looking at the Sobol' indices of the second principal component where the neutrino mass still has almost no impact at all, while the spectral impact becomes actually the dominant parameter.

\begin{figure}
	\includegraphics[width=\columnwidth]{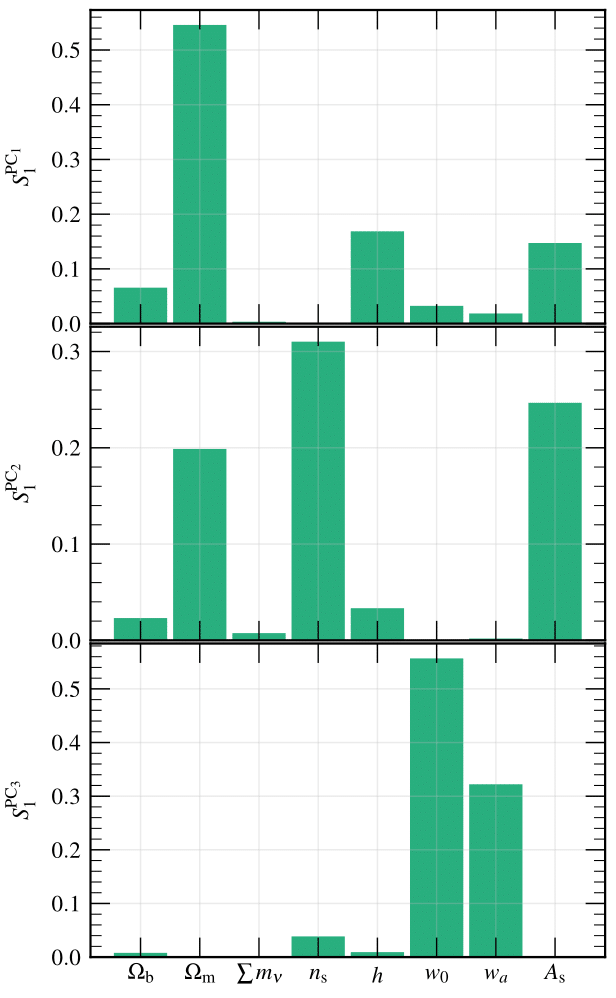}
	\caption{Sobol' analysis plots for the first three principal components. Clearly, {\Omm} is by far the most impactful cosmological parameter of all as its Sobol' index $S_1$ is large for the first two principal components. While the neutrino mass, the spectral index and both {\DE} {\EoS} parameters have only a weak influence on the first order principal component, the spectral index is the most important parameter for \corrOne{PC${}_2$} and $w_0$ and $w_a$ are dominant at third PC level. Neutrino mass becomes only relevant in the seventh principal component (not shown) highlighting the fact that its impact is very small (see discussion in \autoref{subsec:EE1vsEE2}, right panel of \autoref{fig:EE2vsEE1} in particular, and \autoref{sec:numass}).} 
    \label{fig:Sobol}
\end{figure}
The fact that the sum of the neutrino masses is almost entirely negligible when computing the {\NLC} supports our suggestion mentioned in \citet{Knabenhans2019EuclidSpectrum} that to good approximation one can emulate nonlinear power spectra with massive neutrino cosmology by simply computing the corresponding linear power spectrum and multiplying that by an {\NLC} as produced by {\EEone}, i.e. an {\NLC} that does not know anything about massive neutrinos. The test of this hypothesis is deferred to \autoref{subsec:EE1vsEE2} (see \autoref{fig:EE2vsEE1} in particular).

\subsection{Generalisation Performance of \texttt{EuclidEmulator}}
In this section we shall compare {\EEtwo} \corrCarvalho{to} other fast prediction techniques such as {\Halofit} \citep{Bird2012}, {\HMCode} \citep{Mead2016AccurateForces}, {\CosmicEmu} \citep{Lawrence2010,Lawrence2017}, \corrMischa{the very recent emulator based on the BACCO simulation project \citep{Angulo2020}, hereafter referred to as the ``BACCO-emulator'',} and the predecessor {\EEone} \citep{Knabenhans2019EuclidSpectrum} as well as \corrCarvalho{with} {\PKDGRAV} \citep{Potter2016,Potter2016a,Stadel2001} simulations. While the comparisons \corrCarvalho{of {\EEtwo} with {\EEone} and with {\PKDGRAV}}, respectively, can be conducted at the {\NLC}-level, all comparisons with {\Halofit}, {\HMCode} and {\CosmicEmu} are performed at the level of the fully nonlinear power spectrum. To this end we multiply the {\NLC} computed by {\EEtwo} with a linear power spectrum computed by {\CLASS} for the same cosmological parameters.

We compare each pair of predictors in two ways\corrCarvalho{.} On the one hand, we compare them for a set of different cosmologies at redshift $z=0$, while on the other hand we chose a single cosmology equal to the \corrOne{Euclid Reference Cosmology} but with a higher total neutrino mass for comparison at different redshifts $z\leq2$.

For the comparison between {\EEtwo} and {\PKDGRAV} we have used a small validation data set containing three validation cosmologies. 

In order to compare {\EEtwo} to {\EEone}, we primarily focus on two extreme cases: \corrCarvalho{the} \corrOne{Euclid Reference Cosmology} as defined in \citet{Knabenhans2019EuclidSpectrum}, once with massless and once with massive neutrinos.

For all comparisons with {\Halofit}, {\HMCode} and {\CosmicEmu} in this section, we choose the cosmologies from a set of 291 cosmologies, put together by an LHS of size 200, cosmologies along the coordinate axis and along one of the diagonals. From this set, we filter out all cosmologies that are not accepted by any of the emulators. This results in a set of 84 comparison cosmologies \corrMischa{(47 in the case of the comparison to the BACCO-emulator)}.

\subsubsection{Comparison of {\EEtwo} and {\PKDGRAV} simulations}
\label{subsec:ee2topkd}
We start our series of comparisons by checking how well {\EEtwo} is able to approximate simulation data. To this end, we generate a validation set of {\PF} simulations with the same resolution as the training data. The validation set contains only three cosmologies (all unseen by the training process) because the generation of a significantly bigger training set is too expensive. These three cosmologies are all sampled from the ellipsoid inscribed the parameter box with axes given by the limits of each parameter range.

\begin{figure}
	\includegraphics[width=\columnwidth]{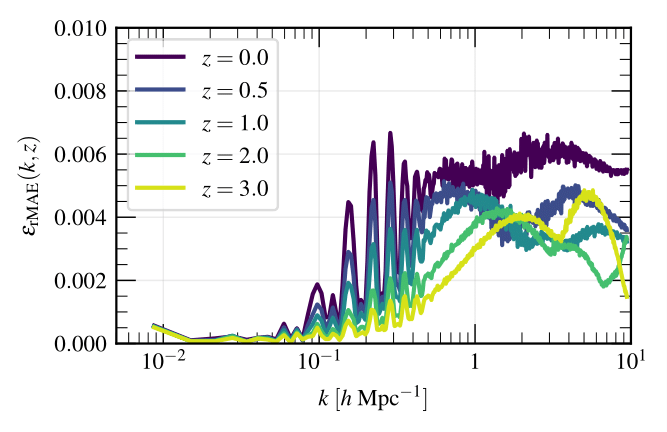}
	\caption{Comparison of {\NLC} factors predicted by {\EEtwo} and ones computed directly from {\PKDGRAV} simulations (averaged over three different cosmologies). The agreement is at the sub-percent level and thus respects the target accuracy.}
    \label{fig:EE2vsPKD}
\end{figure}

We observe in \autoref{fig:EE2vsPKD} that the validation error (given by the relative mean absolute error, rMAE) between emulated and simulated {\NLC} factors is well below $1\%$ for $k$-modes and redshifts of interest to the \corrOne{{\Euclid}} mission, i.e. $k\leq 10 \hompc$ and $z\leq 3$. Of course, as the validation set is very small, there is a substantial uncertainty on this estimate and the rMAE is likely to exceed the $1\%$ limit as one exits the hyperellipsoid inscribed by the parameter box. Yet, the overall error is expected to be dominated by uncertainties in the underlying simulations, especially at very small scales, $k\gtrsim 5\hompc$.

\begin{figure}
	\includegraphics[width=\columnwidth]{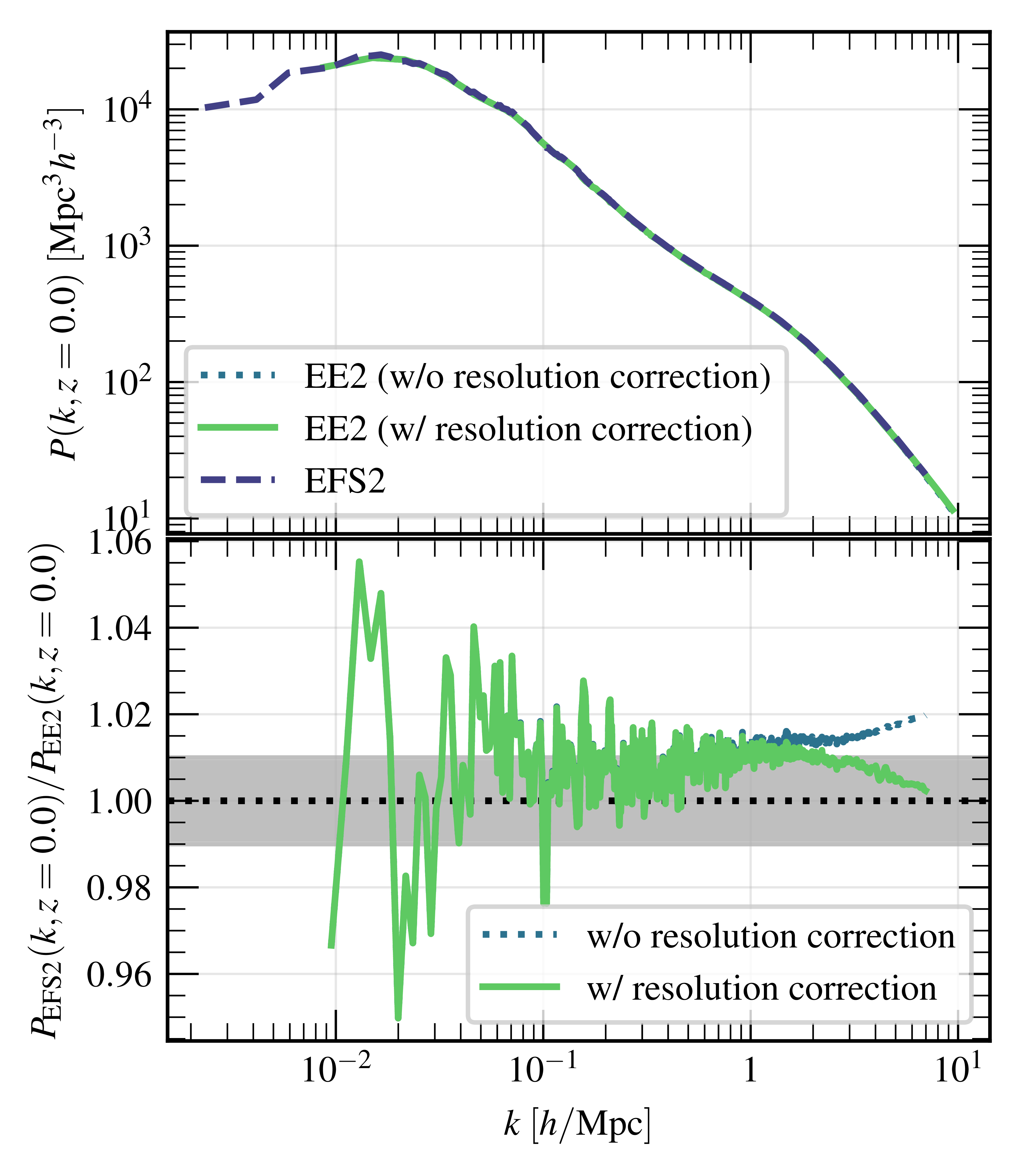}
	\caption{Comparison of full nonlinear power spectra between {\EFStwo} and resolution-corrected product of a {\CLASS} linear power spectrum times an {\EEtwo}-emulated {\NLC} at the \corrOne{Euclid Reference Cosmology} as defined in \autoref{tab:parbox}. The agreement is generally very good and even at the 1\% level or better at small scales. The oscillations at linear scales are expected as {\EFStwo} is not a {\PF} simulation. We also show the comparison to {\EEtwo} when we do not apply the resolution correction factor, which shows a clear 2\% bias at nonlinear scales.}
    \label{fig:EE2vsEFS2}
\end{figure}

In the context of comparing {\EEtwo} to {\PKDGRAV}, it is natural to compare our emulator to the {\EFStwo} simulation. To this end, we evaluate both {\EEtwo} and {\CLASS} at the \corrOne{Euclid Reference Cosmology} defined in \autoref{tab:parbox}. We then produce a nonlinear power spectrum by multiplying the linear power spectrum, the {\NLC} and the {\RCF}. This product is then compared to the {\EFStwo} power spectrum in \autoref{fig:EE2vsEFS2}. Because {\EFStwo} is not a {\PF} simulation, the cosmic variance is clearly visible as oscillations at the level of a few percent at linear scales. Generally, the agreement between {\EEtwo} and {\EFStwo} is at the 1\% level or better for nonlinear $k$ modes. 

\subsubsection{Comparison of EuclidEmulator1 and EuclidEmulator2}
\label{subsec:EE1vsEE2}
\begin{figure*}
	\includegraphics[width=\textwidth]{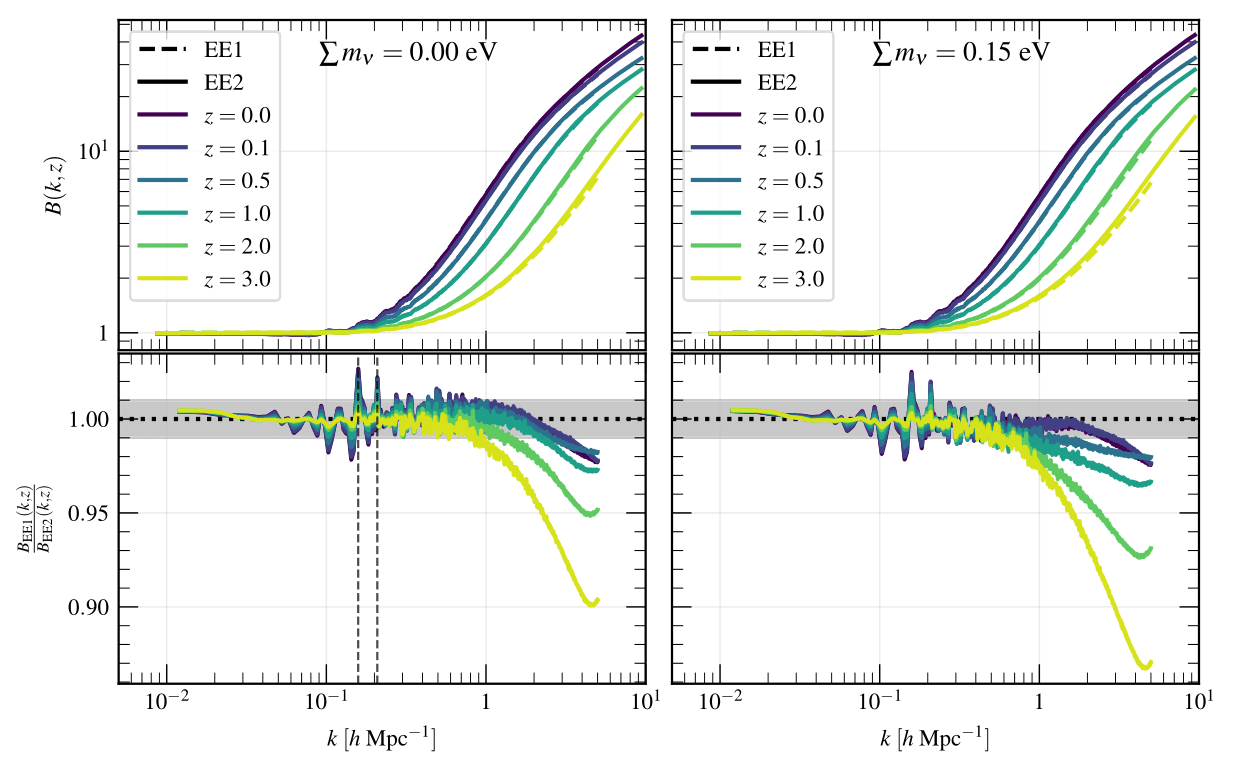}
	\caption{In this figure we compare the {\NLC} prediction of {\EEtwo} to that of {\EEone} for two cosmologies: one for which $\sum m_\nu = 0.0$~eV (left panel) and $\sum m_\nu = 0.15$~eV for the other (right panel). {\As} is the same in both cosmologies, however the value of $\sigma_8$ differs due to the mass difference in the neutrino sector. It is not surprising that the agreement on the largest scales is nearly perfect for both cosmologies as both emulators return values close to unity irrespective of the cosmology by construction. The oscillations around {\BAO} scales are due to cosmic variance. The vertical, black, dashed lines indicate the two $k$-modes where the curves for {\EEone} and {\EEtwo} deviate the most from each other in \autoref{fig:EE1vsEE2vsGRF}. The time-dependent \corrOne{mismatch} on small scales is due to the different mass resolutions of the training of the two emulators. A key point is that this plot shows that the agreement between {\EEone} and {\EEtwo} is not only very good for massless neutrino cosmologies where {\EEone} is supposed to work well by construction, but also for $w_0${\CDM}$+\sum m_\nu$ models which {\EEone} was not designed for. This suggests that {\EEone} is able to predict the nonlinear power spectrum within $1\%$ of accuracy for $z<0.5$ even if the cosmology contains massive neutrinos (of course, the linear power spectrum must be computed taking the neutrino masses into account).}
    \label{fig:EE2vsEE1}
\end{figure*}

It is natural to compare the performance of {\EEtwo} with its predecessor {\EEone} (cf. \autoref{fig:EE2vsEE1}). As both emulators predict the {\NLC}, we can perform the comparison on this level. In a first step we perform this comparison using the version of the \corrOne{Euclid Reference Cosmology} as defined in \citet{Knabenhans2019EuclidSpectrum}, i.e. a cosmology without massive neutrinos.

We observe very good agreement on large scales which is achieved by construction as the variability of the {\NLC} is negligible at these scales. The sub-percent differences at these scales are due to the fact that the simulations volumes of the training simulations underlying both emulators are different.

\begin{figure}
	\includegraphics[width=\columnwidth]{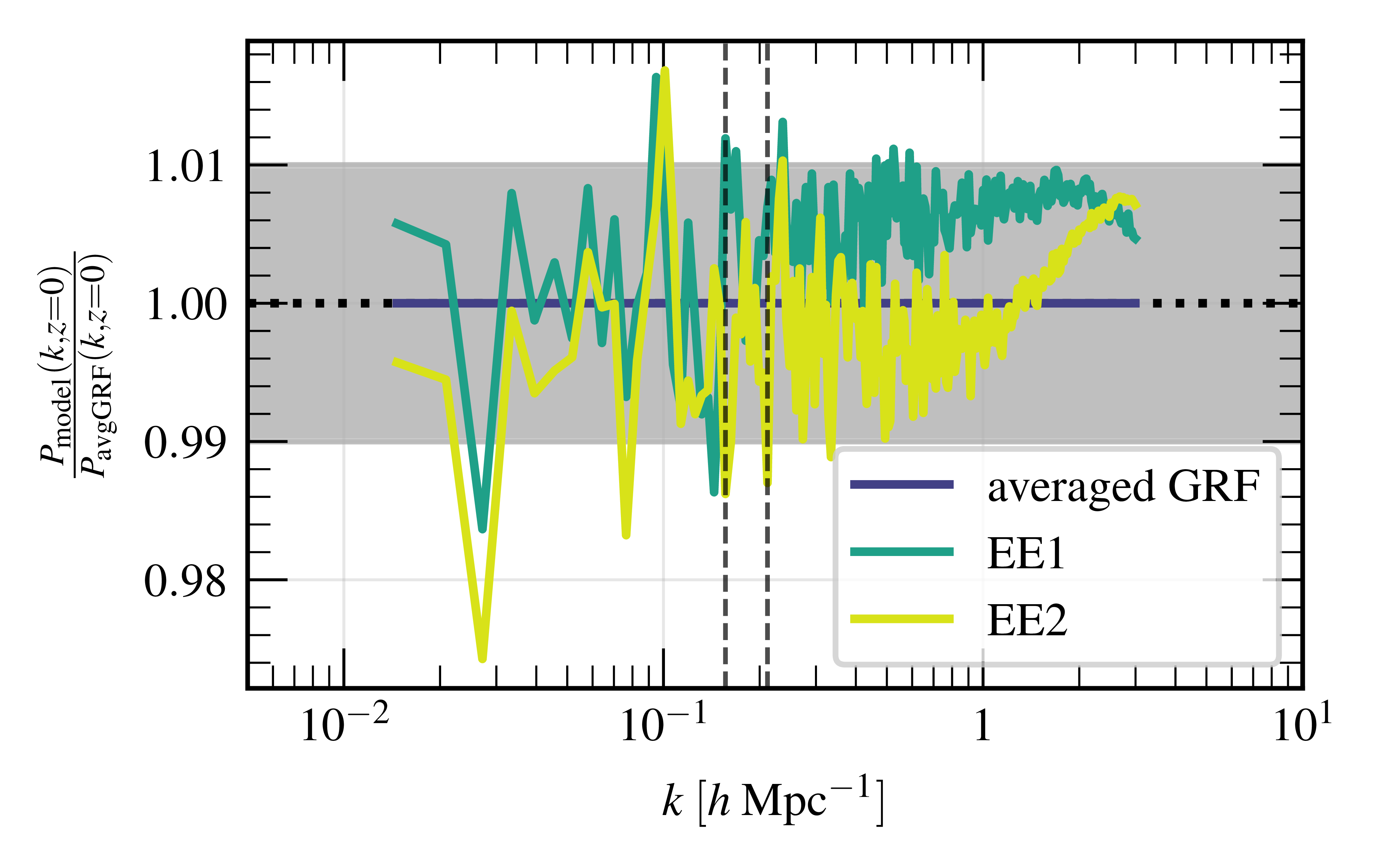}
	\caption{Comparison of {\EEone} and {\EEtwo} to the {\NLC} computed from an ensemble average of fifty Gaussian random field simulations. It can be seen that the cosmic variance is responsible for the oscillatory peaky pattern in \autoref{fig:EE2vsEE1}, as the vertical, black, dashed lines correspond to the two highlighted peaks in that figure.}
    \label{fig:EE1vsEE2vsGRF}
\end{figure}

On intermediate scales around the {\BAO} one observes a peaky pattern at the level of $\lesssim 2.5\%$. We show in \autoref{fig:EE1vsEE2vsGRF} that this can be explained by cosmic variance. We reiterate a point already reported in \citet{Knabenhans2019EuclidSpectrum}: For {\EEone}, cosmic variance is strongest not on large but on intermediate scales. This is because on large scales the cosmic variance is not significantly amplified by nonlinear structure formation. As a consequence the residual (after pairing and fixing) cosmic variance drops out because we compute the {\NLC} for {\EEone} by dividing the nonlinear power spectrum at $z$ by the properly rescaled initial condition. On the other hand, we show in \autoref{fig:VolConv_B} that for {\EEtwo} we choose the volume large enough to render cosmic variance irrelevant. However, on intermediate scales, the residual cosmic variance did not get cancelled out even for {\EEone} as on these scales it is already non-negligibly amplified by nonlinear evolution. As a result, when comparing {\EEone} to {\EEtwo}, one actually divides two signals with oscillatory behaviour on intermediate scales, manifesting itself as oscillations on intermediate scales observed in \autoref{fig:EE2vsEE1}.

Unsurprisingly, {\EEone} underestimates power at small scales compared \corrCarvalho{to} {\EEtwo}. This is simply due to too low a mass resolution of the training simulations of {\EEone}. Here we can confidently report that the \corrPeacock{baseline} in \autoref{fig:EE2vsEE1} given by {\EEtwo} is the (more) correct answer.

Now, as \corrCarvalho{we claimed} above that massive neutrinos \corrCarvalho{do not} have a significant impact on the {\NLC} we should test this hypothesis. To actually do so, we compared predictions of {\EEone} and {\EEtwo} to each other in \autoref{fig:EE2vsEE1}: We evaluated each emulator at the respective cosmology in \autoref{tab:EE1wnu}. The ratio between the two {\NLC} factors is clearly dominated by cosmic sample variance and resolution effects. This suggests that {\EEone} can indeed be used to estimate the {\NLC} to good approximation for $z<0.5$ for $w_0$CDM$+\sum m_\nu$ models. The key point to get the correct answer is to account for the difference in {\Omm}. Given a particular $\mOmm^{\rm EE2} = \mOmb + \mOmcdm + \mOmnu$ for {\EEtwo}, one needs to choose $\mOmm^{\rm EE1} = \mOmb + \mOmcdm$ for {\EEone}, such that $\mOmm^{\rm EE2} = \mOmm^{\rm EE1} + \mOmnu$. As a result, the value for $\sigma_8$ has to be adjusted accordingly. The value $\sigma_8=0.799$, corresponding to $\mAs=2.1\times10^{-9}$ at the ``EE1'' cosmology listed in \autoref{tab:EE1wnu}\corrCarvalho{,} was computed by {\CLASS}.
\begin{table}
    \centering
    \caption{Mutually corresponding cosmologies in order to approximate the {\EEtwo}-based {\NLC} with an {\EEone}-based {\NLC} neglecting massive neutrinos. The relative difference of the {\NLC} factors produced with the respective version of \texttt{EuclidEmulator} are shown in \autoref{fig:EE2vsEE1}, right panel. Notice that $\Omega_{\rm rad}$ is the same for both cosmologies corresponding to $T_{\rm CMB}=2.7255$ K.}
    \begin{tabular}{lll}
            & EE1 & EE2\\ 
          \hline
          $\Omega_{\rm b}$&0.049&0.049\\
          $\Omega_{\rm m}$&0.3154&0.319\\
          $\sum m_\nu$&0.0 eV&0.15 eV\\
          {\ns} &0.96&0.96\\
          $h$&0.67&0.67\\
          $w_0$&\corrOne{$-1.0$}&\corrOne{$-1.0$}\\
          $w_a$&0.0&0.0\\
          $\sigma_8$&0.799&-\\
          {\As} &-&$2.1\times 10^{-9}$\\
    \end{tabular}
    \label{tab:EE1wnu}
\end{table}

\subsubsection{Comparison to \corrPeacock{HALOFIT}}
\begin{figure}
	\includegraphics[width=\columnwidth]{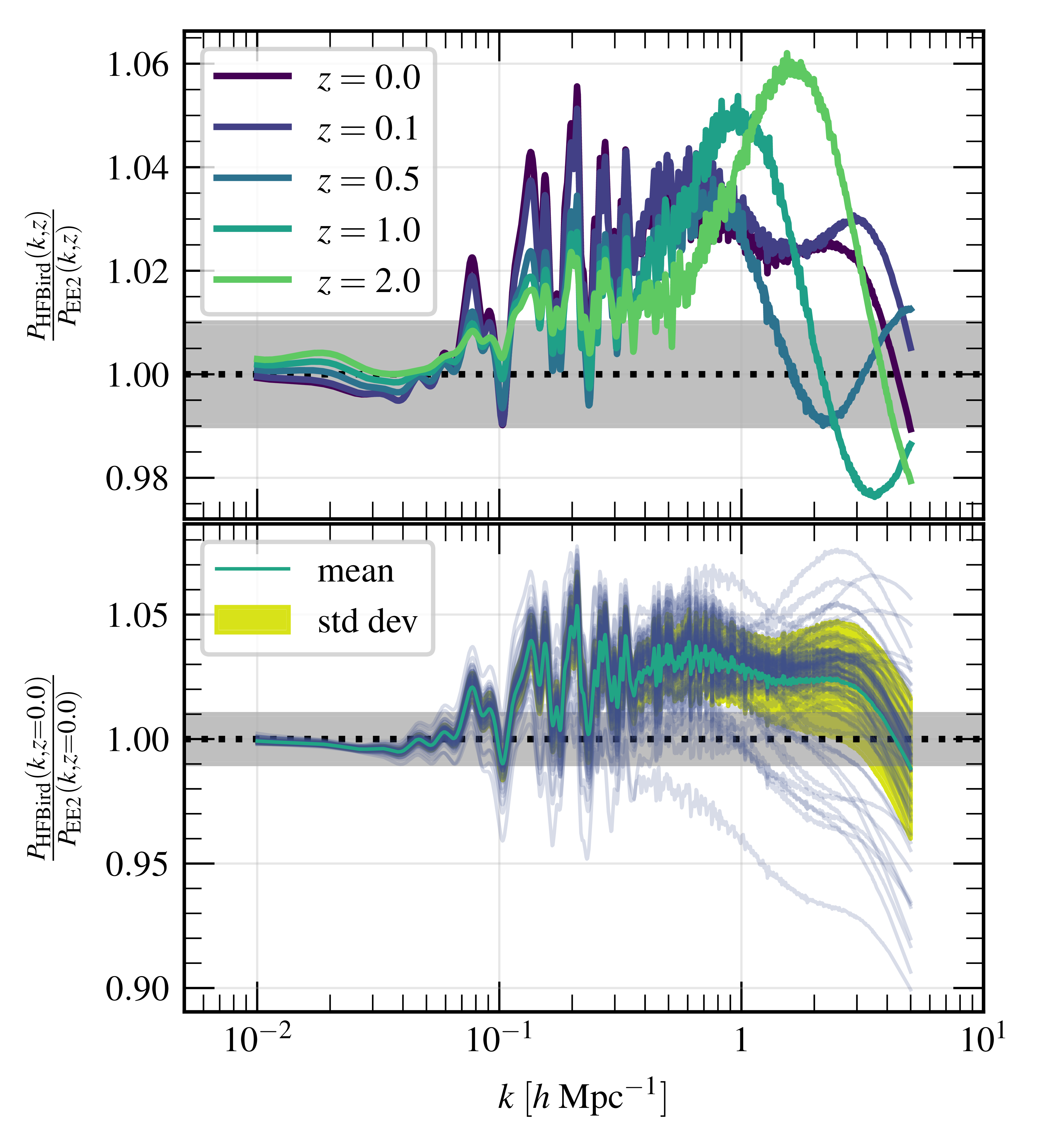}
	\caption{The comparison between {\Halofit} \citep{Bird2012} and {\EEtwo} is consistent with what is found in \citet{Knabenhans2019EuclidSpectrum}. While there are disagreements on intermediate and nonlinear scales, the measured errors stay within the bounds of \corrOne{$5$ to $10\%$} as reported in \citet{Bird2012}.} 
    \label{fig:EE2vsHF}
\end{figure}

{\EEtwo} is compared to the extension of \corrOne{{\Halofit}} by Bird et al. \citep{Bird2012} in \autoref{fig:EE2vsHF}. The comparison across multiple cosmologies shows almost perfect agreement for all cosmologies on large scales. This is expected as {\Halofit} builds on linear theory as does {\EEtwo}. On intermediate scales around \corrOne{{\BAOs}} we find systematic oscillations which are in agreement with what we have found in the corresponding comparison between {\Halofit} by Takahashi et al. \citep{Takahashi2012RevisingSpectrum} and {\EEone} \citep{Knabenhans2019EuclidSpectrum}. \corrViel{While there we attributed those oscillations to {\Halofit}'s inability to capture the {\BAOs} correctly, this may play a less relevant role for this version of {\Halofit}. Rather, the oscillations may be mostly explained by the higher mass resolutions and smaller simulation box sizes used in \citet{Bird2012} compared to the those used in this work. The fact that on average less power is found by {\EEtwo} compared to {\Halofit} (at the level of roughly 3\%) is consistent with the findings presented in Fig. 2 of \citet{Bird2012} where it is reported that \gls{PM}-based neutrino simulations tend to find less power on intermediate to small scales compared to simulations treating neutrinos as particles. This is also why} on small scales we then find an overestimation of power in {\Halofit} relative to {\EEtwo}. The mean including the 1$\sigma$-region stays within the \corrOne{$5$ to $10\%$} error margin, respecting the bounds published in \citet{Takahashi2012RevisingSpectrum, Bird2012}.
The error evolution with redshift looks again very similar to what we have already found for {\EEone}. The systematic oscillations on intermediate scales grow with time while on small scales the disagreement is largest for higher redshifts.

\subsubsection{Comparison to HMcode}
\begin{figure}
	\includegraphics[width=\columnwidth]{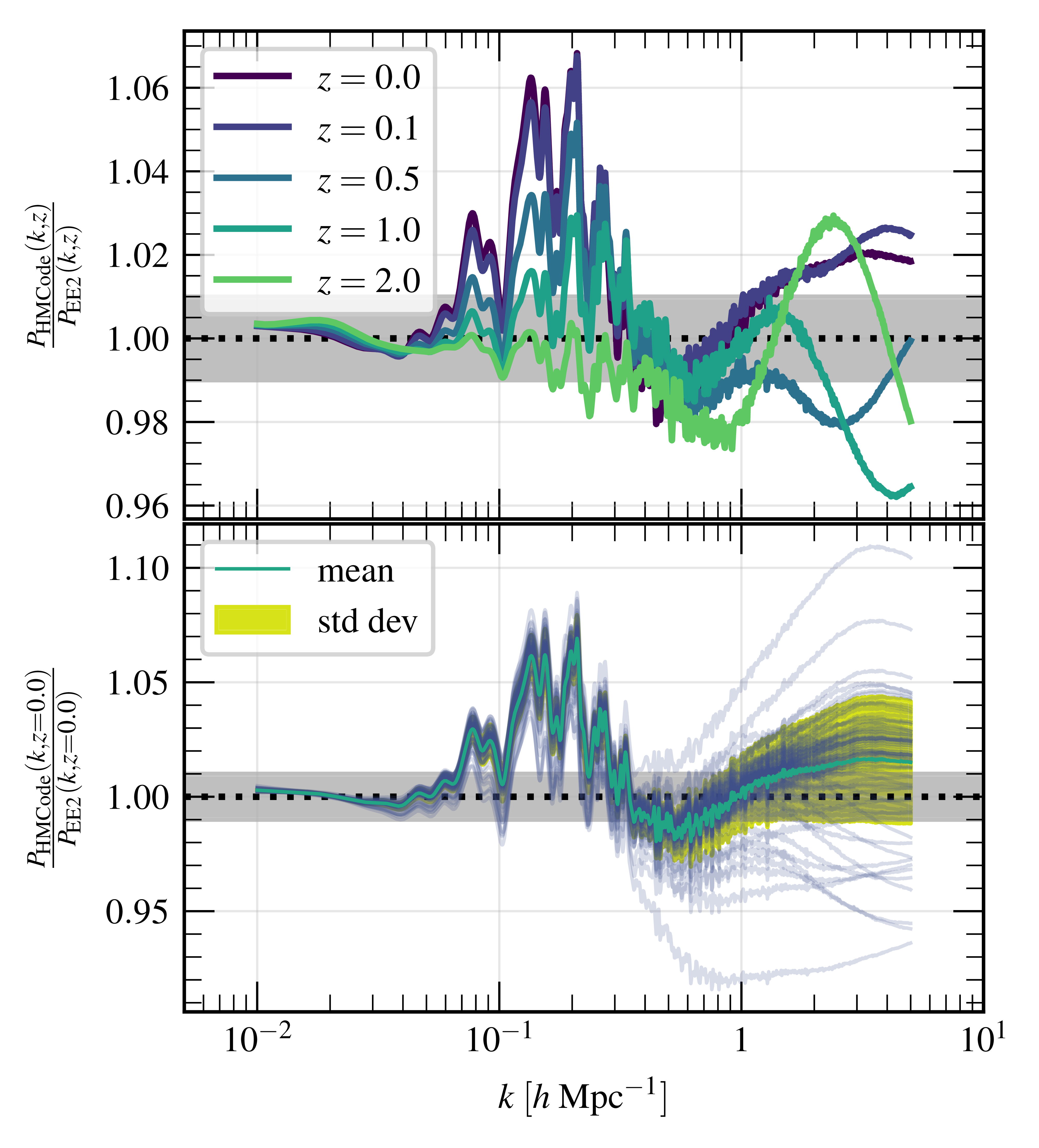}
	\caption{Comparison between {\EEtwo} and {\HMCode} over a set of 84 \corrCarvalho{cosmologies} at $z=0$. The agreement on large scales is nearly perfect while the errors stay within the few percent level over all $k$ as reported in \citet{Mead2016AccurateForces}.}
    \label{fig:EE2vsHM}
\end{figure}

The comparison of {\EEtwo} and {\HMCode} is shown in  \autoref{fig:EE2vsHM}. In \citet{Mead2016AccurateForces} it is reported that {\HMCode} achieves an accuracy of a few percent for cosmologies with massive neutrinos and dynamical {\DE}. We find an agreement at the few percent level both over all tested cosmologies as well as over all redshifts (see \autoref{fig:EE2vsHM}).
Independent of redshift and cosmology the agreement on large scales is virtually perfect. This does not come as a surprise as {\HMCode} is built on top of {\Halofit} which performs almost perfectly on these scales, too. Around {\BAO} scales we find a systematic overprediction of power in {\HMCode} relative to {\EEtwo} (degrading as $z$ increases) which relaxes again at $k\sim 0.6 \hompc$. On small scales, however, the variance in the relative difference is quite large (though always within the few percent limit) both as the cosmology varies as well as over the probed redshift range.

\subsubsection{Comparison to CosmicEmu}
\begin{figure}
	\includegraphics[width=\columnwidth]{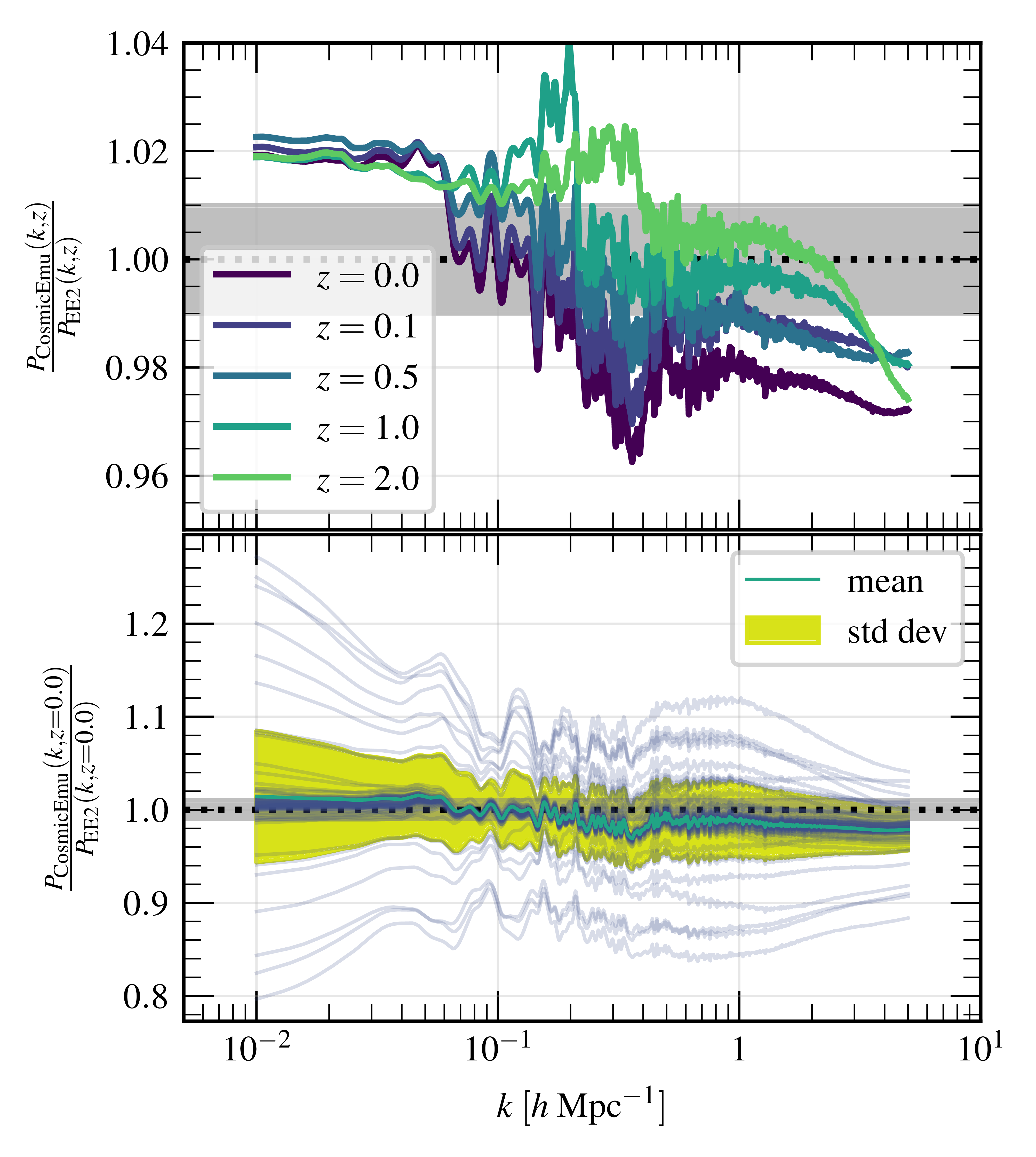}
	\caption{Comparison between {\EEtwo} and {\CosmicEmu} over a set of 84 comparison cosmologies at $z=0$. The mean and even the standard deviation are well below the \corrOne{$10$ to $15\%$} level reported in \citet{Lawrence2017}.}
    \label{fig:EE2vsCE}
\end{figure}

The comparison of {\EEtwo} and {\CosmicEmu} is shown in \autoref{fig:EE2vsCE}. In \citet{Lawrence2017}, they report that for predictions of the 8-parameter model they find an approximation accuracy of \corrOne{$10$ to $15\%$} or better. On average over all probed cosmologies, the comparison error is far below that and it is even relatively constant over the entire $k$-range of interest. Even the standard deviation of the entire set of comparisons is only at the level of five percent over all $k$ (for $z=0$). There are\corrCarvalho{,} however, a few cosmologies for which the comparison is \corrPeacock{significantly poorer}. 
The fact that there is no $k$-region where the comparison is nearly perfect is explained by the fact that {\CosmicEmu} emulates the full nonlinear power spectrum directly while {\EEtwo} emulates the {\NLC} only. It is thus not surprising that there is some generalisation error also on large scales for {\CosmicEmu}, while {\EEtwo} is accurate in this regime by construction. 

\begin{figure}
	\includegraphics[width=\columnwidth]{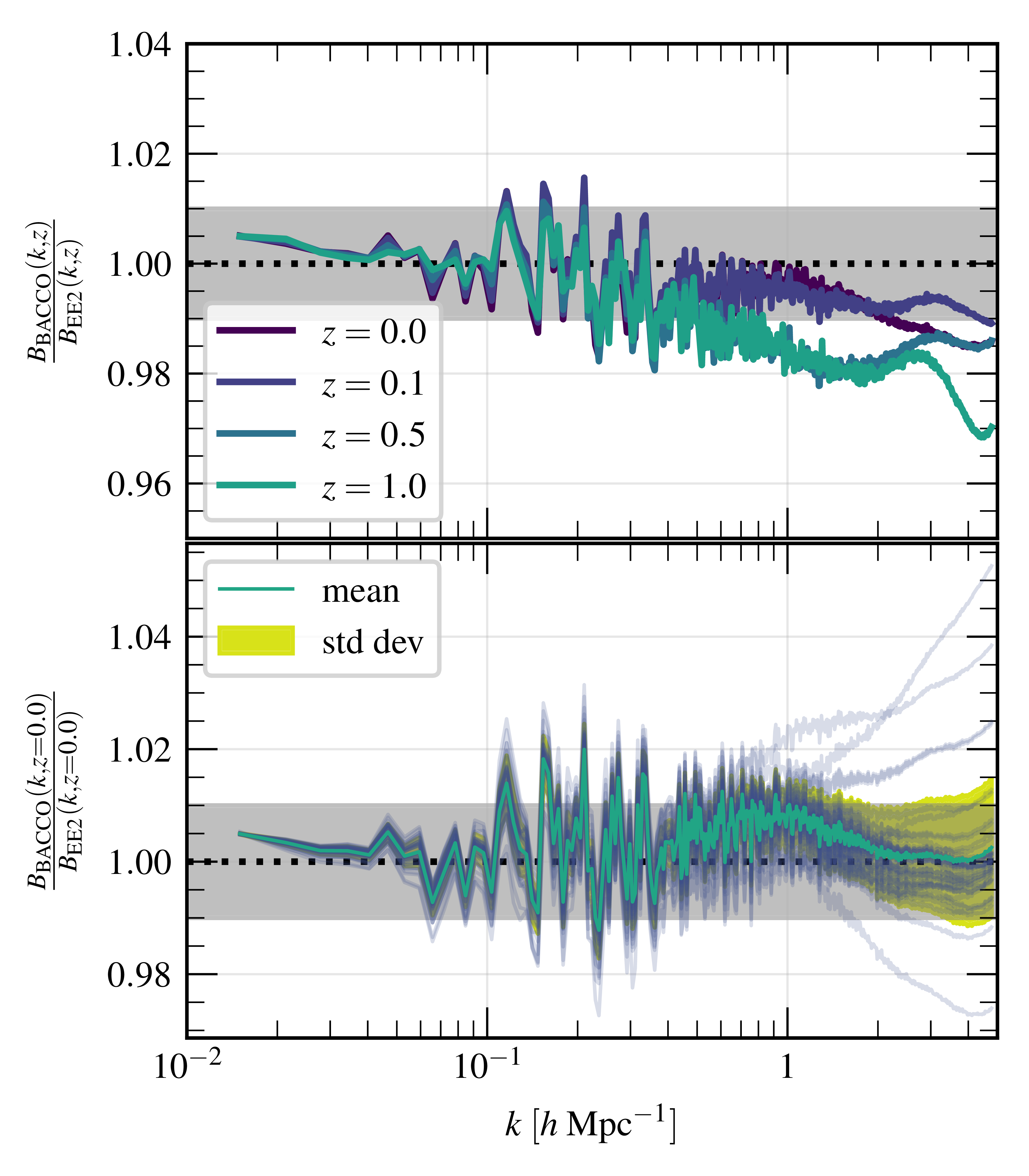}
	\caption{\corrMischa{Comparison between {\EEtwo} and the BACCO emulator over several redshifts. The relative error is mostly smaller than 3\% corrisponing to the expected accuracy of the BACCO emulator reported in \citet{Angulo2020}.}}
    \label{fig:EE2vsBACCO}
\end{figure}

\corrMischa{\subsubsection{Comparison to the BACCO-emulator}
The quantity emulated by the BACCO-emulator \citep{Angulo2020} is also the {\NLC}. For this very reason the comparison between {\EEtwo} and the BACCO-emulator (version 1.1.1) is conducted at the level of the {\NLC} rather than at the fully nonlinear power spectrum level. The result of this comparison is shown in \autoref{fig:EE2vsBACCO}. Clearly, the agreement between these two state-of-the art emulators is extremely good over wide ranges of spatial scales and redshifts. First we discuss the comparison between the two emulators at the Euclid Redshift Cosmology for different redshifts. Notice that the BACCO-emulator allows prediction of the {\NLC} only up to $z=1.5$. For this reason, the comparison at $z=2$ included in the previous comparisons to {\Halofit}, {\HMCode} and {\CosmicEmu} is omitted here. It is found that the agreement at the tested redshifts is mostly at the per cent level, where a suppression of power in the BACCO-emulator is observed relative to {\EEtwo} at small scales. This is explained by the fact that the BACCO-emulator is based on simulation with a resolution parameter of $\linv=3 \hompc$ while the {\EEtwo} {\NLC} were resolution corrected as explained in \autoref{sec:rescorr}.}

\corrMischa{The two emulators were also compared at 47 different cosmologies at redshift $z=0$. The overall agreement is also mostly at the $3\%$ level over the entire $k$-range, where it is reported in \citet{Angulo2020} that the BACCO-emulator is expected to predict the {\NLC} with an accuracy of 3\%. The high-frequency oscillatory pattern at intermediate $k$-scales may be explained by a somewhat poor sampling of the {\BAOs} in the BACCO simulations.}

\section{Exploration of degeneracies in the nonlinear matter power spectrum}
\label{sec:numass}
\begin{figure*}
    \includegraphics[width=\textwidth]{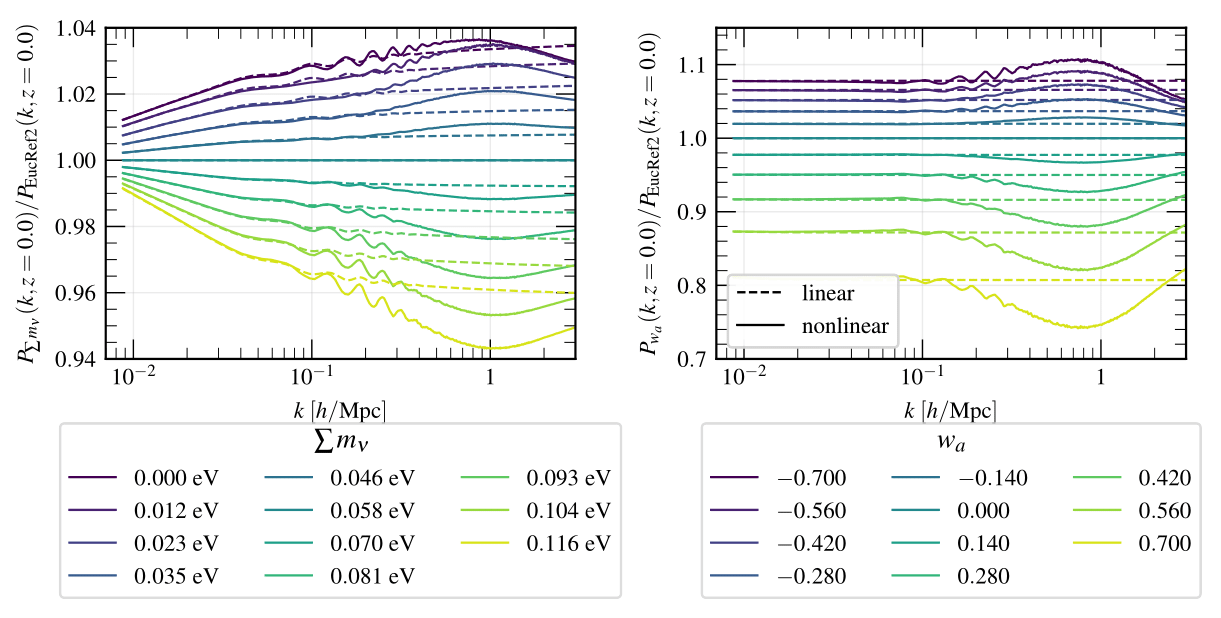}
    \caption{Linear and nonlinear power spectra variation due to varying the cosmological parameters (neutrino mass in the left panel, $w_a$ in the right panel). The power spectra are normalised to the \corrOne{Euclid Reference Cosmology} power spectrum. The {\NLC}s for the nonlinear power spectra have been predicted with {\EEtwo} in all cases. These plots show clearly that the power spectra differ the most at scales around $k\sim 1 \hompc$.}
    \label{fig:variabilityplots}
\end{figure*}

{\EEtwo} is expected to be applied \corrCarvalho{to} parameter forecasts because it is able to very efficiently produce highly accurate predictions of the {\NLC} and hence of the fully nonlinear power spectrum. The \corrOne{{\Euclid}} mission aims at measuring the absolute neutrino mass scale \citep{Laureijs2011EuclidReport} by analysing the effects of neutrinos on cosmic structure formation. Massive neutrinos suppress power particularly at small scales (see e.g. \citealt{Viel2010,Bird2012,Hannestad2012} and others). While this is true also for the linear power spectrum, the effect is largest in the nonlinear power at scales around $k\sim 1\hompc$. The reaction of the linear and nonlinear power spectra to varying the total neutrino mass is shown in \autoref{fig:variabilityplots}. In this figure, the base line is given by the \corrOne{Euclid Reference Cosmology} with $\sum m_\nu = 0.058\;{\rm eV}$. 

\begin{figure}
    \includegraphics[width=\columnwidth]{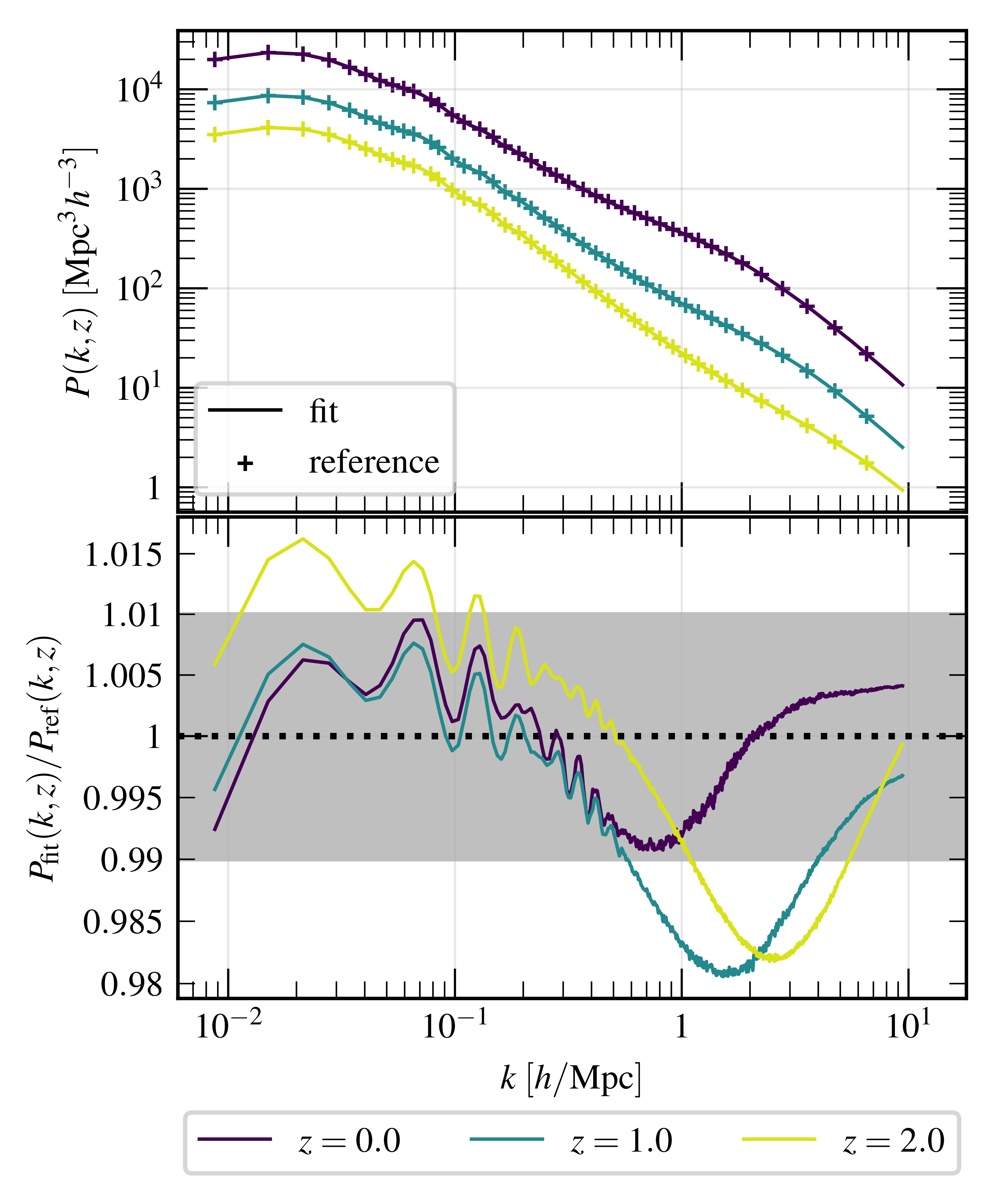}
    \caption{Fit of a nonlinear power spectrum of $w_0w_a${\CDM}$+\sum m_\nu$ cosmology with a cosmology with only massless neutrinos. The purple line in the upper panel is a fit of the reference data to better than $1\%$ at all scales of interest. The different colours correspond to different redshifts. It is evident, that even though the ``fit'' cosmology approximates the reference very well at $z=0$, the degeneracies between the neutrino mass and the other parameters are broken at different redshifts, highlighting the importance of tomographic surveys.}
    \label{fig:nufit}
\end{figure}

While of course a proper Bayesian inference is required to forecast the neutrino mass (as is done e.g. in \citealt{Audren2013NeutrinoErrors}), we shall use {\EEtwo} in order to investigate the uniqueness of the neutrino signal in the nonlinear matter power spectrum. To this end, we use a reference cosmology which has all parameters set identically to the \corrOne{Euclid Reference Cosmology} except the sum of neutrino masses, which is set to $0.15$ eV. We then try to fit the corresponding nonlinear power spectrum with a $w_0w_a${\CDM} cosmology that has only massless neutrinos. Our goal is to fit the reference with an accuracy $\leq 1\%$ on all scales $0.01\hompc\leq k \leq 10 \hompc$. We emphasise that we \textit{do not} perform a proper forecasting by any means, we simply manually adjust all other cosmological parameters but $\sum m_\nu$ until we find a fit. It is worthwhile to note that such a procedure would not be practical without an emulator. The result of this procedure is shown in \autoref{fig:nufit}. We find that the nonlinear power spectra of the two cosmologies defined in \autoref{tab:nufit} agree at a level of better than $1\%$ over the entire $k$ range of interest at $z=0$. 
\begin{table}
    \centering
    \caption{Two cosmologies with nonlinear power spectra that agree to better than $1\%$ over all scales $0.01\hompc\leq k \leq 10 \hompc$ at $z=0$.}
    \begin{tabular}{lcc}
            & reference & fit\\ 
          \hline
          $\Omega_{\rm b}$&0.049&0.049\\
          $\Omega_{\rm m}$&0.3194&0.309\\
          $\sum m_\nu$&0.15 eV&0.00 eV\\
          $\mns$&0.96&0.97\\
          $h$&0.67&0.67\\
          $w_0$&\corrOne{$-1.0$}&\corrOne{$-1.0$}\\
          $w_a$&0.0&0.0\\
          $\mAs$&$2.1\times 10^{-9}$&$2.01\times 10^{-9}$\\
    \end{tabular}
    \label{tab:nufit}
\end{table}
We thus managed to find a cosmology (we call it ``fit'') which is highly degenerate with the reference. The relative difference between the resulting power spectra is below the expected measurement accuracy of the \corrOne{{\Euclid}} mission and hence, \textit{based on this information alone}, \corrOne{{\Euclid}} would not be able to tell these two cosmologies apart.
However, taking the information from higher redshifts into account, the degeneracies are broken. This emphasises the importance of weak lensing tomography for the \corrOne{{\Euclid}} survey in particular and of tomographic surveys in general.

We have further found \corrCarvalho{yet another} two different cosmologies (not shown) whose nonlinear power spectra fit that of the reference cosmology very well only at linear and only at nonlinear wave modes, respectively. This fact makes it \corrPeacock{very clear once again} why modern cosmological surveys need to exploit as much information as possible from both regimes, linear and nonlinear.

\section{Conclusion}
\label{sec:Conclusion}
For this work we have modified {\PKDGRAV} in such a way that {\DM} is not only evolved fully nonlinearly due to self-interaction but also is subject to an additional gravity source due to massive neutrinos, radiation, {\DE} and the metric field perturbations. The latter four species themselves are, however, only evolved linearly. To this end, {\PKDGRAV} has been interfaced with {\CONCEPT} and {\CLASS}. While in older simulations\corrCarvalho{,} as those used in \citet{Knabenhans2019EuclidSpectrum}\corrCarvalho{,} the traditional back-scaling approach has been used for the construction of initial conditions of the N-body simulations, now we employ a novel approach taking advantage of the fully correct linear evolution of particles carried out in Einstein-Boltzmann codes (here {\CLASS}). As a result, {\PKDGRAV} recovers linear theory accurately at all redshifts even in the presence of massive neutrinos.

Moreover, we work with transfer functions in the N-body gauge instead of the more standard synchronous gauge. In this way, results computed with a purely Newtonian N-body code such as {\PKDGRAV} can be interpreted within the framework of general relativity without the need of including general relativistic corrections at the N-body code level.

In a next step we have performed an extensive convergence study with the goal to pin down the smallest volume and the lowest mass resolution necessary in order simulate {\CDM}+baryon {\NLC} factors that have converged at the $1\%$-level all the way up to $k=10 \hompc$. As references, we have used a simulation box of $L=8192 \mpcoh$ for the volume convergence series and a simulation of resolution $\linv=N/L=8\hompc$ for the resolution convergence series. We identify $L=1\gpcoh$ to be just barely enough for the side length of a simulation box necessary to achieve the $1\%$ target accuracy, although this is only true if pairing-and-fixing is used for the construction of the initial conditions. We find that resolution convergence at the targeted level of accuracy is increasingly difficult for higher redshifts. Even at $z=0$ one only achieves the $1\%$ accuracy at $k=10\hompc$ with simulations of \corrOne{$\linv>4\hompc$}, which is beyond our capabilities given the minimal box size. From the convergence series one can further extrapolate that a mass resolution of roughly \corrOne{$\linv\sim6\hompc$} is required to achieve convergence at the $1\%$ level at $k=10\hompc$ at $z\sim 3$. To put this in context we remind the reader that the Euclid Flagship v2.0 simulation ({\EFStwo}), using 4 trillion N-body particles, has resolution parameter of $\linv=4.4\hompc$. 

\corrMischa{In order to correct for the power suppression at small scales resulting from the low mass resolution, we present a way to correct the power spectrum (and equivalently the {\NLC}) curves using a cosmology-independent resolution correction factor, which can be applied in a post processing step. The result of applying this correction to a power spectrum measured in a $\linv=3 \hompc$-simulation is a power spectrum that approximates very closely that obtained from an equivalent simulation with $\linv=8 \hompc$ up to $k\sim 10 \hompc$. }

We have then produced a set of 127 {\PF} simulations of $1 {\rm Gpc}^3/h^3$ with $3000^3$ particles, corresponding to a resolution of $\linv=3\hompc$. This corresponds to a computational cost of roughly 650\,000 node hours which we have invested using the Piz Daint supercomputer located at the Swiss National Scientific Supercomputing \corrPeacock{Centre} (CSCS). At redshift $z=0$ this implies that the simulations are converged at the $2\%$ level at $k=10 \hompc$ (and at $1\%$ up to $k\sim 5 \hompc$). At redshift $z=2.76$ we achieve $1\%$ convergence up to $k\sim 2\hompc$ and $2\%$ up to $k\sim3.5\hompc$. \corrMischa{By applying the resolution correction factor, the convergence is subsequently improved to $\sim 0.5\%$ at $k\sim 10 \hompc$ and $z=0$ and to $\sim 1\%$ at $k\sim 10 \hompc$ and $z=2$ at the cost of introducing an additional source of uncertainty (see \autoref{fig:Rescorr}). We leave the decision about whether or not the resolution correction should be applied to the user of {\EEtwo} by not including it in the training data.}

The key goal of this publication was to construct an emulator which is able to quickly and accurately predict the {\NLC} for $w_0w_a$CDM$+\sum m_\nu$ cosmologies \corrPeacock{up to scales of $k\sim 10 \hompc$}. The emulator takes inputs from within the parameter box defined in \autoref{tab:parbox}. In order to investigate the behaviour of such an emulator and its dependencies on various quantities such as training set size or number of principal components taken into account, we created a mock emulator based on {\Halofit} data. We project that we can achieve a generalisation error of sub-$1\%$ inside the axis-aligned hyperellipsoid inscribed in the parameter box if we exclude a problematic region in the $(w_0, w_a)$-plane in which the first order principal component weight shows exponential behaviour. We exclude this region from the training set by ignoring all cosmologies with $w_a>0.5$, reducing the training set to 108 training examples.

Finally, we construct the actual emulator {\EEtwo} based on the {\PKDGRAV} simulation data containing 108 training cosmologies. We train the emulator using the MATLAB package {\UQLab} within only 9 seconds. The projected error of below $1\%$ \corrPeacock{up to scales of $k\sim 10 \hompc$} is confirmed with a small validation set. \corrPeacock{We stress that for smaller scales {\EEtwo} does not allow the computation of the {\NLC}. At these scales one has to fall back on suitable alternative methods as e.g. {\Halofit}.} Further, {\EEtwo} is compared to multiple other fast predictors such as its predecessor {\EEone} (\autoref{fig:EE2vsEE1}), {\Halofit} (\autoref{fig:EE2vsHF}), {\HMCode} (\autoref{fig:EE2vsHM}) and {\CosmicEmu}  (\autoref{fig:EE2vsCE}). In all comparisons the error bounds as reported on in the corresponding publications have been respected. We have also performed a Sobol' sensitivity analysis (\autoref{fig:Sobol}) which clearly revealed that $w_a$ is a parameter that adds considerable complexity to the underlying model while $\sum m_\nu$ is quasi negligible, at least for the relatively narrow range in $\sum m_\nu$ we have considered.

In first benchmark tests using {\UQLab} we have measured that {\EEtwo} can be evaluated in $\sim0.3$ seconds on a usual laptop. This compares well to the $\sim0.4$ seconds per evaluation of {\EEone} using the python wrapper {\etwopy}. We reiterate here that this implies that the computation of the linear power spectrum by {\CAMB} or {\CLASS} is now the bottleneck in the computation of the fully nonlinear power spectrum.

We have applied {\EEtwo} to investigate degeneracies of the nonlinear matter power spectrum between the total neutrino mass and the other seven cosmological parameters. We have shown that tomographic surveys exploiting both linear and nonlinear scales are critical as it is possible to find different cosmologies with nonlinear matter power spectra agreeing better than $1\%$ at $z=0$ (in our case we have tested a massive neutrino and a massless neutrino cosmology).

Further efforts should be taken in multiple directions\corrCarvalho{.} While the power spectrum (and thus the {\NLC}) clearly belong to the most used summary statistics of cosmic \corrOne{large-scale} structure, \corrOne{higher-order} statistics are becoming more and more used and thus emulators for their prediction are desirable. An example for such a predictor was recently published \citep{Takahashi2019}. A different, more holistic approach is taken in \citet{He2019LearningFormation} where the displacement field is emulated directly, such that any statistic can be derived from the predicted density field. For simulations of resolution as high as the ones used in this work, it is however questionable to what extent such an approach is practical.
Further, in order to \corrOne{assess} more deeply the accuracy of the {\NLC} predictions at small scales, it is not only necessary to estimate the generalisation error and the convergence of the underlying simulations depending on box size and resolution but also to investigate how well different codes agree with each other at the scales under consideration. While such a study has been performed in \citet{Schneider2015} (and augmented by another code in \citealt{Garrison2019AABACUS}) we advocate for new efforts in this direction, as with the new updates to {\PKDGRAV} and developments in other codes the situation may have changed significantly.

Disregarding such uncertainties in the underlying N-body code, at this point we shall summarise the error contributions to {\EEtwo} and their dependence on spatial scales and redshift. At low redshifts, the emulation-only generalisation error is virtually zero by construction on large scales ($k<0.01 \hompc$) such that in this regime the dominant error contribution in the emulator comes from cosmic variance. Based on the results of \citet{Angulo2016}, \corrCarvalho{the cosmic variance} is expected to be sub-percent. At small scales ($k\gtrsim 1\hompc$), cosmic variance is expected to be irrelevant. In this regime, the dominant error contribution (neglecting additional physics such as baryons) is due to emulation itself. The level of the dominant error at high $k$ is estimated to be at the $\sim 0.7\%$ level according to \autoref{fig:EE2vsPKD}. This error is estimated from only a very small sample of validation simulations, however, the error level is also consistent with the estimate in \autoref{fig:learningCurves} and hence we regard this error estimate to be representative at small scales. Estimating the overall error level in the intermediate $k$ range is tricky because several effects contribute errors at a similar level: on the one hand it is evident in \autoref{fig:EE2vsPKD} that an accurate prediction of the {\NLC} around the {\BAO}s is challenging (the observed accuracy is also at the level of $\sim 0.6\%$). At the same time, residual cosmic variance (after pairing-and-fixing) is nonlinearly amplified at these scales. We estimate the error in the intermediate range \corrOne{($0.01\hompc < k < 1\hompc$)} to be at the level of 1\%. The comparison as shown in \autoref{fig:EE2vsEE1} suggests an error at the 2\% level. This may, however, be overly conservative because the cosmic variance in {\EEone} is phase-shifted with respect to {\EEtwo}, leading to an enhancement of errors within this comparison.

The redshift evolution does not greatly change the error contributions discussed above. However, the overall error at intermediate scales is reduced at higher redshifts compared to the low-redshift case. At small scales the resolution effects become the dominant source of error as is visible in \autoref{fig:Rescorr}. As resolution is currently not corrected in a cosmology-dependent manner, the error is expected to be at the level of 1\% at small scales ($k\gtrsim 1\hompc$) and higher redshift ($z \sim 3$).

Last but not least we have seen how the large number of dimensions of the parameter space is really starting to become a major challenge regarding the number of simulations required to arrive at the targeted generalisation error. As more and more dimensions can be expected to be added in the next couple of years, it may be of interest to also compare different emulation strategies \corrCarvalho{to} each other in order to potentially identify the strategy that generalises best based on only very few examples per dimension.

{\EEtwo} is the successor of {\EEone} and will again be published on \url{https://github.com/miknab/EuclidEmulator2}.


\section*{Glossary}
\label{glossary}
\glsfindwidesttoplevelname
\setglossarystyle{alttree}
\printglossary[type=main,title=Codes:]
\printglossary[type=acronym,title=Acronyms:]
\nopagebreak

\section*{Acknowledgements}
\label{ackns}
MK acknowledges support from the Swiss National Science Foundation (SNF) grant 200020\_149848 and the Forschungskredit of the University of Zurich, grant no. K-76102-01-01. Simulations were performed on the PizDaint supercomputer at the Swiss National Scientific supercomputing center CSCS and on the zBox4+ cluster at the University of Zurich. The Euclid Consortium acknowledges the European Space Agency and the support of a number of agencies and institutes that have supported the development of Euclid. A detailed complete list is available on the Euclid web site (\url{http://www.euclid-ec.org}). In particular the Academy of Finland, the Agenzia Spaziale Italiana, the Belgian Science Policy, the Canadian Euclid Consortium, the Centre National d'Etudes Spatiales, the Deutsches Zentrum f\"ur Luft- and Raumfahrt, the Danish Space Research Institute, the Funda\c c\~ao para a Ci\^enca e a Tecnologia, the Ministerio de Economia y Competitividad, the National Aeronautics and Space Ad- ministration, the Netherlandse Onderzoekschool Voor Astronomie, the Norvegian Space Center, the Romanian Space Agency, the State Secretariat for Education, Research and Innovation (SERI) at the Swiss Space Office (SSO), and the United Kingdom Space Agency.




\bibliographystyle{mnras}
\bibliography{EE2paper.bib} 


\appendix
\section{Simulation table}
\label{app:simtable}
Here we summarize all simulations that we have produced specifically for this paper. For each simulation, its unique ID as well as its specifications are listed. The specifications consist of the box size ($L$), the number of particles per side length ($N$), whether it is a {\PF} run (PF yes/no), what order of Lagrangian perturbation theory (LPT) was used to construct the initial conditions, the number of runs, the run time in node hours and on what machine the simulation was executed. Simulations T001 to T127 are the runs that form the actual training set of {\EEtwo} while HRV001-HRV003 were used for the end-to-end test reported in \autoref{subsec:ee2topkd}. The runs VCT1-VT5 were used for the volume convergence test and RCT1-RCT5 for the resolution convergence test (see \autoref{fig:VolConv_B} and \autoref{fig:ResConv}). We used the RES3 and RES8 simulations in order to estimate the variance of the cosmology dependence in the resolution correction factor (see \autoref{fig:Rescorr}). The PF simulation was used in the comparison to the simulations GRF1-GRF50 in order to investigate the cosmic sampling variance in {\PF} simulations (see \autoref{fig:EE1vsEE2vsGRF}). The PV runs were used to estimate the output variance on both boost factor and power spectrum level when one of the parameters $A_{\rm s}$, $w_a$ or $\sum m_\nu$ is varied based on which the parameter box of {\EEtwo} was chosen. The total run time for all simulations sums up to over $700\,000$ node hours.
\begin{table*}
\centering%
\caption{Simulations used for this publication. The table is organized as follows: The training simulations are in the first row followed by the validation simulations in the second and third row. In rows 4 to 13 we list all simulations used in the volume convergence tests (VCT) and the resolution convergence tests (RCT), respectively. Then, in rows 14 and 15 we list the simulations used to investigate the {\RCF} dependence on cosmology. In rows 16 and 17 the simulations used to compare the P+F approach to the traditional ensemble averaging of {\GRF} simulations are listed. Ultimately, we mention the simulations used to estimate the parameter ranges in the rows 18-21.}
\begin{tabular}{lcccccccc}
Simulation identifier&$L$ $[\mpcoh]$&$N$&PF&LPT& number of & total runtime&machine\\ 
&&&&&runs&[node hours]&\\
\hline
T001-T127&1000&3000&yes&1LPT&254&$\sim500\,000$&Piz Daint ($\star$)\\
\hline
\EFStwo&3600&16000&no&1LPT&1&$\sim1\,000\,000$&Piz Daint ($\star$)\\
HRV001-HRV003&1000&3000&yes&1LPT&6&$\sim37\,000$& zBox4+\\
\hline
VCT1&512&170&yes&1LPT&2&10& zBox4+\\
VCT2&1024&342&yes&1LPT&2&24& zBox4+\\
VCT3&2048&682&yes&1LPT&2&118& zBox4+\\
VCT4&4096&1356&yes&1LPT&2&435& zBox4+\\
VCT5&8192&2730&yes&1LPT&2&2780& zBox4+\\
\hline
RCT1&512&512&no&1LPT&1&37& zBox4+\\
RCT2&512&1024&no&1LPT&1&212& zBox4+\\
RCT3&512&1536&no&1LPT&1&919& zBox4+\\
RCT4&512&2048&no&1LPT&1&1987& zBox4+\\
RCT5&512&4046&no&1LPT&1&10\,353& zBox4+\\
\hline
RES3$\_$1-RES3$\_$20&128&384&yes&1LPT&40&$\sim 320$& zBox4+\\
RES8$\_$1-RES8$\_$20&128&1024&yes&1LPT&40&$\sim \corrOne{4800}$&zBox4+\\
\hline
PF&1024&980&yes&1LPT&2&69& zBox4+\\
GRF1-GRF50&1024&980&no&1LPT&50&$\sim\corrOne{3400}$&zBox4+\\
\hline
PV${}_{A_{\rm s}}$&640&1024&no&2LPT&6&$\sim \corrOne{1100}$&zBox4+\\
PV${}_{w_a}$&640&1024&no&2LPT&6&$\sim \corrOne{1100}$&zBox4+\\
PV${}_{\sum_{m_\nu}}$&640&1024&no&2LPT&6&$\sim 670$&zBox4+\\
PV${}_{\rm center}$&640&1024&no&2LPT&1&77&zBox4+\\
\hline
&&&&&\\
${}^\star$ with GPUs&&&&&\\
\label{SimTable}
\end{tabular}
\end{table*}
\section{Error Maps of the \texttt{HALOFIT}-based mock emulator}
\label{app:errmaps}
In this appendix we plot error maps for a two exemplary coordinate planes of the 8D parameter box: the $(\sum m_\nu,h)$- and the $(w_0,w_a)$-plane. The errors are defined as follows:
\begin{equation}
    \varepsilon=\left\vert\underset{k\in[0.01,10.0]\hompc}{\mathrm{max}}\left(\frac{B_{\rm emu}(k,z=0)-B_{\rm sim}(k,z=0)}{B_{\rm sim}(k,z=0)}\right)\right\vert
\end{equation}
The emulator for this investigation was trained with {\Halofit} based on exactly the same 108 cosmologies that were used for the actual, simulation-based {\EEtwo}. We stress that the errors all are measured at $z=0$. The colour bars are ranging from $0\%$ to $8\%$ for both plot panels. The hyperellipsoid inscribed in the parameter box is shown. Notice that the vast majority of cosmologies inside this region features errors at the 2$\%$ level (or even lower). There are, however, also regions with larger errors. This does not contradict the result reported on in \autoref{fig:learningCurves} as the error metric in that figure was averaged over all cosmologies. It is not surprising that particularly validation cosmologies with $w_a>0.5$ feature fairly large errors as in this region there are no training cosmologies. This cut is indicated by a grey, dashed line in all plots with $w_a$ as one of the two dimensions. All parameter planes that have $w_a$ as one of the two dimensions, exhibit larger errors for larger values of $w_a$.

\begin{figure}
\includegraphics[width=\columnwidth]{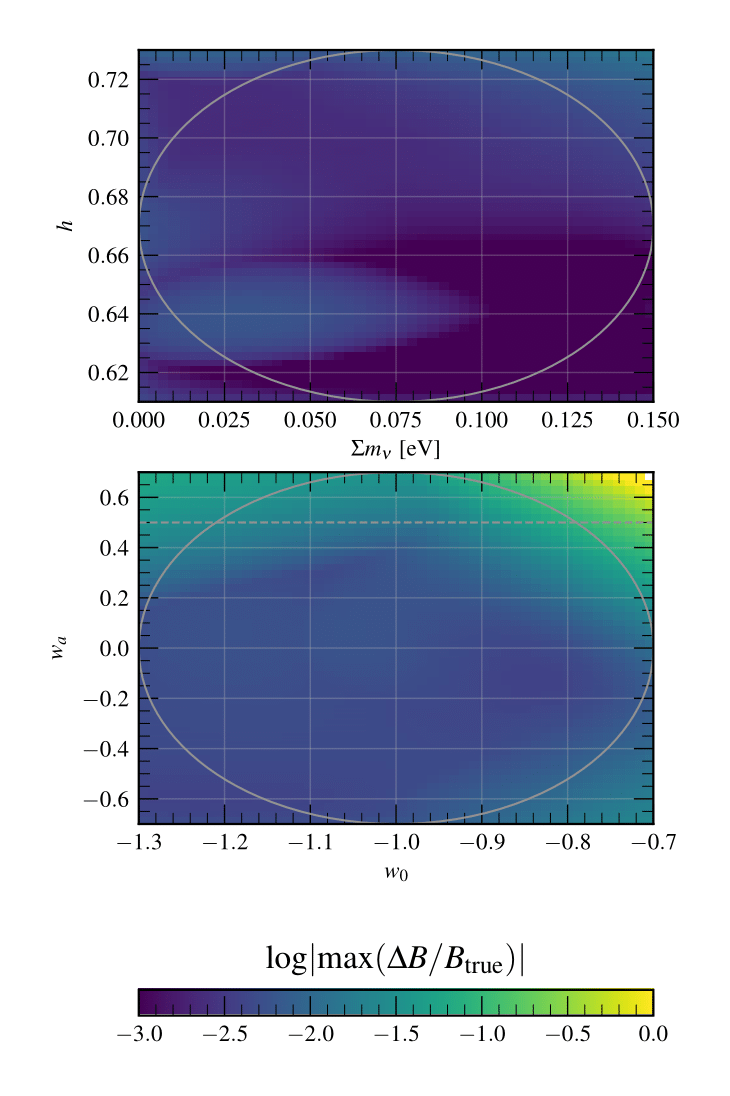}
\caption{Error map of the $(\Sigma m_\nu,h)$-plane (top) and the $(w_0,w_a)$-plane (bottom). These two error maps represent also those of the remaining 26 parameter planes. Most error maps feature only very low errors like the top panel (all errors in the top panel are $\lesssim 1\%$) in this figure. }
\label{fig:ErrorMaps}
\end{figure}


\vspace{0.5in}
\textbf{Affilations}\\
~\\
$^{1}$ Institute for Computational Science, University of Zurich, Winterthurerstrasse 190, 8057 Zurich, Switzerland\\
$^{2}$ Department of Physics and Astronomy, University of Aarhus, Ny Munkegade 120, DK-8000 Aarhus C, Denmark\\
$^{3}$ Chair of Risk, Safety and Uncertainty Quantification, Dept. of Civil Engineering, ETH Zurich, Stefano-Franscini-Platz 5, 8093 Zurich, Switzerland\\
$^{4}$ Institute for Particle Physics and Astrophysics, Dept. of Physics, ETH Zurich, Wolfgang-Pauli-Strasse 27, 8093 Zurich, Switzerland\\
$^{5}$ INAF-Osservatorio Astronomico di Brera, Via Brera 28, I-20122 Milano, Italy\\
$^{6}$ INAF-Osservatorio di Astrofisica e Scienza dello Spazio di Bologna, Via Piero Gobetti 93/3, I-40129 Bologna, Italy\\
$^{7}$ SISSA, International School for Advanced Studies, Via Bonomea 265, I-34136 Trieste TS, Italy\\
$^{8}$ INFN, Sezione di Trieste, Via Valerio 2, I-34127 Trieste TS, Italy\\
$^{9}$ INAF-Osservatorio Astronomico di Trieste, Via G. B. Tiepolo 11, I-34131 Trieste, Italy\\
$^{10}$ Universidad de la Laguna, E-38206, San Crist\'{o}bal de La Laguna, Tenerife, Spain\\
$^{11}$ Instituto de Astrof\'{i}sica de Canarias. Calle V\'{i}a L\`{a}ctea s/n, 38204, San Crist\'{o}bal de la Laguna, Tenerife, Spain\\
$^{12}$ Istituto Nazionale di Astrofisica (INAF) - Osservatorio di Astrofisica e Scienza dello Spazio (OAS), Via Gobetti 93/3, I-40127 Bologna, Italy\\
$^{13}$ Dipartimento di Fisica e Astronomia, Universit\'a di Bologna, Via Gobetti 93/2, I-40129 Bologna, Italy\\
$^{14}$ INFN-Sezione di Bologna, Viale Berti Pichat 6/2, I-40127 Bologna, Italy\\
$^{15}$ INAF-IASF Bologna, Via Piero Gobetti 101, I-40129 Bologna, Italy\\
$^{16}$ Universit\"ats-Sternwarte M\"unchen, Fakult\"at f\"ur Physik, Ludwig-Maximilians-Universit\"at M\"unchen, Scheinerstrasse 1, 81679 M\"unchen, Germany\\
$^{17}$ Max Planck Institute for Extraterrestrial Physics, Giessenbachstr. 1, D-85748 Garching, Germany\\
$^{18}$ IFPU, Institute for Fundamental Physics of the Universe, via Beirut 2, 34151 Trieste, Italy\\
$^{19}$ Department of Astronomy, University of Geneva, ch. d'\'Ecogia 16, CH-1290 Versoix, Switzerland\\
$^{20}$ INFN-Sezione di Roma Tre, Via della Vasca Navale 84, I-00146, Roma, Italy\\
$^{21}$ Department of Mathematics and Physics, Roma Tre University, Via della Vasca Navale 84, I-00146 Rome, Italy\\
$^{22}$ INAF-Osservatorio Astronomico di Roma, Via Frascati 33, I-00078 Monteporzio Catone, Italy\\
$^{23}$ INAF-Osservatorio Astronomico di Capodimonte, Via Moiariello 16, I-80131 Napoli, Italy\\
$^{24}$ Dipartimento di Fisica e Scienze della Terra, Universit\'a degli Studi di Ferrara, Via Giuseppe Saragat 1, I-44122 Ferrara, Italy\\
$^{25}$ INAF, Istituto di Radioastronomia, Via Piero Gobetti 101, I-40129 Bologna, Italy\\
$^{26}$ Institut de Recherche en Astrophysique et Plan\'etologie (IRAP), Universit\'e de Toulouse, CNRS, UPS, CNES, 14 Av. Edouard Belin, F-31400 Toulouse, France\\
$^{27}$ INFN-Sezione di Torino, Via P. Giuria 1, I-10125 Torino, Italy\\
$^{28}$ Dipartimento di Fisica, Universit\'a degli Studi di Torino, Via P. Giuria 1, I-10125 Torino, Italy\\
$^{29}$ INAF-Osservatorio Astrofisico di Torino, Via Osservatorio 20, I-10025 Pino Torinese (TO), Italy\\
$^{30}$ Universit\'{e} C\^{o}te d'Azur, Observatoire de la C\^{o}te d'Azur, CNRS, Laboratoire Lagrange, Bd de l'Observatoire, CS 34229, 06304 Nice cedex 4, France\\
$^{31}$ INAF-IASF Milano, Via Alfonso Corti 12, I-20133 Milano, Italy\\
$^{32}$ Institut de F\'{i}sica d'Altes Energies (IFAE), The Barcelona Institute of Science and Technology, Campus UAB, 08193 Bellaterra (Barcelona), Spain\\
$^{33}$ Instituto de Astrof\'isica e Ci\^encias do Espa\c{c}o, Faculdade de Ci\^encias, Universidade de Lisboa, Tapada da Ajuda, PT-1349-018 Lisboa, Portugal\\
$^{34}$ Institute of Space Sciences (ICE, CSIC), Campus UAB, Carrer de Can Magrans, s/n, 08193 Barcelona, Spain\\
$^{35}$ Institut d'Estudis Espacials de Catalunya (IEEC), 08034 Barcelona, Spain\\
$^{36}$ AIM, CEA, CNRS, Universit\'{e} Paris-Saclay, Universit\'{e} Paris Diderot, Sorbonne Paris Cit\'{e}, F-91191 Gif-sur-Yvette, France\\
$^{37}$ Observatoire de Sauverny, Ecole Polytechnique F\'ed\'erale de Lau- sanne, CH-1290 Versoix, Switzerland\\
$^{38}$ Department of Physics "E. Pancini", University Federico II, Via Cinthia 6, I-80126, Napoli, Italy\\
$^{39}$ INFN section of Naples, Via Cinthia 6, I-80126, Napoli, Italy\\
$^{40}$ Centre National d'Etudes Spatiales, Toulouse, France\\
$^{41}$ Institute for Astronomy, University of Edinburgh, Royal Observatory, Blackford Hill, Edinburgh EH9 3HJ, UK\\
$^{42}$ University of Nottingham, University Park, Nottingham NG7 2RD, UK\\
$^{43}$ European Space Agency/ESRIN, Largo Galileo Galilei 1, 00044 Frascati, Roma, Italy\\
$^{44}$ ESAC/ESA, Camino Bajo del Castillo, s/n., Urb. Villafranca del Castillo, 28692 Villanueva de la Ca\~nada, Madrid, Spain\\
$^{45}$ Univ Lyon, Univ Claude Bernard Lyon 1, CNRS/IN2P3, IP2I Lyon, UMR 5822, F-69622, Villeurbanne, France\\
$^{46}$ University of Lyon, UCB Lyon 1, CNRS/IN2P3, IUF, IP2I Lyon, France\\
$^{47}$ Departamento de F\'isica, Faculdade de Ci\^encias, Universidade de Lisboa, Edif\'icio C8, Campo Grande, PT1749-016 Lisboa, Portugal\\
$^{48}$ Instituto de Astrof\'isica e Ci\^encias do Espa\c{c}o, Faculdade de Ci\^encias, Universidade de Lisboa, Campo Grande, PT-1749-016 Lisboa, Portugal\\
$^{49}$ Aix-Marseille Univ, CNRS, CNES, LAM, Marseille, France\\
$^{50}$ Department of Physics, Oxford University, Keble Road, Oxford OX1 3RH, UK\\
$^{51}$ Department of Physics \& Astronomy, University of Sussex, Brighton BN1 9QH, UK\\
$^{52}$ INFN-Bologna, Via Irnerio 46, I-40126 Bologna, Italy\\
$^{53}$ Department of Physics, P.O. Box 64, 00014 University of Helsinki, Finland\\
$^{54}$ Department of Physics and Helsinki Institute of Physics, Gustaf H\"allstr\"omin katu 2, 00014 University of Helsinki, Finland\\
$^{55}$ Dipartimento di Fisica "Aldo Pontremoli", Universit\'a degli Studi di Milano, Via Celoria 16, I-20133 Milano, Italy\\
$^{56}$ INFN-Sezione di Milano, Via Celoria 16, I-20133 Milano, Italy\\
$^{57}$ Jet Propulsion Laboratory, California Institute of Technology, 4800 Oak Grove Drive, Pasadena, CA, 91109, USA\\
$^{58}$ von Hoerner \& Sulger GmbH, Schlo{\ss}Platz 8, D-68723 Schwetzingen, Germany\\
$^{59}$ Max-Planck-Institut f\"ur Astronomie, K\"onigstuhl 17, D-69117 Heidelberg, Germany\\
$^{60}$ Aix-Marseille Univ, CNRS/IN2P3, CPPM, Marseille, France\\
$^{61}$ Institut de Physique Nucl\'eaire de Lyon, 4, rue Enrico Fermi, 69622, Villeurbanne cedex, France\\
$^{62}$ Universit\'e de Gen\`eve, D\'epartement de Physique Th\'eorique and Centre for Astroparticle Physics, 24 quai Ernest-Ansermet, CH-1211 Gen\`eve 4, Switzerland\\
$^{63}$ Institute of Theoretical Astrophysics, University of Oslo, P.O. Box 1029 Blindern, N-0315 Oslo, Norway\\
$^{64}$ NOVA optical infrared instrumentation group at ASTRON, Oude Hoogeveensedijk 4, 7991PD, Dwingeloo, The Netherlands\\
$^{65}$ Argelander-Institut f\"ur Astronomie, Universit\"at Bonn, Auf dem H\"ugel 71, 53121 Bonn, Germany\\
$^{66}$ Institute for Computational Cosmology, Department of Physics, Durham University, South Road, Durham, DH1 3LE, UK\\
$^{67}$ Mullard Space Science Laboratory, University College London, Holmbury St Mary, Dorking, Surrey RH5 6NT, UK\\
$^{68}$ Space Science Data Center, Italian Space Agency, via del Politecnico snc, 00133 Roma, Italy\\
$^{69}$ INFN-Padova, Via Marzolo 8, I-35131 Padova, Italy\\
$^{70}$ Dipartimento di Fisica e Astronomia ``G.Galilei'', Universit\'a di Padova, Via Marzolo 8, I-35131 Padova, Italy\\
$^{71}$ Departamento de F\'isica, FCFM, Universidad de Chile, Blanco Encalada 2008, Santiago, Chile\\
$^{72}$ Institut d'Astrophysique de Paris, 98bis Boulevard Arago, F-75014, Paris, France\\
$^{73}$ Centro de Investigaciones Energ\'eticas, Medioambientales y Tecnol\'ogicas (CIEMAT), Avenida Complutense 40, 28040 Madrid, Spain\\
$^{74}$ Universidad Polit\'ecnica de Cartagena, Departamento de Electr\'onica y Tecnolog\'ia de Computadoras, 30202 Cartagena, Spain\\
$^{75}$ Infrared Processing and Analysis Center, California Institute of Technology, Pasadena, CA 91125, USA\\
$^{76}$ Jodrell Bank Centre for Astrophysics, School of Physics and Astronomy, University of Manchester, Oxford Road, Manchester M13 9PL, UK\\
$^{77}$ Department of Physics and Astronomy, University College London, Gower Street, London WC1E 6BT, UK\\

\bsp	
\label{lastpage}
\end{document}